\documentclass[12pt]{article}
\usepackage{amsmath}
\usepackage{graphicx}
\usepackage{enumerate}
\usepackage{natbib}
\usepackage{url} 

\usepackage{amssymb}
\usepackage{amsthm}         
\usepackage{amssymb}
\usepackage{mathrsfs}       
\usepackage{algorithm} 
\usepackage{algorithmic}
\usepackage{graphicx}
\usepackage{booktabs}
\usepackage[colorlinks,bookmarksopen,bookmarksnumbered,citecolor=blue,urlcolor=blue,linkcolor=blue]{hyperref}

\usepackage{xr}
\newcommand{\blind}{1}

\newtheorem{theorem}{Theorem}
\newtheorem{assumption}{Assumption}
\newtheorem{definition}{Definition}
\newtheorem{proposition}{Proposition}

\newtheorem{hypothesis}{Hypothesis}

\begin{document}

\def\spacingset#1{\renewcommand{\baselinestretch}%
{#1}\small\normalsize} \spacingset{1}

\date{}

\if1\blind
{
  \title{\bf Time-Varying Home Field Advantage in Football: Learning from a Non-Stationary Causal Process}
  \author{Minhao Qi
    \\
    School of Management, Center for Data Science, \\
    Zhejiang University\\
    Hengrui Cai \\
       Department of Statistics, 
     University of California Irvine \\
     Guanyu Hu \\
     Department of Biostatistics and Data Science\\
     The University of Texas Health Science Center at Houston \\
     and \\
     Weining Shen \\
     Department of Statistics, 
     University of California Irvine \\
     }
  \maketitle
} \fi

\if0\blind
{
  \bigskip
  \bigskip
  \bigskip
  \begin{center}
    {\LARGE\bf Time-Varying Home Field Advantage in Football: Learning from a Non-Stationary Causal Process}
\end{center}
  \medskip
} \fi

\bigskip
\begin{abstract}
In sports analytics, home field advantage is a robust phenomenon where the home team wins more games than the away team. However, discovering the causal factors behind home field advantage presents unique challenges due to the non-stationary, time-varying environment of sports matches.  In response, we propose a novel causal discovery method, DYnamic Non-stAtionary local M-estimatOrs (DYNAMO), to learn the time-varying causal structures of home field advantage.  DYNAMO offers flexibility by integrating various loss functions, making it practical for learning linear and non-linear causal structures from a general class of non-stationary causal processes. By leveraging local information, we provide theoretical guarantees for the identifiability and estimation consistency of non-stationary causal structures without imposing additional assumptions. Simulation studies validate the efficacy of DYNAMO in recovering time-varying causal structures. We apply our method to high-resolution event data from the 2020-2021 and 2021-2022 English Premier League seasons, during which the former season had no audience presence. Our results reveal intriguing, time-varying, team-specific field advantages influenced by referee bias, which differ significantly with and without crowd support. Furthermore, the time-varying causal structures learned by our method improve goal prediction accuracy compared to existing methods.
\end{abstract}

\noindent%
\vfill

\newpage
\spacingset{1.9} 
\section{Introduction}
\label{sec:intro}
Home field advantage is a well-documented phenomenon in sports where home teams win more games than away teams under a balanced home and away schedule \citep{courneya1992home}. Although this phenomenon has been explored in various sports, including both team sports (e.g., football and basketball) and individual sports (e.g., tennis and swimming), the precise mechanisms underlying team-level home advantages remain unclear \citep{schwartz1977home,HomeAdvantage1,goumas2017modelling,marek2020comparison}. A striking illustration comes from football games during the COVID-19 era: in the 2021–2022 English Premier League season (with spectators), home teams won 43.0\% of matches, 9 percentage points more than the away teams. In contrast, when matches were held behind closed doors in 2020-2021, home win rates fell to 37.9\%, which was even lower than 40.3\% of the away teams.

Prior studies investigating the origins of home field advantage have employed survey methods and natural experiments, yielding mixed conclusions.  One strand of the literature attributes the advantage primarily to referee bias \citep{johnston2008referee,endrich2020home, bryson2021causal}, while others emphasize differences in player performance factors—such as familiarity with the playing environment, crowd support, and travel fatigue—as key contributors \citep{pollard2006worldwide,goumas2017modelling}.  However, most analyses have been limited to league-level aggregates, obscuring team-specific dynamics and failing to capture the inherently nonstationary nature of match‐level data. The recent emergence of high-resolution tracking and event data has created fresh opportunities for reliable sports analysis, significantly broadening the statistical methods available in the field \citep{ cervone2016multiresolution,albert2017handbook,hu2021cjs,grieshop2023continuous}. 


In this paper, we use sports-tracking event data to identify factors that influence home field advantage and quantify their time-varying influences. However, this task poses several challenges. First, a football match represents a dynamic, time-varying process that cannot be reduced to a single value or treated as a stationary process. While previous studies often rely on conditional goal differentials to estimate home field advantage \citep{goumas2017modelling,singleton2021big}, we contend that with 22 individuals influencing the outcome, even minor changes can have significant effects on match dynamics. These processes include time-varying strategies set by coaches, fluctuating mental and physical conditions of players, and external factors such as substitutions, all of which resist simplification into a stationary framework. Second, the time-varying patterns of these match processes are difficult to predict with prior knowledge, as home field advantage can vary significantly across teams, leagues, and regions. Finally, the relationships between within-match variables are complex and heterogeneous, and relying solely on regressing match variables on home and away teams' expected goals may result in significant bias.

To address the aforementioned challenges,  we introduce a causal structure learning approach that incorporates local stationarity called \textbf{DYNAMO}, that is, \textit{DYnamic Non-stAtionary local M-estimatOrs for structure learning}. Our premise is based on the key idea that time-varying match processes during proximal times can be similar due to the similar environments created by the 22 players on the field. These match processes can be modeled as locally stationary causal processes, where causal structures are allowed to evolve smoothly over time and can be approximated by their stationary counterparts within the vicinity of each time point. This approach enables us to leverage local information for recovering time-varying causal structures without the need for specific assumptions or adherence to particular classes of time-varying patterns. By utilizing DYNAMO, we can estimate the direct effects of playing on the home field on the in-game variables and identify the pathway leading to the increased expected goals, representing the outcome of home field advantage. 

Below, we summarize the findings of the existing literature and discuss our contributions.

{\it Literature on sports analytics:} Existing research often utilizes linear regression models to analyze game-level summary statistics, treating each match as a single observation \citep{pollard2006worldwide,johnston2008referee,endrich2020home,bryson2021causal}. These models impose strong assumptions on the linearity and homogeneity of home field advantages across match periods, overlooking more complex within-match information. This oversimplification can lead to data under-fitting and provide limited insights into the underlying causal relationships \citep{price2022much}. Moreover, in identifying the factors that influence home field advantages, current mediation analysis often assumes mediator conditional independence given exposure or overlooks potential directed paths among mediators  \citep{shi2022testing}. In football matches, however, determinant factors often influence each other. For example, a referee's decisions may affect the total shots taken by the home team, thereby contributing to the home field advantage.

{\it Literature on causal discovery:} Causal structure learning \citep[see the review in][]{vowels2022d} has emerged as a key tool for uncovering complex causal relationships among variables, offering the potential to understand the home field advantage and its influencing factors. However, several challenges arise due to the dynamic non-stationary environment of the football match processes. Firstly, existing algorithms mostly rely on stationary assumptions \citep{pamfil2020dynotears,gong2022rhino}, which are overly restrictive for time series, as common trends and cycles can introduce non-stationary dependencies \citep{runge2023causal}. Match processes, which include time-varying strategies devised by coaches, the dynamic mental and physical conditions of players, and external factors such as referee's decisions, often do not adhere to the stationary assumption. Secondly, existing algorithms for non-stationary time series either focus on changing causal strengths while maintaining a constant causal structure \citep{huang2019causal}, or make strict assumptions regarding time-varying patterns, such as abrupt changes across regimes \citep{yang2022segment,rahmani2023castor} and periodical structures \citep{gao2023causal}. Violations of these assumptions can lead to erroneous causal discoveries, particularly given the limited understanding of the specific time-varying patterns of causal structures in football matches. \citet{huang2020causal} propose a method using a time surrogate variable to represent non-stationary causal structures, albeit with strict identifiability assumptions. These assumptions include considering confounders as functions of time and incorporating two versions of the faithfulness assumptions.  However, their method primarily focuses on a static graph where time serves as a node, providing limited insights into the time-varying individual graphs that may reflect heterogeneous home field advantages over time.

{\it Literature on locally stationary processes:} The literature on non-stationary processes has received considerable attention for addressing time-varying model structures in time series analysis, especially following the seminal work of \citet{dahlhaus1996kullback} on locally stationary processes. Existing literature explores a broad class of time-series models, including time-varying ARMA, time-varying GARCH, time-varying VAR, and multivariate GARCH \citep{yan2021time,karmakar2022simultaneous,gao2024time}. However, these models often fail to capture the underlying nonlinear causal structures in instantaneous periods. In the context of football matches, where player-influenced environments can exhibit consistent complex patterns during proximal times, assuming linearity in these relationships might be overly restrictive. The occurrence of a goal, for example, involves several actions within a short period, with both instantaneous and lagged factors playing crucial roles. This strand of literature predominantly focuses on statistical models capturing linear relationships during the lagged periods, rather than on causal models that can elucidate nonlinear causal structures in both lagged and instantaneous periods.

{\it Our methodological contributions:} First, we develop a causal discovery method capable of estimating time-varying causal structures without requiring the stationarity assumption for discovering complex structure of home field advantages in football games. DYNAMO offers flexibility by integrating with various loss functions, making it practical for learning non-linear and other application-specific causal structures.  Secondly, we establish the theoretical properties of our method regarding the identifiability of non-stationary causal processes and the consistency of the estimator. By leveraging local information, our framework effectively addresses identifiability problems without relying on faithfulness assumptions and constant Gaussian noise assumptions.

{\it Main findings and impact:} Our study uncovers significant differences in team performance and referee behavior between the 2020–2021 and 2021–2022 seasons, largely driven by the presence or absence of spectators. We identify dynamic patterns of home field advantage and time-varying referee bias, showing that officiating is not uniform but shaped by team style, match timing, and crowd influence. Teams like Liverpool and Arsenal benefit more consistently from favorable decisions when fans are present, while dominant teams like Manchester City appear to trigger early cautioning of opponents. Moreover, non-relegated and higher-ranked teams experienced a shift from opponent fouls in empty stadiums to more yellow cards with crowd return, reflecting the psychological impact of spectators on referees. In contrast, lower-ranked teams saw minimal change.

This research provides nuanced insights into how potential referee bias interacts with team strategy, match dynamics, and crowd influence, contributing to our understanding of home field advantage. These insights may help teams adjust tactics and manage players more effectively. The findings also have implications for referee training and policy discussions aimed at improving fairness.  Integrating these patterns into sports analytics could enhance performance forecasting and enrich fan engagement. Beyond sports, the study offers perspectives on how social and environmental pressures can shape decision-making, with relevance to fields like psychology, law and organizational behavior.

\section{Motivating Data}\label{sec:data}

This study is based on a collaboration with Hudl \& Wyscout, a company that excels at football game scouting and match analysis \citep{price2022much}. The sports event data comprises over 1.6 million within-game events from 760 games during the two English Premier League (EPL) seasons, 2020-2021 and 2021-2022, including passes, shots, dribbles, tackles, penalties, and more. For demonstration, we focus on four representative EPL teams: Manchester City, Liverpool, Arsenal, and Manchester United. During the 2020-2021 season, spectators were not allowed due to COVID-19 restrictions. Manchester City ranked 1st with an 8-point lead over 2nd place Manchester United. Liverpool ranked 3rd, while Arsenal finished 8th. In the 2021-2022 season, with stadiums open to spectators, Manchester City secured the 1st rank, and Liverpool ranked 2nd, with a narrow 1-point gap between them and an 18-point gap to the third-ranked team. Arsenal and Manchester United ranked 5th and 6th, respectively, with Arsenal just 2 points behind the 4th-ranked team and Manchester United trailing by 13 points.

\begin{figure}
    \begin{center}
    \centerline{\includegraphics[width=0.95\textwidth]{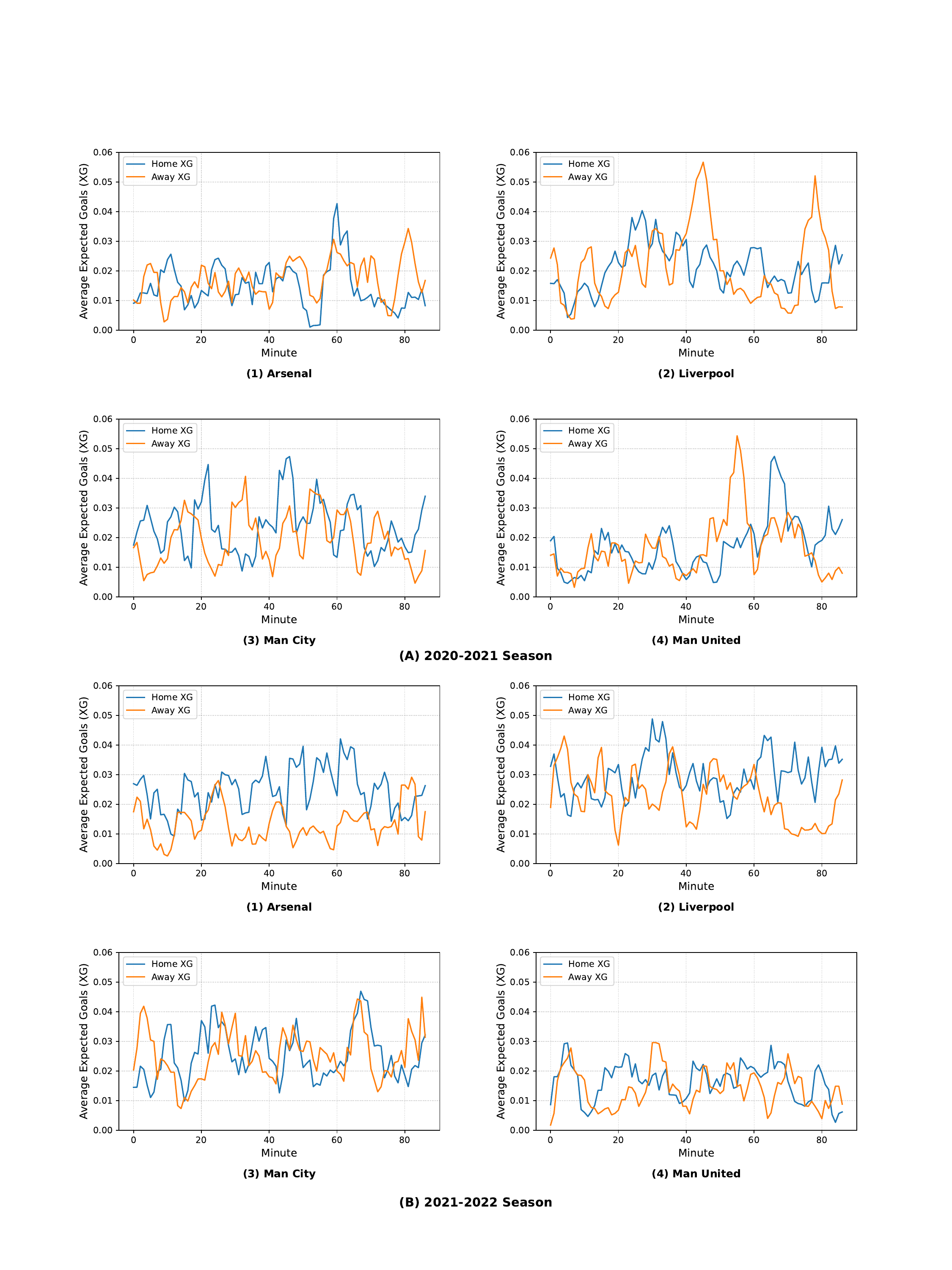}}
    \vspace{-1.2cm}
    \caption{The time-varying average expected goals of 4 EPL teams, Arsenal, Liverpool, Manchester City, and Manchester United. The first panel presents the time-varying average expected goals in both home matches and away matches during the 2020-2021 season, while the second panel presents the time-varying expected goals during the 2021-2022 season.}
    \label{fig:xg4teams}
    \vspace{-0.8cm}
    \end{center}
\end{figure}

Figure~\ref{fig:xg4teams} reveals distinct patterns in expected goals (XG) during match time between home and away matches that differ by season, likely reflecting the impact of spectator presence. The first row presents match results averaged over all games played during the 2020-2021 season, and the second row presents results for the 2021-2022 season. There is a sharp distinction between the two seasons, with the four teams exhibiting distinct time-varying patterns in average XG in home and away matches.  In the 2020–2021 season, played behind closed doors, Arsenal’s home and away XG were overall similar, with home XG only marginally higher between the 55th and 65th minutes. In contrast, during the 2021-2022 season, when crowds were present, Arsenal’s home XG consistently exceeded their away XG, except for a convergence between the 76th and 90th minutes. Liverpool exhibited even more pronounced seasonal differences. In the 2020–2021 season, their home and away XG profiles were nearly identical, with a transient negative home advantage observed between the 40th and 50th minutes. However, in the 2021-2022 season, Liverpool showed a clear home field advantage, most notably near the end of matches, despite starting each half with a lower home XG compared to away matches. For Manchester City, the distribution of XG in home and away matches remained largely consistent throughout both seasons, with the home advantage peaking in the first half. In contrast, Manchester United showed a more complex pattern, with a marked decline in average home XG from the 2020–2021 season to the 2021–2022 season.

To identify the pathway leading to the home advantage, we focus on the following 7 in-game variables as our target variable: \textbf{Total Passes (TP)} as the overall number of passes made; \textbf{Total Shots (TS)} as the overall number of shots taken; \textbf{Pass Accuracy (PA)} as the ratio of successful passes to the total number of passes; \textbf{Shot Accuracy (SA)} as the ratio of shots hitting the ball toward the net to the total shots; \textbf{Opponent's yellow cards (OY)} as the number of yellow cards that the opponent received; \textbf{Opponent's fouls (OF)} as the number of fouls that the opponent committed; \textbf{Expected Goals (XG)} as the average likelihood of scoring a goal, considering the player's position throughout the game.


We also incorporate several with-in-match control variables following previous literature \citep{price2022much}, including key game events such as key passes, dribbles, tackles, etc. To mitigate potential confounding factors related to the opponent's abilities, we use the differences in minute-by-minute mean match statistics for each team as input variables. The detailed data preprocess is presented in the Appendix~D.

\section{Method}\label{sec:method}
In this section, we present a comprehensive framework utilizing local M-estimators for the inference of \textit{linear} or \textit{nonlinear} time-varying causal structures within a non-stationary process $\{\boldsymbol{x}_t\}$. We introduce two methodologies tailored for linear and nonlinear relationships. Then we introduce their computational algorithms and discuss the selection of kernel bandwidth. Finally, we provide the theoretical guarantee of our method.

\subsection{Model}

Let $X = [\boldsymbol{x}_1,\boldsymbol{x}_2,\cdots,\boldsymbol{x}_T]^\top  \in \mathcal{R}^{T\times d} $ be a multivariate time series with $d$ variables at discrete time points $t = 1, \ldots, T$. At each time $t$, we assume that the vector $\boldsymbol{x}_t \in \mathcal{R}^{d}$ can be affected by both instantaneous variables $\boldsymbol{x}_t$ and its lagged version $Y_{t-1} = [\boldsymbol{x}_{t-1}^\top,\boldsymbol{x}_{t-2}^\top,\cdots,\boldsymbol{x}_{t-L}^\top]^\top \in \mathcal{R}^{Ld \times 1}$, where $L$ is the maximum lagged period. Specifically, for each time $t \in [0, T]$, there exists a graph $\mathcal{G}_t = (V_t,\boldsymbol{\theta}_t)$ representing the relationships among $\boldsymbol{x}_t$ and the lagged variables, where $V_t = \{x_t^{[1]},\cdots,x_t^{[d]},\cdots,x_{t-L}^{[1]},\cdots,x_{t-L}^{[d]} \}$ is the set of nodes and $\boldsymbol{\theta}_t$ is the model parameters containing the set of edges. For variables $\{x_{t_0}^{[i]},x_{t_1}^{[j]};t_0 \leq t_1\} \subset V_t$, we define $x_{t_0}^{[i]} \in V_t$ as the parent of $x_{t_1}^{[j]}$ if there exists a directed edge from $x_{t_0}^{[i]}$ to $x_{t_1}^{[j]}$. 

\begin{assumption}[\textbf{Unconfoundness}]
    There are no unobserved confounders.
    \label{asm:unconfoundness}
\end{assumption}

\begin{assumption}[\textbf{Acyclicity}]
    The graph $\mathcal{G}_t$ is a directed acyclic graph (DAG) with no variable being the ancestor of itself at each time $t$. 
    \label{asm:acyclicity}
\end{assumption}
Assumptions \ref{asm:unconfoundness} and \ref{asm:acyclicity} carry the fundamental beliefs of causal sufficiency and stability in causality, and they have been widely used in the literature \citep{pamfil2020dynotears,bello2022dagma,shi2022testing}. The goal of this paper is to estimate the causal structure $\mathcal{G}_t$ of the non-stationary process $\{ \boldsymbol{x}_t \}$ at time $t$, accounting for variations in both the \textit{causal structure} and the \textit{causal strength} across time. 

Specifically, we consider a class of time-varying causal processes as follows
\begin{align}
        \boldsymbol{x}_t :=f(Pa(\boldsymbol{x}_t),\boldsymbol{\epsilon}_t;\boldsymbol{\theta}(\tau_t)),~~~ \text{for} ~ t = 1,\ldots, T,
        \label{eq:non-stationary}
\end{align}
where $\tau_t = t/T$, $Pa(\boldsymbol{x}_t) \subseteq \{\boldsymbol{x}_t,\boldsymbol{x}_{t-1},\ldots,\boldsymbol{x}_{t-L}\} \in \mathcal{R}^{(L+1) \times d}$ is the parent set of $\boldsymbol{x}_t$, $\boldsymbol{\theta}(\tau_t)$ is the model parameter describing the causal structure of $\boldsymbol{x}_t$ that varies across time points $t$, and $\{ \boldsymbol{\epsilon}_t \}$ is a sequence of jointly independent errors. The function $f(\cdot)$ is measurable, with a known form; if linear, each $\boldsymbol{\theta}(\tau_t)$ corresponds uniquely to a graph $\mathcal{G}_t$. For nonlinear $f(\cdot)$, each $\boldsymbol{\theta}(\tau_t)$ encompasses a distinct causal structure $\mathcal{G}_t$. Hence, we refer to $\boldsymbol{\theta}(\tau_t)$ as the causal structure. 

We allow the causal structure to vary across time, such that there exist at least two different time points $t_0$, $t_1$, satisfying $\boldsymbol{\theta}(\tau_{t_0}) \neq \boldsymbol{\theta}(\tau_{t_1}),\mathrm{~} \{t_0,t_1\}\subset[L+1,T]$.
This configuration aligns with the concept of a \textit{non-stationary causal process} as outlined in the literature \citep{gao2023causal}. Thus, the process $\{ \boldsymbol{x}_t \}$ follows shifting causal structures over $t = 1,\ldots, T$. For example, the sequence of $\boldsymbol{x}_1,\boldsymbol{x}_2,\ldots, \boldsymbol{x}_L$ follows causal structures $\boldsymbol{\theta}(\tau_{1}),\boldsymbol{\theta}(\tau_{2}), \ldots, \boldsymbol{\theta}(\tau_{L}) $, respectively. 
Our proposed model encompasses a multitude of existing \textit{time-invariant / stationary } frameworks as specific instances, allowing for the learning of both \textit{linear} and \textit{nonlinear} causal relationships. For example, it encapsulates additive noise models \citep{rolland2022score}, post-nonlinear causal model \citep{gong2022rhino}, non-Gaussian DAG \citep{fujiwara2023causal} and Gaussian DAG \citep{peters2014identifiability}.

\subsection{Stationary Approximation}


Estimation of Eq.~\eqref{eq:non-stationary} is challenging due to the variability in $\boldsymbol{\theta}(\tau_{t})$ over time $t$, and the limitation of having only a single observation per time point. To address this, we approximate $\boldsymbol{x}_t:=f(Pa(\boldsymbol{x}_t),\boldsymbol{\epsilon}_t;\boldsymbol{\theta}(\tau_t))$ of a non-stationary process $\{\boldsymbol{x}_t , t = 1,2,\ldots, T \}$ at a specific time $t$  by constructing a stationary processes $\{\widetilde{\boldsymbol{x}}_{t^{\prime}}(\tau_t), t^{\prime} = 1,2,\ldots,T  \}$. For a fixed $t$, each time $t^{\prime} \in [ 1,2,\ldots,T]$ in this stationary process follows the same causal structure, i.e., $\widetilde{\boldsymbol{x}}_{t^{\prime}}(\tau_t)=f\left(Pa(\widetilde{\boldsymbol{x}}_{t^{\prime}}(\tau_t)),\boldsymbol{\epsilon}_{t^{\prime}};\boldsymbol{\theta}(\tau_t)\right)$. If we can construct this approximated stationary process  $\{ \widetilde{\boldsymbol{x}}_1(\tau_t), \widetilde{\boldsymbol{x}}_2(\tau_t), \ldots, \widetilde{\boldsymbol{x}}_T(\tau_t) \}$ for each $t \in [1,2,\ldots,T]$, we are able to estimate $\theta(\tau_t)$ point-wisely.


To provide identifiable conditions on the model \eqref{eq:non-stationary}, we introduce a set of mild assumptions concerning the non-stationary causal process $\{ \boldsymbol{x}_t \}$, motivated by the prior works on locally stationary process \citep{Dahlhaus2019towards,gao2024time}.

\begin{assumption}[\textbf{Locally Stationary Causality}]
    \label{asm:local_stationary}
    A non-stationary causal process $\{ \boldsymbol{x}_t \}$ in equation \eqref{eq:non-stationary} is a locally stationary causal process if it satisfies:\\
    1. For $\forall \boldsymbol{Z}, \boldsymbol{Z}^\prime \in \mathbb{R}^{(L+1) \times d}$ and $\forall\boldsymbol{\vartheta}$, there exist non-negative sequences $\{\alpha_j (\boldsymbol{\vartheta}) \}_{j=0}^{L}$ such that 
        \begin{equation*} \|f(\boldsymbol{Z},\boldsymbol{\epsilon};\boldsymbol{\vartheta})-f(\boldsymbol{Z}^\prime,\boldsymbol{\epsilon};\boldsymbol{\vartheta})\|_q\leq\sum_{j=0}^{L}\alpha_j(\boldsymbol{\vartheta})\|\boldsymbol{z}_j-\boldsymbol{z}_j^{\prime}\|_q. \notag
        \end{equation*}
        2. For $\forall \tau \in [0,1]$, $\boldsymbol{\theta}(\tau)$ lies in $\Theta_r$, where 
        \begin{equation*}
            \Theta_r:=\{\boldsymbol{\vartheta}\in\Theta_r\mid\sum_{j=0}^{L}\alpha_j(\boldsymbol{\vartheta})<1\}. \notag
        \end{equation*}
        3. There exist nonnegative sequences $\{ \chi_j \}$ with $\sum_{j=0}^{L}\chi_{j}<\infty $ such that for $\forall \boldsymbol{Z} \in \mathbb{R}^{(L+1)\times d}$ and $\forall \boldsymbol{\vartheta}, \boldsymbol{\vartheta}^{\prime} \in \Theta_r$, 
        \begin{equation*}
            \|f(\boldsymbol{Z},\boldsymbol{\epsilon};\boldsymbol{\vartheta})-f(\boldsymbol{Z},\boldsymbol{\epsilon};\boldsymbol{\vartheta}^{\prime})\|_q \leq\|\boldsymbol{\vartheta}-\boldsymbol{\vartheta}^{\prime}\|\sum_{j=0}^{L}\chi_j\|\boldsymbol{z}_j\|_q,\notag
        \end{equation*} 
where $q > 0$ and $\|W\|_q:=(\mathbb{E}|W|^q)^{1/q}$. 
\end{assumption}
The Assumption \ref{asm:local_stationary}  guarantees a weakly dependent stationary approximation of $\{ \boldsymbol{x}_t \}$, for any $t \in [0, T]$, and places a Lipschitz-type condition on $f(\cdot)$ and its parameter space, a condition easily satisfied by a diverse array of models, including those to be studied in Section \ref{Sec:DYNAMO}. To establish the presence of an approximated stationary process  $\widetilde{\boldsymbol{x}}_t(\tau_t)$ derived from the non-stationary process $\boldsymbol{x}_t$, we introduce the following dependence measure. 

\begin{definition}[\textbf{Dependence Measure} in \citet{Dahlhaus2019towards} ]
    \label{def: dependence measure}
    Consider a stationary process $\boldsymbol{e}_t = \boldsymbol{J}(\boldsymbol{\epsilon}_t,\boldsymbol{\epsilon}_{t-1},\cdots)$ with $\boldsymbol{J}(\cdot)$ being a measurable function, the dependence measure is $\delta_q^{\boldsymbol{e}_{t}}(k)=\left\|\boldsymbol{J}(\boldsymbol{\epsilon}_k,\boldsymbol{\epsilon}_{k-1},\ldots\boldsymbol{\epsilon}_1)-\boldsymbol{J}(\boldsymbol{\epsilon}_k,\boldsymbol{\epsilon}_{k-1},\ldots,\boldsymbol{\epsilon}_1^*)\right\|_q,$
    where $\boldsymbol{\epsilon}_1^*$ is an independent copy of $\boldsymbol{\epsilon}_1$, and $\delta_q^{\boldsymbol{e}_{t}}(k)$ quantifies the dependence of $\boldsymbol{e}_t$ on $\boldsymbol{\epsilon}_1$.
\end{definition}

\begin{figure}
    \begin{center}
    \centerline{\includegraphics[width=\textwidth]{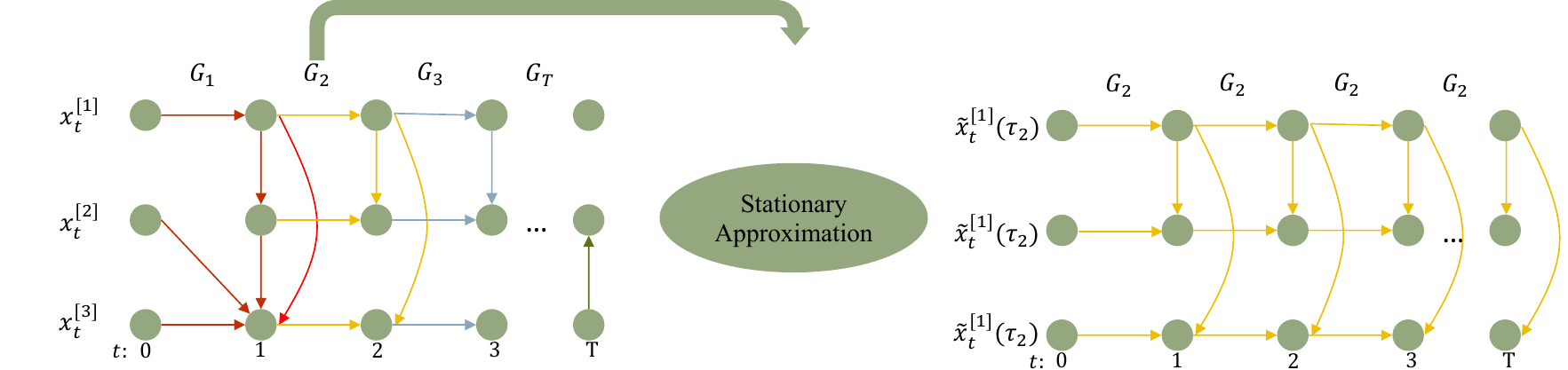}}
    \caption{Stationary approximation of non-stationary causal processes $\{\boldsymbol{x}_t\}$ at time $t = 2$. The left panel presents the non-stationary causal structures $\mathcal{G}_1,\ldots, \mathcal{G}_T$, where different colors of edges represent different causal structures (with lag $L=1$). The right panel displays its stationary approximation $\{\tilde{\boldsymbol{x}}_t(\tau_2)\}$ with a stationary causal structure $\mathcal{G}_2$.}
    \label{fig:locally_stationarity}
    \vspace{-0.4cm}
    \end{center}
\end{figure}

\begin{proposition}[\textbf{Existence of an Approximated Stationary Process}]
    \label{prp:stationary_process}
    Suppose Assumption \ref{asm:local_stationary} holds, for $\forall \tau \in [0,1]$  there exists a stationary process:
    \begin{align}
        \label{eq:stationary_process}
        \widetilde{\boldsymbol{x}}_t(\tau)=f\left(Pa(\widetilde{\boldsymbol{x}}_t(\tau)),\boldsymbol{\epsilon}_t;\boldsymbol{\theta}(\tau)\right),
    \end{align}
    where $\tau = t/T$, $Pa(\widetilde{\boldsymbol{x}}_t(\tau)) \subseteq \{\widetilde{\boldsymbol{x}}_{t}(\tau), \widetilde{\boldsymbol{x}}_{t-1}(\tau),\ldots,\widetilde{\boldsymbol{x}}_{t-L}(\tau)\} \in \mathcal{R}^{(L+1) \times d}$, such that:  
    
    1. $\widetilde{\boldsymbol{x}}_t(\tau)$ admits a causal representation $\widetilde{\boldsymbol{x}}_t(\tau) = \boldsymbol{J}(\tau, \boldsymbol{\epsilon}_t,\boldsymbol{\epsilon}_{t-1},\ldots\boldsymbol{\epsilon}_1)$ with $\boldsymbol{J}(\cdot)$ being a measurable function, and $\sup_{\tau\in[0,1]}\left\|\widetilde{\boldsymbol{x}}_t(\tau)\right\|_r<\infty;$
    
    2. $\sup_{\tau\in [0,1]}\delta_r^{\boldsymbol{\tilde{x}}(\tau)}(t) \rightarrow 0$ as $t \rightarrow \infty$. 
\end{proposition}
Proposition~\ref{prp:stationary_process} demonstrates that for each time $t \in [0,T]$, there exists a bounded stationary process $\{\widetilde{\boldsymbol{x}}_t(\tau)\}$ that possesses an invariant causal structure $\boldsymbol{\theta}(\tau_t)$, identical to the causal structure of non-stationary $\{\boldsymbol{x}_t\}$ at time $t$.  An example with three variables is illustrated in Figure~\ref{fig:locally_stationarity}. Moreover, the dependence of the stationary process $\{\widetilde{\boldsymbol{x}}_t(\tau)\}$ on $\epsilon_1$ tends to 0 as $t$ increases to infinity. The intuition behind the proof is that there exists a Cauchy sequence that converges to $\widetilde{\boldsymbol{x}}_t(\tau)$ and the dependence of $\widetilde{\boldsymbol{x}}_t(\tau)$ on $\epsilon_1$ diminishes to 0 as $t \rightarrow \infty$. The complete proof is provided in Appendix C.

\begin{proposition}[\textbf{Stationary Approximation}]
    \label{prp:stationary_approx}
    Suppose Assumption \ref{asm:local_stationary} holds. Then for any $q > 0$,\\
    1. $\|\widetilde{\boldsymbol{x}}_t(\tau)-\widetilde{\boldsymbol{x}}_t(\tau^{\prime})\|_q=O(|\tau-\tau^{\prime}|)$ for $\forall\tau,\tau^{\prime}\in[0,1]$;\\
    2. $\max_{t\geq1}\|\boldsymbol{x}_t-\widetilde{\boldsymbol{x}}_t(\tau_t)\|_q=O(T^{-1}),$ where $\tau_t = t/T$.
\end{proposition} 
The proof of the Proposition~\ref{prp:stationary_approx} involves computing the difference between causal structures $\boldsymbol{\theta}(\tau)$ and $\boldsymbol{\theta}(\tau^{\prime})$, as well as the differences between the corresponding data. The detailed proofs are provided in Appendix C. Proposition~\ref{prp:stationary_approx} enables the substitution of $\boldsymbol{x}_t$ with the stationary approximation $\widetilde{\boldsymbol{x}}_t(\tau)$, achieving a convergence rate of $|t/T-\tau|+T^{-1}$. While $\widetilde{\boldsymbol{x}}_t(\tau)$ is unobservable, we can approximate the stationary process $\widetilde{\boldsymbol{x}}_t(\tau)$ with a kernel function satisfying the following mild conditions using the observed data.

\begin{assumption}[\textbf{Localizing Kernel}]
    \label{asm:localizing_kernel}
    Let $K(\cdot) \in \mathcal{K}$, where $\mathcal{K}$ is the family of symmetric and positive kernel functions defined on $[-1,1]$ with $\int_{-1}^1K(u)\mathrm{d}u=1$ and bandwidth $h >0$. Moreover, $K(\cdot)$ is Lipschitz continuous on $[-1,1]$. As $(T,h) \rightarrow (\infty,0), Th \rightarrow \infty$.
\end{assumption}
Assumption~\ref{asm:localizing_kernel} comprises a set of mild conditions on the kernel function and the bandwidth, commonly utilized in the literature on locally stationary processes \citep{Dahlhaus2019towards, karmakar2022simultaneous, gao2024time}. For each $\tau_t \in (0,1)$, as $(T,h) \rightarrow (\infty,0), Th \rightarrow \infty$. So we employ a localizing kernel function to approximate the stationary process $\frac{1}{Th}\sum_{l=1}^TK_h(\tau_l-\tau_t)\cdot \boldsymbol{x}_l \to\mathbb{E} \tilde{\boldsymbol{x}}_l(\tau_t)$. Further details are provided in Appendix C. 

\subsection{DYNAMO Method}
\label{Sec:DYNAMO}
Proposition \ref{prp:stationary_approx} guarantees local stationarity for $\boldsymbol{x}_t$, allowing estimating $\boldsymbol{\theta}(\tau_t)$ by using the neighborhood around the given time point $t$, i.e., for any $\tau_{l}$ when $|\tau_{l}-\tau_t| \leq h$. Therefore, we can parameterize $\boldsymbol{\theta}(\tau_t)$. Our \textbf{DYNAMO} model is specified as follows:
\begin{align}
    \mathscr{L}_{t}(\boldsymbol{\vartheta})=\frac{1}{Th}\sum_{l=1}^T\ell(\boldsymbol{x}_l,Y_{l-1};\boldsymbol{\vartheta})K_h(\tau_l-\tau_t),
    \label{eq:constant_likelihood}
\end{align}
where $Y_{l-1} = [\boldsymbol{x}_{l-1}^\top,\boldsymbol{x}_{l-2}^\top,\cdots,\boldsymbol{x}_{l-L}^\top]^\top \in \mathcal{R}^{Ld \times 1}$, and $K_h(\cdot) = K(\cdot/h)$. Accordingly, for any $\tau_t$, $\boldsymbol{\theta}(\tau_t)$ is estimated by $\boldsymbol{\widehat{\theta}}(\tau_t) $ which is the argument minimizer of $\mathscr{L}_t(\boldsymbol{\vartheta})$ for $\boldsymbol{\vartheta}\in\Theta_r$. This framework can be easily adapted to many loss functions $\ell(\boldsymbol{x}_l,Y_{l-1};\boldsymbol{\vartheta})$. Under the linear setting, the causal structures $\boldsymbol{\theta}(\tau_t) = \{W_t, A_t\}$ can be divided into $W_t$ and $A_t$, where $W_t \in \mathcal{R}^{d \times d}$ and $A_t \in \mathcal{R}^{Ld \times d}$ are the instantaneous and lagged adjacent matrices, respectively. We utilize the NOTEARS loss function \citep{zheng2018dags,pamfil2020dynotears}, which is a least square loss with acyclic constraint $H(W_t)$. In reality, the relationships among variables can be highly nonlinear. Under the nonlinear setting, we utilize the NTS-Notear loss function \citep{sun2023nts}, which leverages neural networks to solve the continuous constrained optimization DAG learning problem.

\textbf{Linear DYNAMO } under NOTEAR loss is:
\begin{align}
    \label{eq:linear_DYNAMO}
    \mathscr{L}_{t}^{lr}(W_t,A_t)  =  \frac{1}{Th}\sum_{l=1}^T & \|\boldsymbol{x}_{l}  - W_{t}^\top \boldsymbol{x}_{l} -  A_{t}^\top Y_{l-1}\|_2^2K_h(\tau_l-\tau_t) + \lambda_{1}\|W_t \| +  \lambda_{2}\|A_t\| \\ 
    & +\frac{\rho}{2}H(W_t)^2+ \alpha H(W_t). \notag 
\end{align}

Here, $H(W_t)=\operatorname{tr}(e^{\mathbf{W_t}\circ\mathbf{W_t}})-d$ represents the trace exponential function introduced by \citet{zheng2018dags}, where the symbol $\circ$ denotes
the Hadamard product of two matrices. It equals 0 if and only if $W_t$ is acyclic. Here $\|\boldsymbol{v} \|_{q}^p := (E|\boldsymbol{v}|^q)^{p/q}$ for any $p>0$ and $q>0$, $\lambda_{1}$ and $\lambda_{2}$ control the sparsity of $W_t$ and $A_t$, respectively, while $\rho$ and $\alpha$ jointly control the acyclic constraints of $W_t$.

\textbf{Nonlinear DYNAMO} under NTS-Notear loss is:
\begin{align}
    \mathscr{L}_{t}^{nlr}(\boldsymbol{\theta}(\tau_t)) =  \frac{1}{Th}\sum_{l=1}^T & \|\boldsymbol{x}_{l}-g_t(\boldsymbol{x}_{l},Y_{l-1};\boldsymbol{\theta}(\tau_t))\|_2^2 K_h(\tau_l-\tau_t) +\lambda_1\| \boldsymbol{\theta}(\tau_t)\| \notag \\
     & +\frac{\rho}{2}H(W_t(g_t))^2+ \alpha H(W_t(g_t)),
    \label{eq:nonlinear_DYNAMO}
\end{align}
where $g_t(\cdot)$ is a specific neural network characterized by the parameter $\boldsymbol{\theta}(\tau_t)$, and $W_t(g_t)$ is the corresponding instantaneous graph based on the neural network $g_t(\cdot; \boldsymbol{\theta}(\tau_t))$. The specific structure of $g_t(\cdot)$ and $W_t(g_t)$ are presented in Appendix B.

The solving algorithms of DYNAMO are highly dependent on the base learner, NOTEAR, in our linear DYNAMO, and NTS-Notear in nonlinear DYNAMO. The algorithm for linear DYNAMO is described in Algorithm 1 of Appendix A, while the solving algorithm for nonlinear DYNAMO is presented in Algorithm~2 of Appendix  A. 

For solving linear DYNAMO, we introduce the decomposition $W_t = W_{t+} - W_{t-}$, ensuring that $W_{t+} \geq \boldsymbol{0}$ and $W_{t-} \leq \boldsymbol{0}$, to employ standard solvers such as L-BFGS-B \citep{zhu1997algorithm}. This allows us to recast the original problem into a quadratic form with twice the number of elements.

We recommend choosing the Epanechnikov kernel $K(u) = \frac{3}{4}(1-u^2)I(|u| \leq 1)$ due to its popularity in time-varying models \citep{yan2021time,gao2024time}. This kernel offers several advantages, including compact support, robustness to outliers, and efficient computational forms. Alternatively, Gaussian kernels $K(u) = 1/\sqrt{2\pi}\exp(-{u^2}/{2})$ can also be employed in our model. 

The determination of the lagged period $L$ is based on the existing literature \citep{pamfil2020dynotears}. Specifically, we consider the magnitude of the weights in the estimated lagged causal structures and choose $L$ as the number of largest lagged matrices that contain entries significantly above 0 in magnitude.

\subsection{Selection of Kernel Bandwith}
In practice, the global optimal kernel bandwidth is hindered by various unknown quantities, as explained by \citet{karmakar2022simultaneous} in their discussion on the theoretical aspects of the global optimum. We propose a ``quasi-$k$-fold-cross-validation'' method for determining the bandwidth $h$, inspired by the ``quasi-leave-one-out'' approach proposed by \citet{richter2019cross}. Our modification is motivated by computational efficiency considerations specific to our model, as the original ``quasis-leave-one-out'' method proves to be computationally slow in our setting.  We first define a ``quasi-$k$-fold-cross-validation'' local likelihood as $\mathscr{L}_{-T_k}^{t}(\boldsymbol{\vartheta})=\frac{1}{Th}\sum_{l \notin T_k}\ell(\boldsymbol{x}_l,Y_{l-1};\boldsymbol{\vartheta}),$
where $T_k$ represents the $k$-th fold subset, $\ell(\boldsymbol{x}_l,Y_{l-1};\boldsymbol{\vartheta})$ is the loss function in Eq.~\eqref{eq:linear_DYNAMO} or Eq.~\eqref{eq:nonlinear_DYNAMO}, and $ \mathscr{L}_{-T_k}^{t}(\boldsymbol{\vartheta})$ denotes the local likelihood without the contribution of data in subset $T_k$. 
We then choose the optimal $\widehat{h}$ for each time point $t$ via minimizing the cross-validation function $\mathscr{L}_{CV}^{t}(h)$ in equation~\eqref{eq:bandwidth_sl}. The specific algorithm for selecting bandwidth is presented in  Algorithm~3 of Appendix A.
\begin{align}
    \label{eq:bandwidth_sl}
    \mathscr{L}_{CV}^{t}(h) :=\sum_{k=1}^{K} \sum_{l \in T_k}\ell(\boldsymbol{x}_l,Y_{l-1};\widehat{\boldsymbol{\theta}}_{-T_k}(\tau_t | h)), \quad \text{where} \quad \widehat{\boldsymbol{\theta}}_{-T_k}(\tau_t | h)=\underset{\boldsymbol{\vartheta}\in\Theta_r}{\operatorname*{\mathrm{argmin}}}    \mathscr{L}^{t}_{-T_k}(\boldsymbol{\vartheta}).
\end{align}

\subsection{Theoretical Analysis}
\label{sec:identifiability}

One of the principal challenges in causal discovery within non-stationary causal processes is demonstrating structural identifiability. We address this challenge by providing general conditions on identifiability for non-stationary processes, with only a mild Assumption~\ref{asm:local_stationary}. Unlike previous approaches, we do not require the strict faithfulness assumptions \citep{spirtes2001causation} and are not limited by assumptions of constant Gaussian noises. Our method offers a general framework that can be applied to a wide class of identifiable models, including Gaussian DAG \citep{peters2014identifiability}, Non-Gaussian DAG \citep{lanne2017identification}, Additive Noise Models \citep{peters2014causal}, and Post-Nonlinear Models \citep{gong2022rhino}. 

The novelty of our proof lies in demonstrating that the DYNAMO loss $\mathscr{L}_{t}(\boldsymbol{\vartheta})$ at time $t$, as defined in Eq.~\eqref{eq:constant_likelihood}, can be effectively approximated by the corresponding loss function $L_{t}(\boldsymbol{\vartheta})$ under stationary causal processes $\{\tilde{\boldsymbol{x}}_t\}$. Since stationary time series $\{\tilde{\boldsymbol{x}}_t \}$ shares the same causal structure as the non-stationary process $\{\boldsymbol{x}_t \}$ at time $t$, we demonstrate that causal structure within non-stationary time series $\{\boldsymbol{x}_t \}$  at each time $t$ is identifiable as long as its corresponding stationary causal structures are identifiable. The overall proof of DYNAMO's identifiability is shown in Appendix C.



\begin{theorem}[\textbf{Identifiability of DYNAMO}]
    \label{Thm:identify_dynamo}
    Assuming DYNAMO satisfies the Assumption \ref{asm:unconfoundness}, \ref{asm:acyclicity}, \ref{asm:local_stationary}, and \ref{asm:localizing_kernel}, then DYNAMO defined in Eq.~\eqref{eq:non-stationary} is structurally identifiable for nonstationary causal processes.
\end{theorem}

\begin{theorem}[\textbf{Consistency of DYNAMO}]
    \label{Thm:consistency_dynamo}
    Assuming DYNAMO satisfies Assumption \ref{asm:unconfoundness}, \ref{asm:acyclicity}, \ref{asm:local_stationary}, and \ref{asm:localizing_kernel}, and the loss function $\mathscr{L}_{t}(\boldsymbol{\vartheta})$ satisfies the smoothness assumption C.3 in Appendix C, then in both linear and nonlinear settings, the estimated $\boldsymbol{\widehat{\theta}}(\tau_t)$ minimizing the loss in Eq.~\eqref{eq:non-stationary} converges to the ground truth $ \boldsymbol{\theta}(\tau_t)$ with the probability approaching 1, as $(T,h) \rightarrow (\infty,0)$, $Th \rightarrow \infty$, and $Th^7 \rightarrow 0$.
\end{theorem}

Our proof proceeds in two steps. Firstly, we demonstrate that $\boldsymbol{\widehat{\theta}}(\tau_t) \rightarrow \boldsymbol{\widehat{\theta}}_0(\tau_t) $, where the $\boldsymbol{\widehat{\theta}}_0(\tau_t)$ is the unique minimizer of the loss function $L_{t}(\boldsymbol{\vartheta})$ under the stationary assumption. Secondly, leveraging the consistency under the stationarity assumption, we establish the consistency of DYNAMO estimators for non-stationary causal processes. The proof and detailed conditions are provided in Appendix C. 

\section{Simulation Study}\label{sec:simu}

In this section, we empirically evaluate our method, DYNAMO, with non-stationary causal processes under both linear and nonlinear contexts. We conduct a comparative analysis between DYNAMO and other state-of-the-art methods, including DYNOTEARS \citep{pamfil2020dynotears}, NTS-NOTEAR \citep{sun2023nts}, PCMCI+ \citep{runge2020discovering} and CD-NOD \citep{huang2020causal}.  We utilize  Structural Hamming Distance (SHD, a smaller value is better) and $F_1$ score (a larger value is better) as metrics to assess the effectiveness of these methods in recovering the overall graphs. Each setting is replicated over 20 random seeds.

We generate data based on the structural equation model outlined in Eq.~\eqref{eq:non-stationary}. To introduce progressively evolving causal structures, we initialize a fundamental graph structure using the Erdos–Renyi (ER) model and adjust the \textit{non-zero} weight $w_{ij}(\tau_t)$  as a function of $t$ and a threshold $\gamma$. Specifically, the weight function is defined as $  w_{ij}(\tau_t) = \cos [ (\delta_{ij}+t/\Phi)\pi ] \cdot I(\cos [ (\delta_{ij}+t/\Phi)\pi ] > \gamma)$, where $\delta_{ij}$ is a randomly drawn value ranging from 0 to 1, $\gamma$ denotes the sparsity threshold, and $\Phi$ determines the changing speed. A smaller $\Phi$ corresponds to a faster alteration in the causal structure. The threshold $\gamma$ ensures that our causal structure $\boldsymbol{\theta}(\tau_t)$ not only evolves gradually in terms of causal strength, but also transforms the entire causal structure as $t$ progresses from 0 to $T$. To generate nonlinear relationships, we build a nonlinear model incorporating tanh and sigmoid activation functions. The detailed process of generating the data is presented in Appendix E. The simulation results are shown in Figure \ref{fig:main_simulation}.

\begin{figure}[!t]
    \begin{center}
\centerline{\includegraphics[width=0.8\textwidth]{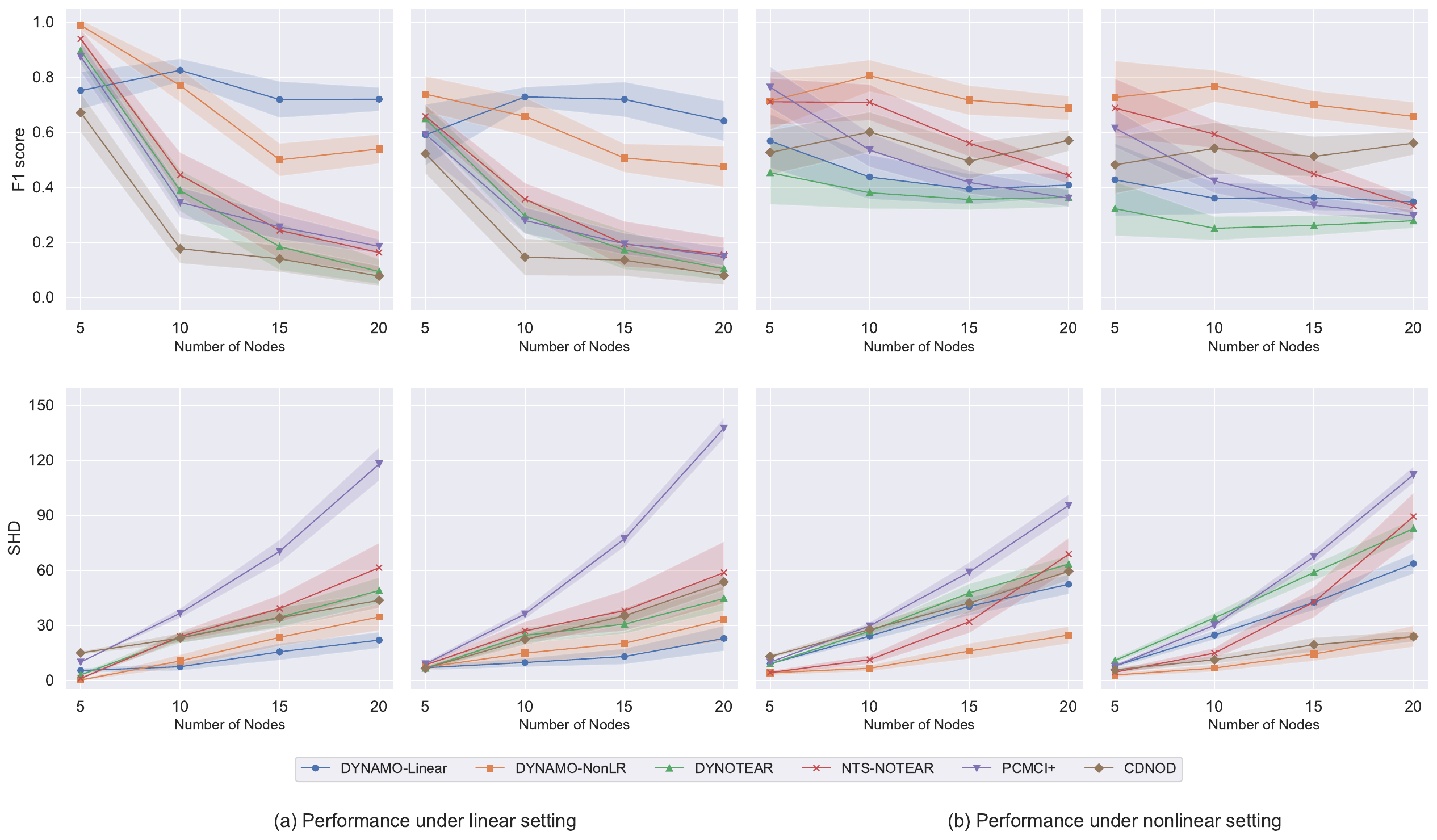}}
   \caption{Comparison study of linear and nonlinear non-stationary data with $T$ = 500 at time 60 over 20 replications. The first two columns represent the simulation results under linear settings, and the last two columns represent simulation results under nonlinear settings. For each setting, the left panel shows instantaneous graphs, while the right panel presents lagged graphs.}
    \label{fig:main_simulation}
    \vspace{-0.8cm}
    \end{center}
\end{figure}

Figure \ref{fig:main_simulation} presents the SHD and F1 scores for different methods based on 20 replications. Our methods exhibit superior performance in both linear and nonlinear scenarios, with the DYNAMO estimators notably improving upon their base learner. In linear settings (first two columns), both linear DYNAMO and nonlinear DYNAMO outperform other methods, achieving the highest F1 scores and lowest SHD across different node sizes. The advantages of our proposed model remain consistent with increasing node size.  In nonlinear settings (the last two columns), the nonlinear DYNAMO surpasses other methods, while the performance of linear DYNAMO consistently outperforms its base learner DYNOTEAR. The underperformance of linear DYNAMO in nonlinear settings and nonlinear DYNAMO in linear settings can be attributed to the linear base learner's inability to capture the nonlinear structures and the nonlinear base learner's tendency to overfit in linear datasets. Additional simulation results, including running times, diverse time points, time lengths, varying time-varying speeds, more lagged periods, another noise type, and hyper-parameter selection are detailed in Appendix E.

\section{Discovering Time-varying Home Field Advantages in Football Games}
\label{sec:real_data}

We utilize the linear and nonlinear DYNAMO models to analyze the factors contributing to home field advantage and their influences throughout the entire match. For convenience, we first focus on four representative teams, Manchester City, Liverpool, Arsenal, and Manchester United. A discussion of the other 16 teams is given at the end of Section \ref{sec53}. We build time-series datasets representing each team's average performance disparity between home and away matches at every minute. The difference in expected goals (XG) conditioned on other within-match variables between home and away matches indicates home field advantage. In this section, we refrain from presenting complex causal structures and solely focus on edges indirectly and directly influencing XG. Using XG instead of simple match results provides a more detailed understanding of a team's performance. Simple results like wins or losses only reflect the outcome, without considering the quality of chances created or conceded. XG measures the likelihood that each shot will become a goal based on factors like shot location and type, offering a better view of the actual performance of a team. With XG, we can assess whether a team was truly dominant or just lucky. For example, a 1-0 win may hide poor chance creation, while a loss might obscure numerous high-quality opportunities. This makes XG a more reliable tool for performance analysis than just relying on results.

\subsection{Hypothesis Development}

Home field advantage primarily arises from two factors: referee bias and disparities in player performances, with the latter being attributed to environmental familiarity and crowd support \citep{SportsPerformance,bryson2021causal}. In addition to its direct impact, the referee bias can also indirectly affect home field advantage through team performance by potentially increasing penalties against the visiting team \citep{buraimo201012th}. During a match, the effects of referee bias and players' performances on home field advantages may not remain consistent due to dynamic influencing factors and the intrinsic dynamics of the match process. The enthusiasm and vocal support of the crowd vary over time, reaching peaks at the beginning and end of the game and during critical events in the match process \citep{hill2022roar}. Furthermore, as matches progress, players endure physical and mental fatigue, which can affect their performance and home field advantage \citep{reilly2008muscle}.   Based on the above analysis, we propose the following hypothesis. 

\begin{hypothesis}
    \label{hyp:home_advantage}
    Player performances and referee bias can contribute significantly to home field advantage, and their impacts may vary across different match periods.
    \vspace{-0.2cm}
\end{hypothesis}

\begin{figure}[ht]
    \begin{center}
     \vspace{-0.2cm}\centerline{\includegraphics[width=0.9\textwidth]{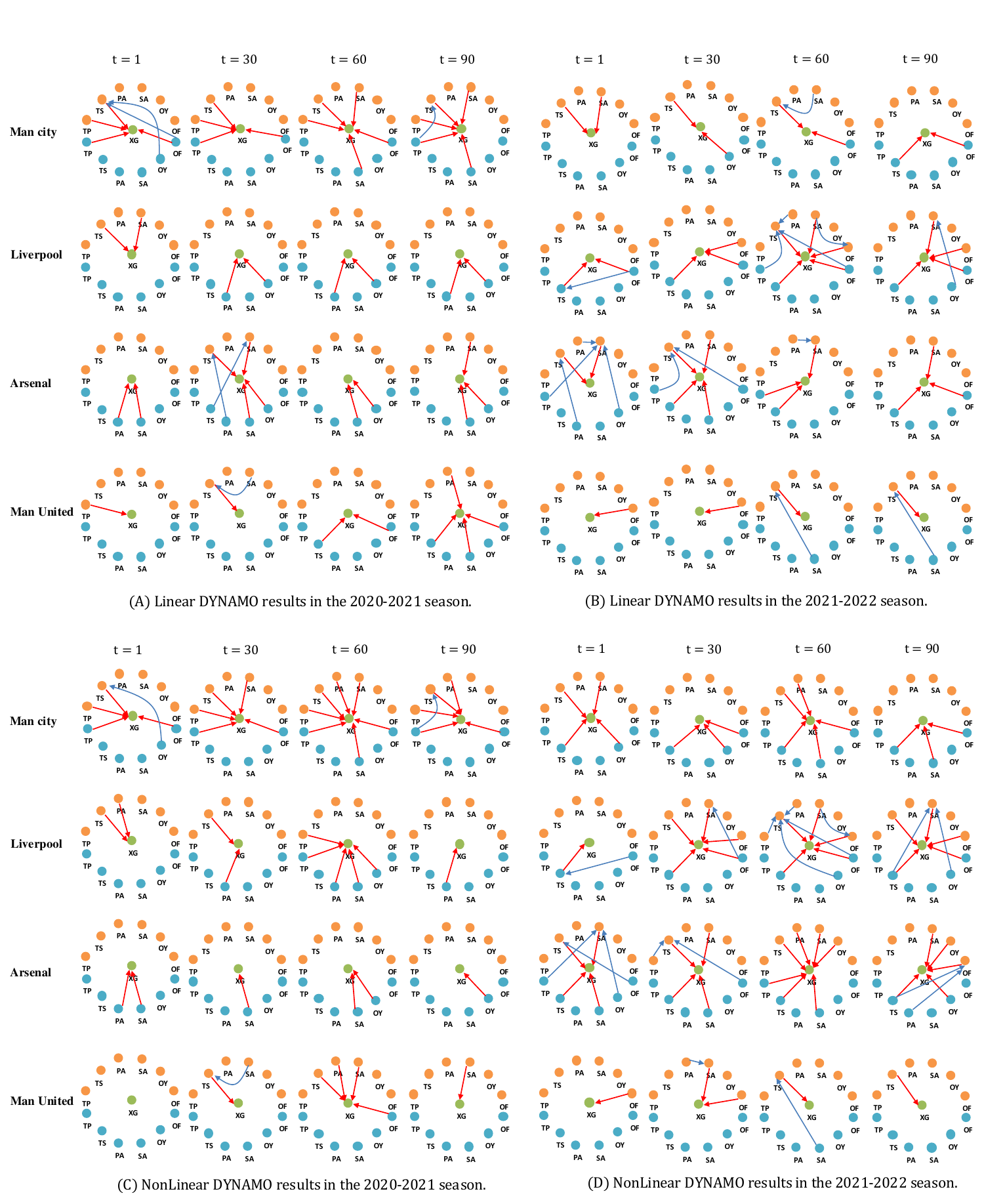}}
    \caption{\footnotesize Model results for 4 representative teams, Manchester City, Liverpool, Arsenal and Manchester United.  Variables: TP - Total Passes, TS - Total Shots, PA - Pass Accuracy, SA - Shot Accuracy, OY - Opponent's yellow cards, OF - Opponent's Fouls. Orange notes represent contemporary variables at time $t$, while blue notes denote the lagged variable at time $t-1$. Orange arrows denote the direct effects on the XG, and blue arrows denote the indirect effects on the XG. Panel A and Panel B present the results of linear DYNAMO for the 2020-2021 season and the 2021-2022 season, respectively. Panel C and Panel D present the model results of nonlinear DYNAMO during the 2020-2021 season and 2021-2022 season, respectively. }
    \label{fig:main_results}
    \vspace{-0.8cm}
    \end{center}
\end{figure}

We validate Hypothesis~\ref{hyp:home_advantage} for specific teams by analyzing time-varying causal structures using both linear and nonlinear DYNAMO models. We select key variables such as the number of total passes (TP), the number of total shots (TS), pass accuracy (PA), shot accuracy (SA), the opponent's yellow cards (OY), and the opponent's fouls (OF), along with several within match variables following previous literature \citep{endrich2020home,bryson2021causal}. If a direct link exists from TP, TS, PA, or SA to XG, it suggests that specific player performance contributes to the home field advantage. Similarly, a direct link from OY or OF to XG indicates the presence of referee bias effects. We employ linear and nonlinear DYNAMO models to estimate time-varying causal structures and compared our methods with DYNOTEARS and CDNOD. Additionally, we conduct robustness checks for our linear and nonlinear DYNAMO models with different kernel bandwidths and lagged periods. Detailed information regarding robustness checks and hyper parameter selection is provided in Appendix D.

\subsection{Time-varying Home Field Advantage and Model Validation}

Figure \ref{fig:main_results} illustrates the time-varying home field advantages for four representative teams, as estimated by linear DYNAMO and nonlinear DYNAMO with a lag period of one. Panel A presents the results of linear DYNAMO for the 2020-2021 Premier League season, during which no audience was allowed due to COVID-19 restrictions. Panel B shows the results of linear DYNAMO for the 2021-2022 Premier League season when the stadiums were reopened to spectators. Panel C and Panel D present the model results of nonlinear DYNAMO during the 2020-2021 season and 2021-2022 season, respectively. The analysis focuses on four key time points: the beginning of the match, the middle of the first half, the middle of the second half, and the end of the match.

Although ground-truth causal structures are unavailable, we validate and interpret the results of our method through three approaches. First, we compare our linear DYNAMO and nonlinear DYNAMO with existing methods DYNOTEARS and CD-NOD, operating under the assumption that accurate learning of the true causal structures will yield closely aligned outputs for both linear and nonlinear DYNAMO. Second, we examine and compare the results of our DYNAMO methods between the 2020-2021 and 2021-2022 seasons. The literature suggests that home advantage during the 2021-2022 season should be greater than in the 2020-2021 season, as crowd support significantly influences home advantage \citep{bryson2021causal,singleton2023decade}. Third, we use our DYNAMO model to predict the expected goals (XG) at each minute and compare these predictions to the actual goals scored. If our method effectively learns the causal structures, the predictions should be closer to the real scores than the original XG, as they account for dynamic causal relationships.

Remarkably, our linear DYNAMO and nonlinear DYNAMO offer valuable insights into identifying the factors influencing home field advantage across the four match points. In comparison, DYNOTEARS and CD-NOD struggle to capture the evolving dynamic effectively, as evidenced in Appendix D. CD-NOD struggles to consistently identify the factors influencing XG differences, while DYNOTEARS exhibits abrupt and sometimes contradictory changes in causal structures. The linear DYNAMO and nonlinear DYNAMO consistently estimate similar causal structures for each team, with nonlinear DYNAMO capturing more intricate relationships. Both models reveal time-changing patterns throughout the match period, with each team exhibiting a core causal structure that remains relatively stable across time, alongside temporal structures that change over time.

When comparing the model results for the 2020-2021 season with those for the 2021-2022 season in Figure \ref{fig:main_results}, distinct patterns emerge for these four teams with and without crowd support. For example, during the 2021-2022 season, Liverpool consistently benefits from lagged total shots across the four time points, while Arsenal enjoys increased shot accuracy at home. However, during the 2020-2021 season without crowd support, Liverpool only benefits from lagged total shots in the first half of the match, while fewer variables contribute to Arsenal's expected goals. Robustness tests with different bandwidths further confirm the stability of our results, as depicted in Appendix D.

We use our DYNAMO model to predict XG at each minute and compare these predictions to the actual goals scored. Table~\ref{tab:performance_comparison} presents the mean squared error (MSE) of our DYNAMO method, DYNOTEARS, and the original XG. The results show that predictions based on linear DYNAMO outperform both DYNOTEARS and simple XG predictions. This suggests that incorporating time-varying causal structures improves the alignment between estimated XG and actual scores, thereby validating the utility of our method.

\begin{table}[H]
    \centering
    \caption{Goal Prediction Performance Comparison (Mean Squared Error)}
    \begin{tabular}{ccccccc}
        \toprule
         & \multicolumn{3}{c}{2020-2021 Season} & \multicolumn{3}{c}{2021-2022 Season} \\
        \cmidrule(lr){2-4} \cmidrule(lr){5-7}
        Team & {DYNAMO} & {DYNOTEAR} & {XG} & {DYNAMO} & {DYNOTEAR} & {XG} \\
        \midrule
        Arsenal     & \textbf{0.00046} & 0.00061 & 0.00063 & \textbf{0.00079} & 0.00106 & 0.00111 \\
        Liverpool   & \textbf{0.00061} & 0.00073 & 0.00075 & \textbf{0.00080} & 0.00080 & 0.00081 \\
        Man City    & \textbf{0.00072} & 0.00095 & 0.00101 & \textbf{0.00062} & 0.00091 & 0.00095 \\
        Man United  & \textbf{0.00055} & 0.00062 & 0.00067 & \textbf{0.00053} & 0.00060 & 0.00064 \\
        \bottomrule
    \end{tabular}
    \label{tab:performance_comparison}
\end{table}


\subsection{Referee Bias and Crowd Support Effect}\label{sec53}
Expanding on this discovery, the specific patterns of \textit{referee bias} not only highlight the unique styles and strategies of each team but also suggest a deeper relationship between game dynamics and officiating. For Manchester City, referee bias in 2020-2021 and 2021-2022 showed a first-half preference for penalizing opponents with yellow cards and a second-half focus on fouls. The bias towards issuing yellow cards to their opponents in the first half could be interpreted as a tactical response by referees to City’s aggressive style of play and their ability to control the game early on. This early bias may deter opposing teams from engaging in reckless challenges, leading to fewer yellow cards in the later stages of the match, when fouls instead become the primary factor influenced by referees. The shift in referee decisions from yellow cards to fouls between halves reflects a nuanced understanding of how Manchester City's tempo and pressure affect not just the opponents, but also the officiating tendencies.


Liverpool's home advantage patterns vary sharply with or without crowd support. During the 2020-2021 season, their referee bias patterns were not stable throughout the match period, while during the 2021-2022 season, they experience steady and escalating referee bias throughout the match, underscoring the profound impact of their home environment, particularly at Anfield \citep{hill2023makes}. The atmosphere in the stadium, often described as intimidating for visiting teams and officials alike, can influence referees to favor the home side, especially as the match progresses and crowd intensity rises. The continuity of referee bias throughout the game implies that Liverpool's style, which is characterized by relentless pressure and intensity, maintains a psychological influence on both the opponents and the officials, amplifying their home field advantage.

Arsenal's referee bias pattern varied across seasons, with second-half favoritism in 2020-2021 and performance-driven bias in 2021-2022, especially with spectators. Without an audience, home advantage came through increased opponent yellow cards, while with spectators, referee decisions early and late in matches were influenced by prior fouls, enhancing Arsenal's control and exploiting opponent fatigue.


In contrast, Manchester United’s experience of referee bias is related to their overall match fitness and competitiveness. During the 2020-2021 season, where they ranked 2nd, they experienced referee bias during the second half, while during the 2021-2022 season, they experienced referee bias during the first half. The fact that referee bias does not persist into the second half of the 2021-2022 season may indicate that their early intensity fades as the match progresses, leaving them more vulnerable to opponents who are less penalized for fouls in the later stages. This pattern suggests that referee decisions might not only reflect home field dynamics but also mirror the physical and psychological state of the teams as the match unfolds.

We also extend our analysis to the remaining 16 teams. Based on their average rankings over the two seasons, we classify these teams into four groups: higher-ranked, middle-ranked, lower-ranked, and relegated teams. Detailed classification and model estimates are reported in the Appendix~D. Our findings reveal several noteworthy patterns across both team groups and match periods. First, many non-relegated teams experienced favorable referee bias as more opponent fouls in the spectator-free 2020–2021 season, which shifted to more yellow cards issued to opponents after spectators returned in 2021–2022.  This transition suggests that the presence of spectators and associated crowd pressure may influence referees to penalize away teams more harshly. Second, higher- and middle-ranked teams saw little referee bias early in matches without spectators, but nearly half exhibited early-match bias once crowds returned in 2021–2022. This shift may be attributed to the influence of crowd atmosphere and vocal fan support at the beginning of games. Lastly, for lower-ranked and relegated teams, the differences in referee bias between seasons with and without spectators are less pronounced. This may be attributed to their relatively limited crowd support, which may reduce the influence of spectators on referee decisions compared to higher-ranked teams.

These findings highlight how referee bias interacts with both team strategy and the evolving conditions of a match, offering new insights into the multifaceted nature of home field advantage. Understanding these dynamics could provide teams with strategic opportunities to exploit or mitigate these biases, depending on the timing and nature of their gameplay.


\section{Conclusion}
\label{sec:conc}
In this work, we propose a novel causal structure learning method, DYNAMO, to effectively estimate the time-varying home field advantage. We establish the theoretical properties of DYNAMO regarding the identifiability and consistency of non-stationary causal processes. Empirical evaluation validates the superior performance of DYNAMO in recovering time-varying causal structures. Applying our method to high-resolution event data from the 2020-2021 and 2021-2022 English Premier League seasons provides valuable insights into the time-varying, team-specific home field advantages. Our findings highlight significant variations in field advantages influenced by referee bias, which differ notably between seasons with and without crowd support. Compared to competing methods, our approach effectively captures local stationarity, and the time-varying causal structures learned improve the accuracy of goal prediction. 

There are several possible directions for further investigation. Firstly, DYNAMO's effectiveness and computational efficiency largely depend on the choice of a base learner. In real-world datasets, the linearity or nonlinearity
of the underlying structure is often unknown. It would be worthwhile to explore strategies for devising more suitable base learners capable of identifying both linear and nonlinear causal relationships more efficiently. Secondly, while our method uncovers the time-varying home-field advantage at the team level, it would be even more insightful to investigate how playing at home influences individual player performance. However, this is a challenging task, as players operate in a complex environment characterized by potential unmeasured confounders and spatio-temporal interferences caused by interactions with teammates and opponents. Recent advances in reinforcement learning offer promising tools for addressing these challenges \citep{liu2022uncertainty}. For example, reinforcement learning methods have been used to evaluate offline policies in complex environments with spatiotemporal interferences and unmeasured confounders \citep{dai2024causal, yu2024two}. Exploring player-level home-field advantages using reinforcement learning techniques could provide deeper insights into individual contributions and dynamics.

\setlength{\baselineskip}{24pt}
\bibliographystyle{chicago}

\bibliography{Bibliography-MM-MC.bib}
\end{document}


\def\spacingset#1{\renewcommand{\baselinestretch}%
{#1}\small\normalsize} \spacingset{1}


\if1\blind
{
  \title{\bf Supplement to "Time-Varying Home Field Advantage in Football: Learning from a Non-Stationary Causal Process"}
\author{Minhao Qi
    \\
    School of Management, Center for Data Science, \\
    Zhejiang University\\
    Hengrui Cai \\
     Department of Statistics,
     University of California Irvine \\
     Guanyu Hu \\
     Department of Biostatistics and Data Science\\
     The University of Texas Health Science Center at Houston \\
     and \\
     Weining Shen \\
     Department of Statistics, 
     University of California Irvine \\
     }
  \maketitle
} \fi

\if0\blind
{
  \bigskip
  \bigskip
  \bigskip
  \begin{center}
    {\LARGE\bf Supplement to "Time-Varying Home Field Advantage in Football: Learning from \\ a Non-Stationary Causal Process"}
\end{center}
  \medskip
} \fi

\bigskip



This supplementary material contains (1) solving algorithms for linear DYNAMO, nonlinear DYNAMO, and bandwidth selection; (2) the detailed structures of NTS-NOTEAR; (3) technical proofs;  (4) further real data analysis including robustness checks and hyperparameters; (5) simulation extensions; and (6) limitations.

\section{Solving Algorithms}
\label{appx:solving_algorithms}

\begin{algorithm}[H]
   \caption{Solving DYNAMO under linear setting}
   \label{alg:DYNAMO-LC}
   \begin{algorithmic}
   \STATE {\bfseries Input:} data $X$, bandwidth $h$, time-stamp $t$, acyclicity constraint $\rho, \alpha$, constraint upper bound $\rho_{max}$, acyclicity tolerance $\eta_{tol}$, progress rate $c$ and $q$.  
   \STATE {\bfseries Initiate:} Random guess instantaneous adjacent matrix $W_t$ and lagged adjacent matrix $A_t$.
   \FOR{maximum amount of iterations}
   \WHILE{ $\rho < \rho_{max}$}
   \STATE $\eta \leftarrow H(W_t)$;~ $\mathcal{L}_m \leftarrow \mathscr{L}_{t}^{lr}(W_t,A_t|X,h,t,\rho)$;
   \STATE $W_t,A_t$ $\leftarrow$ L-BFGS-B.update$(\mathcal{L}_m)$; ~ $\eta^{*} \leftarrow H(W_t)$;
   \IF{$\eta^{*} > c \eta$}
   \STATE $\rho \leftarrow q\rho$;
   \ENDIF
   \ENDWHILE
    \IF{$\eta^{*} < \eta_{tol}$}
    \STATE Break;
    \ENDIF
   \ENDFOR
   \RETURN $W_t, A_t$.
\end{algorithmic}
\end{algorithm}

\begin{algorithm}[H]
   \caption{Solving DYNAMO under nonlinear setting}
   \label{alg:DYNAMO-nonlinear}
   \begin{algorithmic}
   \STATE {\bfseries Input:} data $X$, bandwidth $h$, time-stamp $t$, acyclicity constraint $\rho, \alpha$, constraint upper bound $\rho_{max}$, acyclicity tolerance $\eta_{tol}$, progress rate $c$ and $q$.
   \STATE {\bfseries Initiate:} Random guess $\boldsymbol{\theta}(\tau_t)$.
   \FOR{maximum amount of iterations}
   \WHILE{ $\rho < \rho_{max}$}
   \STATE $W_t \leftarrow \boldsymbol{\theta}(\tau_t)$; ~$\eta \leftarrow H(W)$;~ $\mathcal{L}_m \leftarrow \mathscr{L}_{t}^{nlr}(\boldsymbol{\theta}(\tau_t)|X,h,t,\rho) $;
   \STATE $\boldsymbol{\theta}(\tau_t)$ $\leftarrow$ L-BFGS-B.update$(\mathcal{L}_m)$; ~  $W_t \leftarrow \boldsymbol{\theta}(\tau_t)$; ~ $\eta^{*} \leftarrow H(W_t)$;
   \IF{$\eta^{*} > c \eta$}
   \STATE $\rho \leftarrow q\rho$;
   \ENDIF
   \ENDWHILE
    \IF{$\eta^{*} < \eta_{tol}$}
    \STATE Break;
    \ENDIF
   \ENDFOR
   \RETURN $\boldsymbol{\theta}(\tau_t)$.
\end{algorithmic}
\end{algorithm}

\begin{algorithm}[H]
   \caption{Bandwidth Selection}
   \label{alg:bandwidth}
   \begin{algorithmic}
   \STATE {\bfseries Input:} data $X$, bandwidth list $h$-$List$, time $t$, Number of folds $K$.
   \FOR{$h$ {\bfseries in } $h$-$List$}
   \STATE Randomly divide the data $X$ into K folds.
   \FOR{$k$ {\bfseries in} range(K)}
   \STATE Use the $k$-th fold data as the test data $V$, and the remaining data as the training data $D$.
   \STATE {\bfseries Estimate} $\widehat{\boldsymbol{\theta}}_{-T_k}(\tau_t | h)$ by training data $D$ using $\mathscr{L}_{-T_k}^{t}(\boldsymbol{\vartheta})$.
   \STATE {\bfseries Calculate} the CV loss $\sum_{l \in T_k}\ell(\boldsymbol{x}_l,Y_{l-1};\widehat{\boldsymbol{\theta}}_{-T_k}(\tau_t | h))$ by test data $V$.
   \ENDFOR
   \STATE{\bfseries Sum} the CV loss of bandwidth $h$.
   \ENDFOR
    \STATE{\bfseries Select} the $h$ that minimizes the CV loss $\mathscr{L}^{t}_{CV}(h)$.
   \RETURN $h$.
\end{algorithmic}
\end{algorithm}

\section{Detailed Structures of NTS-NOTEAR}
\label{appx:nonlinear_model}

In this paper, we utilize the NTS-NOTEAR loss as the base learner of our nonlinear DYNAMO. We recommend readers to refer to \citet{sun2023nts} for a detailed introduction. The NTS-NOTEAR utilizes $d$ Convolutional neural networks (CNNs) to jointly model the data, where the $j$-th CNN predicts the expectation of the target variable $x_t^{[d_j]}$ at each time step $t \geq L+1$ given lagged and contemporary variables:
\begin{align*}
    \mathbb{E}[x_t^{[d_j]} | Pa(x_t^{[d_j]})] = \text{CNN}_j(\boldsymbol{x}_t^{-[d_j]},Y_{t-1}),
\end{align*}
where $Pa(x_t^{[d_j]}) \subseteq \{\boldsymbol{x}_t,\boldsymbol{x}_{t-1},\ldots,\boldsymbol{x}_{t-L}\} \in \mathcal{R}^{(L+1) \times d}$ is the parent set of $x_t^{[d_j]}$ that is defined by the trained CNNs, $Y_{t-1} = [\boldsymbol{x}_{t-1},\boldsymbol{x}_{t-2},\cdots,\boldsymbol{x}_{t-L}]^\top \in \mathcal{R}^{L \times d}$, and $\boldsymbol{x}_t^{-[d_j]}$ denotes the $\boldsymbol{x}_t$ excluding $x_t^{[d_j]}$. Following \citet{zheng2020learning}, the weighted adjacency matrices $W_t$ and $A_t$ can be calculated by the rationale that a function $g_t(X)$ is independent of input $x_t^{[d0]}$ if and only if $\|\partial g_t(X) / x_t^{[d0]}\|_2 = 0$.  Hence, we can use the partial derivatives as the weighted adjacency matrices.

Let $\boldsymbol{\theta}_j \in \boldsymbol{\theta} = (\boldsymbol{\theta}_1, \ldots, \boldsymbol{\theta}_d) $ denote the parameters for the $j$-th CNN. Based on the Theorem 1 in \cite{sun2023nts}, the function class $F$ of 1D CNNs that are independent of $x_t^{[d_j]}$ is equivalent to the class $F_0$ of 1D CNNs where the $j$-th kernel parameters of the \textbf{first} layer CNN kernels are all zeros.  Hence, we can further leverage the 
kernel weight parameters for input contemporary variable $x_t^{[d_i]}$ in the \textbf{first} convolutional layer of the $j$-th CNN,
$\phi_t^{i,j} \subset \boldsymbol{\theta}_j$, to represent $W_t^{ij}$. Similarly, we leverage the 
kernel weight parameters for input lagged variable $Y_{t-1}^{[d_i]}$ in the first convolutional layer of the $j$-th CNN,
$\phi_t^{i-d,j} \subset \boldsymbol{\theta}_j$, to represent $A_t^{ij}$, specifically, we have
\begin{align*}
    W_t^{ij}(g_t(\cdot)) &= \| \phi_t^{i,j} \|_2 \quad \text{for} ~ i \leq d,\\
    A_t^{ij}(g_t(\cdot)) &= \| \phi_t^{i-d,j} \|_2 \quad \text{for} ~ d < i \leq (L+1) \times d.
\end{align*}

The specific structure of $g_t(\cdot)$ is
\begin{align*}
    g_t(\boldsymbol{x}_{l},Y_{l-1};\boldsymbol{\theta}(\tau_t)) = [\text{CNN}_1(\boldsymbol{x}_{l},Y_{l-1}; \boldsymbol{\theta}(\tau_t)),\ldots,\text{CNN}_d(\boldsymbol{x}_{l},Y_{l-1}; \boldsymbol{\theta}(\tau_t))].
\end{align*}

\section{Technical Proofs}

\begin{table}[!hb]
    \centering
    \caption{The table of notations.}
    \begin{tabular}{ll}
        \bottomrule
        \textbf{Notation} & \textbf{Description} \\
        \hline
        T & The time length of the dataset \\
        d & The number of variables \\
        L & The maximum lagged period \\
        $\mathcal{G}_t = (V_t,\boldsymbol{\theta}_t(\tau_t))$ & Causal graph with nodes set $V_t$ and adjacent matrix (edges) $\boldsymbol{\theta}_t$  \\
        $X = \{ \boldsymbol{x}_t(\tau_t)\}$  &  A multivariate non-stationary time series matrix $X \in \mathcal{R}^{T \times d}$ \\
        $\boldsymbol{x}_t = \boldsymbol{x}_t(\tau_t)$ & Realization of non-stationary $X$ at time $t$, $\boldsymbol{x}_t(\tau_t) \in \mathcal{R}^d$ \\
        $x_t^{[j]} = x_t^{[j]}(\tau_t)$ &  The $j$th item of $\boldsymbol{x}_t$ \\
        $Y_{t-1}(\tau_t)$ &  $[\boldsymbol{x}_{t-1},\boldsymbol{x}_{t-2},\cdots,\boldsymbol{x}_{t-L}]^\top \in \mathcal{R}^{L \times d}$ \\
        $X_t(\tau_t)$ & $[\boldsymbol{x}_t(X_t(\tau_t)),Y_{t-1}(X_t(\tau_t))]^\top \in \mathcal{R}^{(L+1)\times d}$ \\
        $\tilde{X}(\tau) = \{ \tilde{\boldsymbol{x}}_t(\tau)\}$  &  A multivariate stationary time series matrix $\tilde{X} \in \mathcal{R}^{T \times d}$ \\
        $\tilde{\boldsymbol{x}}_t(\tau)$ & Realization of stationary causal process $\tilde{X}(\tau)$ at time $t$, $\boldsymbol{x}_t(\tau) \in \mathcal{R}^d$ \\
        $\tilde{x}_t^{[j]}(\tau)$ &  The $j$th item of $\boldsymbol{x}_t(\tau)$ \\
        $\tilde{Y}_{t-1}(\tau)$ &  $[\tilde{\boldsymbol{x}}_{t-1}(\tau),\tilde{\boldsymbol{x}}_{t-2}(\tau),\cdots,\tilde{\boldsymbol{x}}_{t-L}(\tau)]^\top \in \mathcal{R}^{L \times d}$ \\
        $\tilde{X}_t(\tau)$ & $[\tilde{\boldsymbol{x}}_t(\tau),\tilde{Y}_{t-1}(\tau)]^\top \in \mathcal{R}^{(L+1)\times d}$ \\
        $K_h(\cdot) = K(\cdot/h)$ & A kernel function $K(\cdot)$ with bandwidth $h$ \\
        $\|W \|_q$ & $\|W\|_q:=(\mathbb{E}|W|^q)^{1/q}$ \\
        $\nabla^k_{\boldsymbol{\vartheta}}f(\cdot)$ & The $k$-th order derivative of function $f(\cdot)$ on $\boldsymbol{\vartheta}$ \\
        \bottomrule
    \end{tabular}
    \label{tab:notations}
\end{table}

Before proceeding with our proof, we introduce some notations: uppercase letter $X$ denotes a matrix, bold lowercase letter represents a vector, and non-bold lowercase letter denotes a scalar. For example, $X$ is a multivariate time series matrix, $\boldsymbol{x}_t$ represents a row vector of $X$ at time $t$, and $x_t^{[j]}$ is the $j$th item in $\boldsymbol{x}_t$. The symbol $\|\cdot \|$ denotes the Euclidean norm of a vector or the spectral norm for a matrix. The symbol $\circ$ denotes
the Hadamard product of two matrices. Let $q > 0$ and $\|W\|_q:=(\mathbb{E}|W|^q)^{1/q}$. $K_h(\cdot) = K(\cdot/h)$, where $K(\cdot)$ and $h$ stand for a nonparametric kernel function and a bandwidth respectively. Let $\nabla^k_{\boldsymbol{\vartheta}}f(\cdot)$ denote the $k$-th order derivative of function $f(\cdot)$ on $\boldsymbol{\vartheta}$.

\label{appx:proofs}
\subsection{Proofs of Proposition 1}
The proofs of Proposition~1are motivated by \cite{Dahlhaus2019towards,gao2024time}. 
We first define a stationary process $\tilde{\boldsymbol{x}}_{T,t}(\tau)$ as follows:
\begin{align*}
    \tilde{\boldsymbol{x}}_{T,t}(\tau) = f(\tilde{\boldsymbol{x}}_{T,t}(\tau),\tilde{\boldsymbol{x}}_{T,t-1}(\tau),\ldots,\tilde{\boldsymbol{x}}_{T,t-L}(\tau); \boldsymbol{\theta}(\tau)), \quad t > -T,
\end{align*}
for $t > -T $. If $ t < -T $, then $\tilde{\boldsymbol{x}}_{T,t}(\tau) = \boldsymbol{0}$. Additionally, we define
\begin{align*}
    \tilde{\boldsymbol{x}}_{t}(\tau) = f(\boldsymbol{\epsilon}_t,\boldsymbol{\epsilon}_{t-1},\ldots,\boldsymbol{\epsilon}_{t-L}; \boldsymbol{\theta}(\tau)).
\end{align*}

The objective of this proof is to demonstrate that as the $T \rightarrow \infty$, a stationary process $\tilde{\boldsymbol{x}}_{t}(\tau)$ exists such that $\tilde{\boldsymbol{x}}_{T,t}(\tau) \rightarrow \tilde{\boldsymbol{x}}_{t}(\tau)$. 

\textbf{Proofs of (1).}
\begin{align*}
    & \|\tilde{\boldsymbol{x}}_{T,0}(\tau) -\tilde{\boldsymbol{x}}_{T+1,0}(\tau) \|_q \\
    = & \|f(\tilde{\boldsymbol{x}}_{T,0}(\tau),\ldots,\tilde{\boldsymbol{x}}_{T,-L}(\tau); \boldsymbol{\theta}(\tau)) - f(\tilde{\boldsymbol{x}}_{T+1,0}(\tau),\ldots,\tilde{\boldsymbol{x}}_{T+1,-L}(\tau); \boldsymbol{\theta}(\tau)) \|_q \\
    \leq & \sum_{j=0}^{L} \alpha_j(\boldsymbol{\theta}(\tau)) \|\tilde{\boldsymbol{x}}_{T,-j}(\tau) - \tilde{\boldsymbol{x}}_{T+1,-j}(\tau) \|_q \\
    \leq & \rho(\tau) \sup_{j \in [0,L]} \|\tilde{\boldsymbol{x}}_{T-j,0}(\tau) - \tilde{\boldsymbol{x}}_{T+1-j,0}(\tau) \|_q,
\end{align*}
where $\rho(\tau) = \sum_{j=0}^{L} \alpha_j(\boldsymbol{\theta}(\tau))$.The first step follows from the definition of $\tilde{\boldsymbol{x}}_{T,0}(\tau)$, the second step follows from the Assumption~3, the third step follows from the $\sum_{j=0}^{L} \alpha_j(\boldsymbol{\theta}(\tau)) < 1$, and the last step follows from the definition of the stationary process $\tilde{\boldsymbol{x}}_{T,0}(\tau)$.

Let $u_T = \|\tilde{\boldsymbol{x}}_{T,0}(\tau) -\tilde{\boldsymbol{x}}_{T+1,0}(\tau) \|_q$, and $v_t = \max_{T\geq t} u_T$. Since $v_t$ is a nonincreasing sequence, $v_t \leq \rho(\tau)v_{t-L}$, then we have $u_T \leq v_T \leq \rho(\tau)^{T/L}u_0$. Consequently, the sequence $\{\tilde{\boldsymbol{x}}_{T,0}(\tau) \}_{T \in \mathbb{N}}$ constitutes a Cauchy sequence in the $\mathbb{L}^{q}$ space, converging to $ \tilde{\boldsymbol{x}}_{0}(\tau)$ due to the completeness of this space. The $\tilde{\boldsymbol{x}}_{0}(\tau)$ is measurable with respect to the $\sigma-$field generated by $\{\boldsymbol{\epsilon}_t\}_{t \leq 0}$. Therefore, there exists a measurable function $\boldsymbol{J}(\cdot)$ such that $\tilde{\boldsymbol{x}}_{0}(\tau) = \boldsymbol{J}(\tau,\boldsymbol{\epsilon}_0,\boldsymbol{\epsilon}_{-1},\cdots)$. Progressing $t$ time points for $\tilde{\boldsymbol{x}}_{0}(\tau)$ yields $\tilde{\boldsymbol{x}}_{t}(\tau) = \boldsymbol{J}(\tau,\boldsymbol{\epsilon}_t,\boldsymbol{\epsilon}_{t-1},\cdots)$. Therefore, we can readily conclude that $\tilde{\boldsymbol{x}}_{T,t}(\tau)  \rightarrow \tilde{\boldsymbol{x}}_{t}(\tau) $ as $T \rightarrow \infty$ in the $\mathbb{L}^q$ space. As a limit of strictly stationary process in $\mathbb{L}^q$, $\tilde{\boldsymbol{x}}_{t}(\tau)$ is also a stationary process, satisfying $\sup_{\tau \in [0,1]} \|\tilde{\boldsymbol{x}}_{t}(\tau)  \|_q < \infty$. Additionally, $\tilde{\boldsymbol{x}}_{t}(\tau) = \boldsymbol{J}(\tau,\boldsymbol{\epsilon}_t,\boldsymbol{\epsilon}_{t-1},\cdots)$ serves as the limit in $\mathbb{L}^q$ of $\tilde{\boldsymbol{x}}_{T,t}(\tau) = \boldsymbol{J}_T(\tau,\boldsymbol{\epsilon}_t,\boldsymbol{\epsilon}_{t-1},\cdots)$.

\vspace{0.2cm}


\textbf{Proofs of (2).}

We define $\tilde{\boldsymbol{x}}_{t}^*(\tau) = \boldsymbol{J}(\tau,\boldsymbol{\epsilon}_t,\boldsymbol{\epsilon}_{t-1},\cdots,\boldsymbol{\epsilon}_{1}^*)$ and $\tilde{\boldsymbol{x}}_{T,t}^*(\tau) = \boldsymbol{J}_T(\tau,\boldsymbol{\epsilon}_t,\boldsymbol{\epsilon}_{t-1},\cdots,\boldsymbol{\epsilon}_{1}^*)$. The only distinction between $\tilde{\boldsymbol{x}}_{t}^*(\tau)$ and $\tilde{\boldsymbol{x}}_{t}(\tau)$ lies in the last components, which are $\boldsymbol{\epsilon}_{1}$ in the former case and $\boldsymbol{\epsilon}_{1}^*$ in the latter case, respectively. 

Furthermore, we define $u_t = \|\tilde{\boldsymbol{x}}_{T,t}(\tau) - \tilde{\boldsymbol{x}}_{T,t}^*(\tau) \|_q$. By definition, $u_t = 0$ if $t < 0$, and $u_0 = \|\tilde{\boldsymbol{x}}_{T,0}(\tau) - \tilde{\boldsymbol{x}}_{T,0}^*(\tau) \|_q = \mathcal{O}(\|\boldsymbol{\epsilon}_{1} - \boldsymbol{\epsilon}_{1}^* \|_q) = \mathcal{O}(1)$. For $t > 0$, Assumption~3 implies that
\begin{align*}
    u_t = \|\tilde{\boldsymbol{x}}_{T,t}(\tau) - \tilde{\boldsymbol{x}}_{T,t}^*(\tau) \|_q \leq \sum_{j=0}^{L} \alpha_j(\boldsymbol{\theta}(\tau)) \|\tilde{\boldsymbol{x}}_{T,t-j}^*(\tau) - \tilde{\boldsymbol{x}}_{T,t-j}^*(\tau)\|_q.
\end{align*}
We have $u_t \leq u_0$ for $\forall t >0$.  Therefore, we have $\|\tilde{\boldsymbol{x}}_{T,t}(\tau) - \tilde{\boldsymbol{x}}_{T,t}^*(\tau) \| \rightarrow 0$ as $T \rightarrow \infty$.

For $t > 0$, as $t \rightarrow \infty$ , $T \rightarrow \infty$. Based on the result of the proof (1), we have $\|\tilde{\boldsymbol{x}}_{t}(\tau) - \tilde{\boldsymbol{x}}_{T,t}(\tau) \|_q \rightarrow 0$ and $\|\tilde{\boldsymbol{x}}^*_{t}(\tau) - \tilde{\boldsymbol{x}}^*_{T,t}(\tau) \|_q \rightarrow 0$ as $T \rightarrow \infty$. Therefore, we have
\begin{align*}
    \|\tilde{\boldsymbol{x}}_{t}(\tau) - \tilde{\boldsymbol{x}}^*_{t}(\tau)  \|_q \leq \|\tilde{\boldsymbol{x}}_{t}(\tau) - \tilde{\boldsymbol{x}}_{T,t}(\tau) \|_q + \|\tilde{\boldsymbol{x}}_{T,t}(\tau) - \tilde{\boldsymbol{x}}^*_{T,t}(\tau) \|_q + \|\tilde{\boldsymbol{x}}^*_{T,t}(\tau) - \tilde{\boldsymbol{x}}^*_{t}(\tau) \|_q.
\end{align*}
 Since $\|\tilde{\boldsymbol{x}}_{t}(\tau) - \tilde{\boldsymbol{x}}^*_{t}(\tau)  \|_q \rightarrow 0$ as $t \rightarrow \infty$ holds for $\forall \tau \in [0,1]$, we can conclude that $\sup_{\tau \in [0,1]}\|\tilde{\boldsymbol{x}}_{t}(\tau) - \tilde{\boldsymbol{x}}^*_{t}(\tau)  \|_q \rightarrow 0$ as $t \rightarrow \infty$.

\subsection{Proofs of Proposition~2}
The proofs of Proposition~2 are motivated by \cite{Dahlhaus2019towards,gao2024time}.

\textbf{Proofs of (1).}
\begin{align*}
    & \|\widetilde{\boldsymbol{x}}_t(\tau)-  \widetilde{\boldsymbol{x}}_t(\tau^{\prime})\|_q   \\
    & = \|f(\widetilde{\boldsymbol{x}}_t(\tau),\widetilde{\boldsymbol{x}}_{t-1}(\tau),\ldots,\widetilde{\boldsymbol{x}}_{t-L}(\tau); \boldsymbol{\theta}(\tau)) - f(\widetilde{\boldsymbol{x}}_t(\tau^\prime),\widetilde{\boldsymbol{x}}_{t-1}(\tau^\prime),\ldots,\widetilde{\boldsymbol{x}}_{t-L}(\tau^\prime); \boldsymbol{\theta}(\tau^\prime)) \|_q \\
    & \leq  \|f(\widetilde{\boldsymbol{x}}_t(\tau),\widetilde{\boldsymbol{x}}_{t-1}(\tau),\ldots,\widetilde{\boldsymbol{x}}_{t-L}(\tau); \boldsymbol{\theta}(\tau)) - f(\widetilde{\boldsymbol{x}}_t(\tau^\prime),\widetilde{\boldsymbol{x}}_{t-1}(\tau^\prime),\ldots,\widetilde{\boldsymbol{x}}_{t-L}(\tau^\prime); \boldsymbol{\theta}(\tau)) \|_q \\
    & + \|f(\widetilde{\boldsymbol{x}}_t(\tau^\prime),\widetilde{\boldsymbol{x}}_{t-1}(\tau^\prime),\ldots,\widetilde{\boldsymbol{x}}_{t-L}(\tau^\prime); \boldsymbol{\theta}(\tau)) - f(\widetilde{\boldsymbol{x}}_t(\tau^\prime),\widetilde{\boldsymbol{x}}_{t-1}(\tau^\prime),\ldots,\widetilde{\boldsymbol{x}}_{t-L}(\tau^\prime); \boldsymbol{\theta}(\tau^\prime)) \|_q \\
    & \leq \sum_{j=0}^{L} \alpha_j(\boldsymbol{\theta}(\tau)) \|\widetilde{\boldsymbol{x}}_{t-j}(\tau)-\widetilde{\boldsymbol{x}}_{t-j}(\tau^{\prime}) \|_q + |\boldsymbol{\theta}(\tau) - \boldsymbol{\theta}(\tau^\prime) | \sum_{j=0}^{L} \chi_j |\widetilde{\boldsymbol{x}}_t(\tau^{\prime})| \\
    & \leq \sum_{j=0}^{L} \alpha_j(\boldsymbol{\theta}(\tau)) \|\widetilde{\boldsymbol{x}}_{t-j}(\tau)-\widetilde{\boldsymbol{x}}_{t-j}(\tau^{\prime}) \|_q + M|\tau - \tau^\prime| \sum_{j=0}^{L} \chi_j |\widetilde{\boldsymbol{x}}_t(\tau^{\prime})|, \\
\end{align*}
where the first step follows from the definition in Proposition~1, the second step follows from Triangle Inequality, the third step follows from Assumption~3, and finally the fourth step holds since there exists an integer $\exists M$ that $M|\tau - \tau^\prime| \geq |\boldsymbol{\theta}(\tau) - \boldsymbol{\theta}(\tau^\prime) |$ following Assumption~3.  

Since $\widetilde{\boldsymbol{x}}_{t-j}(\tau)$ and $\widetilde{\boldsymbol{x}}_{t-j}(\tau^{\prime})$ are stationary process,  we have
\begin{align}
    \label{eq:station_bound}
    \|\widetilde{\boldsymbol{x}}_t(\tau)- \widetilde{\boldsymbol{x}}_t(\tau^{\prime})\|_q \leq M[1-\sum_{j=0}^{L} \alpha_j(\boldsymbol{\theta}(\tau))]^{-1}|\tau - \tau^\prime| \sum_{j=0}^{L} \chi_j |\widetilde{\boldsymbol{x}}_t(\tau^{\prime})| =  O(|\tau-\tau^{\prime}|).
\end{align}

\textbf{Proofs of (2).}

\begin{align*}
    & \|\boldsymbol{x}_t(\tau_t)-  \widetilde{\boldsymbol{x}}_t(\tau_t)\|_q   \\
    & = \|f(\boldsymbol{x}_t,\boldsymbol{x}_{t-1},\ldots,\boldsymbol{x}_{t-L}; \boldsymbol{\theta}(\tau_t)) - f(\widetilde{\boldsymbol{x}}_t(\tau_t),\widetilde{\boldsymbol{x}}_{t-1}(\tau_t),\ldots,\widetilde{\boldsymbol{x}}_{t-L}(\tau_t); \boldsymbol{\theta}(\tau_t)) \|_q \\
    & \leq \sum_{j=0}^{L} \alpha_j(\boldsymbol{\theta}(\tau)) \|\boldsymbol{x}_{t-j}-\widetilde{\boldsymbol{x}}_{t-j}(\tau_t) \|_q  \\
    & \leq \sum_{j=0}^{L} \alpha_j(\boldsymbol{\theta}(\tau)) \|\boldsymbol{x}_{t-j}-\widetilde{\boldsymbol{x}}_{t-j}(\tau_{t-j}) \|_q  + \sum_{j=0}^{L} \alpha_j(\boldsymbol{\theta}(\tau)) \|\widetilde{\boldsymbol{x}}_{t-j}(\tau_{t-j}) -\widetilde{\boldsymbol{x}}_{t-j}(\tau_t) \|_q  \\
    & \leq \sum_{j=0}^{L} \alpha_j(\boldsymbol{\theta}(\tau)) \|\boldsymbol{x}_{t-j}-\widetilde{\boldsymbol{x}}_{t-j}(\tau_{t-j}) \|_q  + M \sum_{j=0}^{L} \frac{j}{T} \alpha_j(\boldsymbol{\theta}(\tau)), \\
\end{align*}
where the first step follow from the definition of $\boldsymbol{x}_t(\tau)$ and $ \widetilde{\boldsymbol{x}}_t(\tau_t)$, the second step follows from Assumption~3, the third step follows from Triangle Inequality, and the fourth step follows from the Eq.~\eqref{eq:station_bound}.

We can show that $\|\boldsymbol{x}_1(\tau)-  \widetilde{\boldsymbol{x}}_1(\tau_1)\|_q = O(T^{-1})$. Based on the second assumption in Assumption 3, we have
\begin{align*}
        \|\boldsymbol{x}_t(\tau)- \widetilde{\boldsymbol{x}}_t(\tau_t)\|_q \leq \sum_{j=0}^{L} \alpha_j(\boldsymbol{\theta}(\tau)) O(T^{-1})  + M \sum_{j=0}^{L} \frac{j}{T} \alpha_j(\boldsymbol{\theta}(\tau)) = O(T^{-1}).
\end{align*}


\subsection{Law of Large Numbers}

\begin{proposition}[\textbf{Law of Large Numbers}]
    \label{prp:stationary_approximation}
    Assuming DYNAMO satisfies the Assumption 3 and 4. Then based on Proposition~2 for each $\tau_t \in (0,1)$, as $(T,h) \rightarrow (\infty,0), Th \rightarrow \infty$,
        \begin{align*}
            \frac{1}{Th}\sum_{l=1}^TK_h(\tau_l-\tau_t)\cdot \boldsymbol{x}_l \to\mathbb{E} \tilde{\boldsymbol{x}}_l(\tau_t).
        \end{align*}\vspace{-0.4cm}
\end{proposition}
Proposition~\ref{prp:stationary_approximation} shows the law of large numbers governing non-stationary processes $\{ \boldsymbol{x}_t \}$. The proof stems from the premise that a locally stationary process can be approximated by a stationary process, and the discrepancy between any pair of stationary processes is bounded, as shown in Proposition~2.

\textbf{Proof}. Before starting the proof, we introduce the following Lemma

\begin{lemma}
    \label{lem:appendix}
    Assume that $\boldsymbol{z}_t$ is a stationary and ergodic process with $\mathbb{E} |\boldsymbol{z}_1| < \infty$. Assume that $\tau \in (0,1)$. Let $h=h_T \rightarrow 0$ such that $Th \rightarrow \infty$. Then the following convergence holds in $\mathbb{L}^d$:
    \begin{align*}
        \frac{1}{Th}\sum_{t=1}^TK_h(t/n - \tau)\cdot \boldsymbol{z}_t \to\mathbb{E} \boldsymbol{z}_1.
    \end{align*}
\end{lemma}

To prove Proposition~\ref{prp:stationary_approximation}, we start with
\begin{align*}
    & \frac{1}{Th}\sum_{l=1}^TK_h(\tau_l-\tau_t)\cdot \| \boldsymbol{x}_l - \tilde{\boldsymbol{x}}_l(\tau_t)\| \\
    \leq & \frac{1}{Th}\sum_{l=1}^TK_h(\tau_l-\tau_t)\cdot ( \| \boldsymbol{x}_l - \tilde{\boldsymbol{x}}_l(\tau_l)\| + \| \tilde{\boldsymbol{x}}_l(\tau_l) - \tilde{\boldsymbol{x}}_l(\tau_t)\| ) \\
    \leq & \sup_l |K_h(\tau_l-\tau_t)| \cdot (\sup_{l=1,\ldots,T} \|\boldsymbol{x}_l - \tilde{\boldsymbol{x}}_l(\tau_l) \| + \sup_{|\tau_l -\tau_t| \leq h/2} \| \tilde{\boldsymbol{x}}_l(\tau_l) - \tilde{\boldsymbol{x}}_l(\tau_t)\| ) \\
    \leq & \sup_l |K_h(\tau_l-\tau_t)| \cdot (O(T^{-1}) + O(|\tau_l-\tau_t|)),
\end{align*}
where the first step follows a triangle inequality, the second step follows the Assumption~4 of Kernel Function $K_h(\cdot)$, and the third step follows from the Proposition~2. 

Then based on the Lemma~\ref{lem:appendix}, we have
\begin{align*}
    \frac{1}{Th}\sum_{l=1}^TK_h(\tau_l-\tau_t)\cdot \boldsymbol{x}_l \to \frac{1}{Th}\sum_{l=1}^TK_h(\tau_l-\tau_t)\cdot \tilde{\boldsymbol{x}}_l(\tau_t)  \to \mathbb{E} \tilde{\boldsymbol{x}}_l(\tau_t),
\end{align*}
where the last step follows since $\tilde{\boldsymbol{x}}_l(\tau_t)$ is a stationary process.

\subsection{Proofs of Theorem~1}

The proof of the identifiability of DYNAMO can be divided into two steps. Firstly, we establish general conditions indicating that the DYNAMO loss $\mathscr{L}_{t}(\boldsymbol{\vartheta})$ in Eq.~3 can be effectively approximated by loss function $L_{t}(\boldsymbol{\vartheta})$ under stationary time series $\{\tilde{\boldsymbol{x}}_t\}$, where
\begin{align}
    \label{eq:identify_stationary}
    L_{t}(\boldsymbol{\vartheta})=\frac1T\sum_{l=1}^T\ell(\tilde{\boldsymbol{x}}_l,\tilde{Y}_l;\boldsymbol{\vartheta}).
\end{align}
Secondly, we establish the identifiability of stationary time series $\{\tilde{\boldsymbol{x}}_t \}$ in both linear and nonlinear scenarios. Since stationary time series $\{\tilde{\boldsymbol{x}}_t \}$ shares the same causal structure of non-stationary process $\{\boldsymbol{x}_t(\tau_t) \}$ at time $t$, we demonstrate that time-varying causal structures within non-stationary time series $\{\boldsymbol{x}_t(\tau_t) \}$ can be identified with only the mild assumption described in Assumption~3.

The overall proof of DYNAMO's identifiability is supported by the lemmas presented below.  We first introduce a class of loss functions that facilitates the approximation of the DYNAMO loss $\mathscr{L}_{t}(\boldsymbol{\vartheta})$ by the stationary loss $L_{t}(\boldsymbol{\vartheta})$.

\begin{definition}[The class $\tilde{\mathcal{L}}_{L+1}(M,C)$ in \cite{Dahlhaus2019towards}]
    \label{def:l_class}
    A function $g: \mathbb{R}^{L+1} \times \Theta \to \mathbb{R}$ is in the class $\tilde{\mathcal{L}}_{L+1}(M,C)$ with $C = (C_z,C_\theta)$ and constant $C_z, C_\theta \geq 0$ and $M \geq 0$ if for all $z \in \mathbb{R}^{p+1}$, $\theta \in \Theta$. It holds that $g(\cdot, \theta) \in \mathcal{L}_{L+1}(M,C_z)$ and $g(z,\cdot) \in \mathcal{L}_d(0, C_\theta(1+|z|_1^{M+1}))$, where a function $g: \mathbb{R}^{r} \to \mathbb{R}$ is in the class of $\mathcal{L}_{r}(M,C)$ if $M \geq 0$ and
    \begin{align*}
        \sup_{y\neq y^{\prime}}\frac{|g(y)-g(y^{\prime})|}{|y-y^{\prime}|_1\cdot(1+|y|_1^M+|y^{\prime}|_1^M)}\leq C,
    \end{align*}
    where $|y|_1:=\sum_{i=1}^r|y_i|$.
\end{definition}
Following the above definition, satisfying Assumption 3 naturally leads to the loss function $\ell \in \tilde{\mathcal{L}}_{L+1}(M,C)$. Hence, both the linear and nonlinear DYNAMO loss in Eq. 4 and Eq.~5 belong to the class $\tilde{\mathcal{L}}_{L+1}(M,C)$. Next, we demonstrate that the $\mathscr{L}_{t}(\boldsymbol{\vartheta})$ approximates the estimation of the loss function $L_{t}(\boldsymbol{\vartheta})$ within stationary causal processes as $Th \rightarrow \infty$.

\begin{lemma}
    \label{lem:loss_approximation}
    Assuming DYNAMO satisfies the Assmuption 3, and 4. Based on Proposition~1 and Proposition~\ref{prp:stationary_approximation}, for all $t \in (0,T)$ with $h\rightarrow0$ and $Th \rightarrow \infty$,
    \begin{align*}
        \sup_{\vartheta\in\Theta}|\mathscr{L}_{t}(\boldsymbol{\vartheta})-L_{t}(\boldsymbol{\vartheta})|\overset{p}{\operatorname*{\to}}0.
    \end{align*}
\end{lemma}
The proof of Lemma~\ref{lem:loss_approximation} relies on the convergence of that $\mathscr{L}_{t}(\boldsymbol{\vartheta})$ to $L_{t}(\boldsymbol{\vartheta})$ with probability 1 as $Th \rightarrow \infty$, as well as on the continuity of both functions with respect to  $\boldsymbol{\vartheta}$. The detailed proof is shown in the Appendix~\ref{appx:proofs}. Consequently, given the identifiability of stationary causal structures, the identifiability of locally stationary processes follows. Next, we delineate the conditions under which the causal structure of stationary time series data can be identified.

\begin{lemma}
    \label{lem:stationary_identify}
    The causal structure of the stationary process $\{ \tilde{\boldsymbol{x}}_t\}$ in Eq.~(2) is identifiable under the $L_{t}(\boldsymbol{\vartheta})$ under a broad class of identifiable models.
\end{lemma}

Based on the Lemma~\ref{lem:loss_approximation} and \ref{lem:stationary_identify},the proof of Theorem~\ref{Thm:identify_dynamo} is straightforward, as $\mathscr{L}_{t}(\boldsymbol{\vartheta})$ and $L_{t}(\boldsymbol{\vartheta})$ in Eq.~\eqref{eq:identify_stationary} estimate the same causal structure at time $t$. Consequently, the identifiability of stationary causal processes $\{ \tilde{\boldsymbol{x}}_t\}$ naturally extends to encompass the identifiability of the non-stationary causal processes $\{\boldsymbol{x}_t(\tau_t) \}$ at time $t$.

\subsubsection{Proofs of Lemma~\ref{lem:loss_approximation}}
The proof of Lemma~\ref{lem:loss_approximation} is divided into two parts. In the first part, we aim to establish $\mathscr{L}_t(\vartheta) \overset{p}{\operatorname*{\to}} L_t(\vartheta)$. In the second part, we demonstrate the continuity of both $\boldsymbol{\theta} \mapsto L_t(\vartheta)$ and $\boldsymbol{\theta} \mapsto \mathscr{L}_t(\vartheta)$. This enables us to prove $        \sup_{\vartheta\in\Theta}|\mathscr{L}_{t}(\boldsymbol{\vartheta})-L_{t}(\boldsymbol{\vartheta})|\overset{p}{\operatorname*{\to}}0$. For each $\hat{\boldsymbol{\theta}}(\tau) \in \Theta_r$, we can disregard the term $\tau$, as the subsequent proof applies uniformly for every fixed $\tau \in (0,1)$.

The objective of the first part is to prove that as $h \to 0$ and $Th \to \infty$,
\begin{align*}
    \frac{1}{Th}\sum_{l=1}^TK_h(\tau_l-\tau_t) \cdot \ell(\boldsymbol{x}_l,Y_{l-1},\hat{\boldsymbol{\theta}})  - \frac{1}{T}\sum_{l=1}^T \ell(\tilde{\boldsymbol{x}}_l,\tilde{Y}_{l-1},\hat{\boldsymbol{\theta}}) \overset{p}{\operatorname*{\to}} 0.
\end{align*}

Based on Lemma~\ref{lem:appendix}, as $h \to 0$ and $Th \to \infty$, we have
\begin{align*}
    \frac{1}{Th}\sum_{t=1}^TK_h(t/n - \tau)\cdot \ell(\tilde{\boldsymbol{x}}_l,\tilde{Y}_{l-1},\hat{\boldsymbol{\theta}}) \to \frac{1}{T}\sum_{l=1}^T \ell(\tilde{\boldsymbol{x}}_l,\tilde{Y}_{l-1},\hat{\boldsymbol{\theta}}).
\end{align*}

Therefore, our task simplifies to proving $\frac{1}{Th}\sum_{l=1}^TK_h(\tau_l-\tau_t) \|\ell(\boldsymbol{x}_l,Y_{l-1},\hat{\boldsymbol{\theta}}) - \ell(\boldsymbol{x}_l,Y_{l-1},\hat{\boldsymbol{\theta}})   \|_q \overset{p}{\operatorname*{\to}} 0$. Based on the Assumption~3 and Definition~\ref{def:l_class}, we have
\begin{align*}
    & \frac{1}{Th}\sum_{l=1}^TK_h(\tau_l-\tau_t) \|\ell(\boldsymbol{x}_l,Y_{l-1},\hat{\boldsymbol{\theta}}) - \ell(\tilde{\boldsymbol{x}}_l,\tilde{Y}_{l-1},\hat{\boldsymbol{\theta}})   \|_q   \\
    = & \frac{1}{Th}\sum_{l=1}^TK_h(\tau_l-\tau_t) \|\ell(X_l,\hat{\boldsymbol{\theta}}) - \ell(\tilde{X_l},\hat{\boldsymbol{\theta}})\| \\
    \leq & \frac{1}{Th}\sum_{l=1}^TK_h(\tau_l-\tau_t) \cdot (\|X_l -\tilde{X_l}\| \cdot (1+\|X_l\|^M + \|\tilde{X_l}\|^M )) \rightarrow  0 ,\\
\end{align*}
where the first step follows from $X_l = [\boldsymbol{x}_l,Y_{l-1}]$ and $\tilde{X}_l = [\tilde{\boldsymbol{x}}_l,\tilde{Y}_{l-1}]$ , the second step follows from the Definition~\ref{def:l_class}, and the last step follows from the Assumption~3 and Proposition~\ref{prp:stationary_approximation}.

In the second part, we focus on proving the continuity of $\boldsymbol{\theta} \mapsto L_t(\vartheta)$ and $\boldsymbol{\theta} \mapsto \mathscr{L}_t(\vartheta)$. The function $\boldsymbol{\theta} \mapsto L_t(\vartheta)$ is continuous since
\begin{align*}
    | L_t(\boldsymbol{\theta}) - L_t(\boldsymbol{\theta}^{'})) |  \leq \| \frac1T \sum_{l=1}^T (\ell(\tilde{\boldsymbol{x}}_l,\tilde{Y}_l;\boldsymbol{\theta}) - \ell(\tilde{\boldsymbol{x}}_l,\tilde{Y}_l;\boldsymbol{\theta}^{'}) \| 
    \leq \sup_{l \in [0,T]} \boldsymbol{C}_{\boldsymbol{\theta}} \cdot \|\boldsymbol{\theta} -  \boldsymbol{\theta}^{'} \| \cdot (1+ \| X_t \|_{M+1}^{M+1} ),
\end{align*}
where the definition holds due to Definition~\ref{def:l_class}. Now, we need to demonstrate the stochastic equicontinuity of $\mathscr{L}_t(\vartheta)$. We define $h: \mathbb{R}^{(L+1)\times d} \to \mathbb{R}, h(X_t) = C_{\boldsymbol{\theta}}(1+\|X_t \|_1^{M+1})$. It is evident that
\begin{align*}
    | \mathscr{L}_t(\boldsymbol{\theta}) - \mathscr{L}_t(\boldsymbol{\theta}^{'})) |  & \leq \| \frac1T \sum_{l=1}^T K_h(\tau_l-\tau_t)\cdot (\ell(\boldsymbol{x}_l,Y_l;\boldsymbol{\theta}) - \ell(\boldsymbol{x}_l,Y_l;\boldsymbol{\theta}^{'}) \| \\
    & \leq  \|\boldsymbol{\theta} -  \boldsymbol{\theta}^{'} \| \cdot \frac{1}{Th}\sum_{l=1}^T K_h(\tau_l-\tau_t) \cdot h(X_t).
\end{align*}
Note that $h \in \mathcal{L}_{L+1}(M,C)$ with some constant $C > 0$. Based on the Proposition~2, for all $\tau \in (0,1)$ we have:
\begin{align*}
    \frac{1}{Th}\sum_{l=1}^T K_h(\tau_l-\tau_t) \cdot h(X_t) \overset{p}{\operatorname*{\to}} \int |K| dx \cdot \mathbb{E}h(\tilde{X_t}) =: c(t).
\end{align*}
Choosing $\delta = \frac{\eta}{2c(t)}$ yields
\begin{align*}
    \mathbb{P} & (\sup_{\|\boldsymbol{\theta} - \boldsymbol{\theta} \| \leq \eta} |\mathscr{L}_t(\boldsymbol{\theta}) - \mathscr{L}_t(\boldsymbol{\theta}^{'})) | > \eta ) \\
    & \leq \mathbb{P}(|\frac{1}{Th}\sum_{l=1}^T |K_h(\tau_l-\tau_t)| \cdot h(X_t) -c(u) | > c(u)) \to 0,
\end{align*}
as $h \to 0$ and $Th \to \infty$. Hence, this gives
\begin{align*}
    \sup_{\vartheta\in\Theta}|\mathscr{L}_{t}(\boldsymbol{\vartheta})-L_{t}(\boldsymbol{\vartheta})|\overset{p}{\operatorname*{\to}}0.
\end{align*}

\subsubsection{Proofs of Lemma~\ref{lem:stationary_identify}}

In this section, we provide the identifiability of a broad class of models for stationary causal processes $\widetilde{\boldsymbol{x}}_t(\tau)$, such as \textit{Gaussian DAG} \cite{peters2014identifiability}, \textit{Non-Gaussian DAG} \cite{lanne2017identification}, \textit{Additive Noise Models} \cite{peters2014causal}, and \textit{Post-Nonlinear Models} \cite{gong2022rhino}, among others. The general form of our model is as follows:
\begin{align}
    \widetilde{\boldsymbol{x}}_t(\tau)=f\left(\widetilde{\boldsymbol{x}}_t(\tau),\widetilde{Y}_{t-1}(\tau),\boldsymbol{\epsilon}_t;\boldsymbol{\theta}(\tau)\right).
    \label{eq:stationary_dag}
\end{align}

\textbf{Gaussian DAG} refers to causal structures in Eq.~\eqref{eq:stationary_dag} with errors $\epsilon_t$ that are jointly independent and normally distributed with constant variances. The identifiability in Gaussian DAG $\{\widetilde{\boldsymbol{x}}_t(\tau)\}$ can be readily obtained by Theorem 1 of \cite{peters2014identifiability}. 

\textbf{Non-Gaussian DAG} refers to causal structures in Eq.~\eqref{eq:stationary_dag} with non-Gaussian errors $\epsilon_t$. The identifiability in Non-Gaussian DAG is a well-known consequence of Marcinkiewicz’s theorem on the cumulants
of the normal distribution \cite{shimizu2011directlingam,lanne2017identification}. For detailed proofs of Non-Gaussian DAG, we recommend referring to \citet{lanne2017identification}.

\textbf{Additive-Noise Models} refer to causal structures in Eq.~\eqref{eq:stationary_dag} with the following form:
\begin{align}
    \widetilde{\boldsymbol{x}}_t(\tau)=f\left(\widetilde{\boldsymbol{x}}_t(\tau),\widetilde{Y}_{t-1}(\tau);\boldsymbol{\theta}(\tau)\right) + \boldsymbol{\epsilon}_t.
    \label{eq:additive_noise}
\end{align}

\textbf{Post-Nonlinear models} refer to causal structures in Eq.~\eqref{eq:stationary_dag} with the following form:
\begin{align}
    \widetilde{\boldsymbol{x}}_t(\tau)=v \left( f\left(\widetilde{\boldsymbol{x}}_t(\tau),\widetilde{Y}_{t-1}(\tau)\right) + \boldsymbol{\epsilon}_t \right).
    \label{eq:post_nonlinear}
\end{align}

For Additive-Noise Models and Post-Nonlinear models, we require two mild additional assumptions to establish the identifiability: Causal Minimality, and Well-defined Density.

\begin{assumption}[Causal Minimality in \citet{gong2022rhino}]
\label{asm:causal_minimal}
    Consider a distribution $p$ and a DAG $G$, we say this distribution
satisfies causal minimality w.r.t $G$ if it is Markovian w.r.t. $G$ but not to any proper subgraph of $G$.
\end{assumption}

\begin{assumption}[Well-defined Density in \citet{gong2022rhino}]
\label{asm:well_density}
    We assume that the joint likelihood induced by the structural equation models (SEM)
(Eq.~\eqref{eq:stationary_dag}) is absolutely continuous w.r.t a Lebesgue or counting measure, and is 3rd-order differentiable. For all possible $G$, $|\log p(\tilde{\boldsymbol{x}}_1,\ldots,\tilde{\boldsymbol{x}}_T; G) | < \infty$. 
\end{assumption}

Following \citet{gong2022rhino}, we break down the entire proof into three steps:
\begin{enumerate}
    \item Establish a general condition for the bivariate time series model.
    \item Demonstrate that the SEMs in Eq.~\eqref{eq:additive_noise} and Eq.~\eqref{eq:post_nonlinear} satisfy the conditions outlined in step 1.
    \item Extend the identifiability established in the previous step to the multivariate case.
\end{enumerate}

\textbf{Step 1.} Our model Eq.~\eqref{eq:stationary_dag} shares the same SEM structure with the model in Eq.~16 of \citet{gong2022rhino} in the bivariate case. Hence, we can directly apply Theorem 3 from \citet{gong2022rhino} to determine the identifiability conditions for bivariate time series.

\begin{theorem}[Theorem 3 in \citet{gong2022rhino}]
    Assuming Assumptions 1, 2, \ref{asm:causal_minimal}, \ref{asm:well_density}  are satisfied, given a bivariate temporal process $\{ \tilde{x}_t^{[i]},\tilde{x}_t^{[j]} \}$ that is governed by the SEM Eq.~\ref{eq:transition_model} with source model Eq.~\ref{eq:source_model}, then the SEM for the bivariate temporal process $\{ \tilde{x}_t^{[i]},\tilde{x}_t^{[j]} \}$ is structurally identifiable if the following conditions are true
    \begin{enumerate}
        \item Source model Eq.~\ref{eq:source_model} is structurally identifiable for all $i = 1,\ldots, d$ and $s \in [0,S]$.
        \item The transition model Eq.~\ref{eq:transition_model} is \textbf{bivariate identifiable} w.r.t the instantaneous parents. Namely, if graph $G$ induced conditional distributions $p(\tilde{x}_t^{[i]},\tilde{x}_t^{[j]}|Pa_G^{i,j}(<t))$ , then $\nexists G^\prime$ such that $G = G^\prime$ and the induced conditional $\Bar{p}(\tilde{x}_t^{[i]},\tilde{x}_t^{[j]}|\overline{Pa}_{G^\prime}^{i,j}(<t))$ for all $t \in [S+1, T]$ .
    \end{enumerate}
\end{theorem}
where $Pa_G^{i,j}(<t))$ is the union of the lagged parents of $\tilde{x}_t^{[i]}$ and $\tilde{x}_t^{[j]}$ under $G$, and $\overline{Pa}_{G^\prime}^{i,j}(<t))$ is the lagged parents under $G^\prime$. The source model is 
\begin{align}
    \widetilde{x}_t^{[i]}= f_{i,t}\left(Pa_G^i(<t),Pa_G^i(t),\epsilon_t^i\right),
    \label{eq:source_model}
\end{align}
for $t \in [S+1,T]$. The transition model is
\begin{align}
    \widetilde{x}_s^{[i]}= f_{s,t}\left(Pa_G^i,\epsilon_s^i\right),
    \label{eq:transition_model}
\end{align}
for $s \in [0, S]$.

\textbf{Step 2.} We establish the Additive Noise model and Post-Nonlinear model are bivariate identifiable w.r.t the instantaneous parents in the transition model Eq.~\eqref{eq:transition_model}. 

We assume $\tilde{x}_t^{[i]} \rightarrow \tilde{x}_t^{[j]}$ represents as an instantaneous relationship in $G$, and $\tilde{x}_t^{[j]} \rightarrow \tilde{x}_t^{[i]}$ in $G^\prime$. Here, $\boldsymbol{h}$ represents the entire history $\tilde{x}_{0:t-1}^{[i]} \cup \tilde{x}_{0:t-1}^{[j]}$. Thus, $h_G^i = Pa_G^i(<t) \in \boldsymbol{h}$, $h_G^j = Pa_G^j(<t) \in \boldsymbol{h}$, $\bar{h}_{G^\prime}^i = \overline{Pa}_{G^\prime}^i(<t) \in \boldsymbol{h}$, and $\bar{h}_{G^\prime}^j = \overline{Pa}_{G^\prime}^j(<t) \in \boldsymbol{h}$. 

For \textbf{Additive Noise models}, we have
\begin{align*}
    & \tilde{x}_t^{[j]} = f(h_G^j,\tilde{x}_t^{[j]}) + \epsilon_t^j, \quad \text{under}~ G,\\
    & \tilde{x}_t^{[i]} = \bar{f}(\bar{h}_{G^\prime}^i,\tilde{x}_t^{[i]}) + \epsilon_t^i, \quad \text{under}~ G^\prime.
\end{align*}
The equations above align with Eq.~2 and Eq.~3 in \citet{hoyer2008nonlinear}. Consequently, we can utilize Theorem 1 of \citet{hoyer2008nonlinear} to establish the bivariate identifiability of Additive Noise models.

For \textbf{Post-Nonlinear models}, we have
\begin{align*}
    & \tilde{x}_t^{[j]} = v(f(h_G^j,\tilde{x}_t^{[j]}) + \epsilon_t^j), \quad \text{under}~ G,\\
    & \tilde{x}_t^{[i]} = \bar{v}(\bar{f}(\bar{h}_{G^\prime}^i,\tilde{x}_t^{[i]}) + \epsilon_t^i), \quad \text{under}~ G^\prime.
\end{align*}

The equations above correspond to Eq.~2 and Eq.~3 in \citet{zhang2009identifiability}. Therefore, we can establish the bivariate identifiability of Post-Nonlinear models using Theorem 1 from \citet{zhang2009identifiability}, which covers bivariate identifiability in most cases, except for 5 special cases (see Table 1 in \cite{zhang2009identifiability} for details). Moreover, Theorem 4 in \citet{gong2022rhino} extends the bivariate identifiability to more general Post-Nonlinear models, history-dependent Post-Nonlinear models.

\textbf{Step 3.} The generalized model in Eq.~\eqref{eq:stationary_dag} shares the same structure as Eq.~16 in \citet{gong2022rhino}. Therefore, we can directly apply Theorem 3 from \citet{gong2022rhino}, which establishes that if the SEM is bivariate identifiable w.r.t. instantaneous parents for each pair of nodes,  and the source model is identifiable, then the model in Eq.~\eqref{eq:stationary_dag} under stationary process $\{ \widetilde{\boldsymbol{x}}_t(\tau) \}$ is identifiable in multivariate cases. 

Finally, we complete the proofs of identifiability of stationary process $\{ \widetilde{\boldsymbol{x}}_t(\tau) \}$ in Eq.~\eqref{eq:stationary_dag}.

\subsection{Proofs of Theorem~\ref{Thm:consistency_dynamo}}
In this section, we aim to calculate the convergence rate of $\|\boldsymbol{\widehat{\theta}}(\tau_t) - \boldsymbol{\theta}(\tau_t) \|$, where $\boldsymbol{\theta}(\tau_t)$ contains the ground truth causal structures and $\boldsymbol{\widehat{\theta}}(\tau_t)$ is the minimizer of DYNAMO loss in Eq.~\eqref{eq:constant_likelihood}. Before diving into the proof details, it is necessary to introduce an additional smoothness assumption.

\begin{assumption}[\textbf{Smoothness}]
    \label{asm:smoothness}
    $f(\cdot; \boldsymbol{\vartheta})$ is twice continuously differentiable with respect to $\boldsymbol{\vartheta}$. And there exist non-negative sequences $\{ \chi_j \}$ with $\sum_{j=1}^{L+1}\chi_{j}<\infty $ such that for any $ Z,Z^\prime \in \mathbb{R}^{(L+1)\times d}$ and any $\vartheta, \vartheta^\prime \in \Theta_r$ with $k = 1,2$:
        \begin{align}
            & |\nabla^k_{\boldsymbol{\vartheta}}f(Z,\boldsymbol{\epsilon};\boldsymbol{\vartheta})-\nabla^k_{\boldsymbol{\vartheta}}f(Z,\boldsymbol{\epsilon};\boldsymbol{\vartheta}^{\prime})|\leq|\boldsymbol{\vartheta}-\boldsymbol{\vartheta}^{\prime}|\sum_{j=1}^{L+1}\chi_j|\boldsymbol{z}_j|,\notag \\
            & |\nabla^k_{\boldsymbol{\vartheta}}f(Z,\boldsymbol{\epsilon};\boldsymbol{\vartheta})-\nabla^k_{\boldsymbol{\vartheta}}f(Z^\prime,\boldsymbol{\epsilon};\boldsymbol{\vartheta})|\leq|\sum_{j=1}^{L+1}\chi_j|\boldsymbol{z}_j-\boldsymbol{z}^\prime_j|.\notag
        \end{align}
\end{assumption}

Assumption \ref{asm:smoothness} imposes Lipschitz-type conditions on the first and second-order derivatives of $f_t(\cdot)$ to ensure the smoothness. Based on the Definition~\ref{def:l_class} , it also indicates that $\nabla^k_{\boldsymbol{\vartheta}}\ell \in \tilde{\mathcal{L}}_{L+1}(M,C)$ with $k = 1,2$.

We define $\boldsymbol{\widehat{\theta}}_0(\tau_t)$ as the unique minimizer of $L_{t}(\boldsymbol{\vartheta})$ in Eq.~\eqref{eq:stationary_dag} under stationary causal processes. Hence, we can divide the convergence rate of $\|\boldsymbol{\widehat{\theta}}(\tau_t) - \boldsymbol{\theta}(\tau_t) \|$ into two parts:
\begin{align*}
    \|\boldsymbol{\widehat{\theta}}(\tau_t) - \boldsymbol{\theta}(\tau_t) \| \leq \|\boldsymbol{\widehat{\theta}}(\tau_t) - \boldsymbol{\widehat{\theta}}_0(\tau_t) \| + \|\boldsymbol{\widehat{\theta}}_0(\tau_t) - \boldsymbol{\theta}(\tau_t) \|,
\end{align*}
where the first part $\|\boldsymbol{\widehat{\theta}}(\tau_t) - \boldsymbol{\widehat{\theta}}_0(\tau_t) \|$ denotes the differences between our DYNAMO estimator and its stationary counterparts at time $t$, and the second part $\|\boldsymbol{\widehat{\theta}}_0(\tau_t) - \boldsymbol{\theta}(\tau_t) \|$ denotes the convergence rate in a stationary causal processes. 

\textbf{Convergent rate of $\|\boldsymbol{\widehat{\theta}}(\tau_t) - \boldsymbol{\widehat{\theta}}_0(\tau_t) \|$. } Lemma~\ref{lem:loss_approximation} has shown that  $\mathscr{L}_{t}(\boldsymbol{\vartheta})$ converges to $L_{t}(\boldsymbol{\vartheta})$ with probability 1 as $Th \rightarrow \infty$. We can further calculate its convergent rate based on Theorem 5.4 from \citet{Dahlhaus2019towards} and Theorem 2.1 from \citet{gao2024time}, as Assumption \ref{asm:local_stationary} and \ref{asm:smoothness} are equivalent to their assumptions. These theorems indicate that for any $t \in [0,T]$, $\|\boldsymbol{\widehat{\theta}}(\tau_t) - \boldsymbol{\widehat{\theta}}_0(\tau_t) \| = O(1/\sqrt{Th})$ if $Th \rightarrow \infty$ and $Th^7 \rightarrow 0$. 

Hence, we can complete this proof as long as the $\|\boldsymbol{\widehat{\theta}}_0(\tau_t) - \boldsymbol{\theta}(\tau_t) \| \rightarrow 0$ as $Th \rightarrow \infty$.

\textbf{Convergent rate of  $\|\boldsymbol{\widehat{\theta}}_0(\tau_t) - \boldsymbol{\theta}(\tau_t) \|$.} $\|\boldsymbol{\widehat{\theta}}_0(\tau_t) - \boldsymbol{\theta}(\tau_t) \|$ denotes the convergent rate of the estimated causal graphs to the ground true causal graphs under stationarity assumptions. This convergent speed largely depends on the base learner and its underlying identifiable models. In this paper, we leverage NOTEARS loss and NTS-NOTEARS in linear and nonlinear DYNAMO, respectively, which can be applied in a wide class of identifiable models, including Gaussian DAG, Non-Gaussian DAG, Additive Noise Models, and Post-Nonlinear Models.

For Gaussian DAG, Theorem 5.1 from \citet{cai2023learning} demonstrates that under Gaussian DAG with NOTEAR loss, $\|\boldsymbol{\widehat{\theta}}_0(\tau_t) - \boldsymbol{\theta}(\tau_t) \| = O(\sqrt{\log T/T})$. For other models including Non-Gaussian DAG, Additive Noise Models, and Post-Nonlinear Models, existing literature \cite{gong2022rhino,gao2022idyno} has demonstrated the consistency under stationarity assumptions. 

Hence, we complete the proofs of Theorem~\ref{Thm:consistency_dynamo} by showing that as $(T,h) \rightarrow (\infty,0)$, $Th \rightarrow \infty$, and $Th^7 \rightarrow 0$,
\begin{align*}
    \|\boldsymbol{\widehat{\theta}}(\tau_t) - \boldsymbol{\theta}(\tau_t) \| = \|\boldsymbol{\widehat{\theta}}(\tau_t) - \boldsymbol{\widehat{\theta}}_0(\tau_t) \| + \|\boldsymbol{\widehat{\theta}}_0(\tau_t) - \boldsymbol{\theta}(\tau_t) \| \rightarrow 0.
\end{align*}

\section{Real Data Analysis}
\label{appx:real_data}
Except for the seven key variables, we incorporate several with-in-match control variables following previous literature \citep{price2022much} in our dataset, including metrics such as the number of key passes leading to a shot, the number of tackles, the overall number of throw-ins, the overall number of aerial balls, the number of attempted dribbles, the number of successful dribbles, cumulative count of touches, encompassing passes, shots, dribbles, traps, and tackles, the number of opponent's successful tackles, the number of opponent's interceptions and the number of opponent's clearances. We construct several multivariate time series datasets based on the within-game event data for the four representative teams to explore the dynamic factors contributing to the home field advantages. To mitigate potential confounding factors related to the opponent's abilities, we use the differences in minute-by-minute mean match statistics for each team as input variables. Specifically, the match data for each team was segmented into the home and away parts, and the mean differences between home and away matches were calculated for each team.
We define the dataset $X_t^i \in \mathcal{R}^{T \times d}$, representing the home-away difference statistics of team $i$ at time $t \in \{1,\ldots,T \}$ with $d$ variables, using the formula $X_t^i = \frac{2}{N_i} \sum_{j=1}^{N_i} X_{t}^{ij} \{I(loc_j = 1) - I(loc_j = 0)\}$,
where $I(\cdot)$ is the indicator function, $X_{t}^{ij}  \in \mathcal{R}^{T \times d}$ denotes the match results of the $j$-th match of team $i$ at time $t$, $N_i$ is the number of matches that team $i$ has played in a single season, and $loc_j = 1$ signifies the match being played at home, while $loc_j = 0$ denotes an away match. In the context of association football games, each team faces the same opponent twice, once at home and once away. The value $N_i = 38$ is constant for each team $i$. This formulation effectively controls for potential confounding effects arising from variations in the opponents' abilities \citep{price2022much}. 

To standardize the match duration across games and minimize the influence of varying added time periods on match outcomes, we exclude minutes beyond the 90-minute. The summary statistics of the final data for each team in the 2020-2021 season and the 2021-2022 season are presented in Table~\ref{tab:descriptive2020} and Table~\ref{tab:descriptive2021}, each variable representing the average differences between home and away matches. On average, these four teams except Manchester City have higher expected goals in home matches than in away matches. As anticipated, these four teams exhibit higher passes, shots, pass accuracy, and shot accuracy in home matches. It is evident that the average differences during the 2021-2022 season are larger than those observed during the 2020-2021 season, highlighting a more pronounced home advantage in the later season with the existence of crowd support.

\begin{table}
\begin{center}
\caption{Summary statistics for each team during the 2020-2021 season}
\begin{tabular}{rrrrrrrrr}
\hline
  Variable & Team & Mean & Median & SD & Team & Mean & Median & SD\\\hline
TP & Arsenal& 0.206 & 0.211 & 1.318 & Liverpool & 0.866 & 1.105 & 1.486  \\
TS & Arsenal & 0.011 & 0.000 & 0.105 & Liverpool  & 0.018 & 0.000 & 0.149 \\
PA & Arsenal & 0.013 & 0.026 & 0.011 & Liverpool & 0.053 & 0.041 & 0.011 \\
SA & Arsenal & -0.000 & 0.000 & 0.058 & Liverpool & 0.002 & 0.000 & 0.067 \\
OY & Arsenal & 0.002 & 0.000 & 0.045 & Liverpool  & -0.001 & 0.000 & 0.039 \\
OF & Arsenal & 0.002 & 0.000 & 0.097 & Liverpool  & -0.010 & 0.000 & 0.083 \\
XG & Arsenal & -0.003 & -0.005 & 0.021 & Liverpool & 0.000 & 0.002 & 0.029 \\
TP  & Man City & 0.602 & 0.579 & 1.932 & Man United & 0.707 & 0.684 & 1.666\\
TS  & Man City & 0.029 & 0.053 & 0.129 & Man United & 0.020 & 0.000 & 0.120 \\
PA  & Man City & 0.026 & 0.034 & 0.012 & Man United & 0.044 & 0.049 & 0.011 \\
SA  & Man City & 0.005 & 0.000 & 0.070 & Man United & -0.002 & 0.000 & 0.064\\
OY  & Man City & -0.001 & 0.000 & 0.039 & Man United & -0.007 & 0.000 & 0.042\\
OF  & Man City & -0.014 & 0.000 & 0.114 & Man United & -0.015 & 0.000 & 0.099\\
XG  & Man City & 0.004 & 0.003 & 0.031 & Man United & 0.001 & 0.002 & 0.025\\

\hline
\end{tabular}
\label{tab:descriptive2020}
\end{center}
\end{table}

\begin{table}
\begin{center}
\caption{Summary statistics for each team during the 2021-2022 season}
\begin{tabular}{rrrrrrrrr}
\hline
  Variable & Team & Mean & Median & SD & Team & Mean & Median & SD\\\hline
TP & Arsenal& 0.391 & 0.263 & 1.417 & Liverpool & 0.865 & 1.000 & 1.585  \\
TS & Arsenal & 0.057 & 0.053 & 0.123 & Liverpool  & 0.040 & 0.000 & 0.16 \\
PA & Arsenal & 0.034 & 0.028 & 0.013 & Liverpool & 0.050 & 0.056 & 0.011 \\
SA & Arsenal & 0.013 & 0.000 & 0.060 & Liverpool & 0.004 & 0.000 & 0.084 \\
OY & Arsenal & 0.003 & 0.000 & 0.045 & Liverpool  & -0.001 & 0.000 & 0.041 \\
OF & Arsenal & 0.008 & 0.000 & 0.112 & Liverpool  & -0.002 & 0.000 & 0.082 \\
XG & Arsenal & 0.011 & 0.006 & 0.029 & Liverpool & 0.007 & 0.006 & 0.028 \\
TP  & Man City & 0.116 & 0.158 & 1.718 & Man United & 0.347 & 0.474 & 1.417\\
TS  & Man City & 0.003 & 0.000 & 0.171 & Man United & 0.028 & 0.000 & 0.126 \\
PA  & Man City & 0.014 & 0.029 & 0.012 & Man United & 0.021 & 0.018 & 0.012 \\
SA  & Man City & 0.007 & 0.000 & 0.078 & Man United & 0.005 & 0.000 & 0.064\\
OY  & Man City & -0.001 & 0.000 & 0.041 & Man United & -0.005 & 0.000 & 0.032\\
OF  & Man City & -0.008 & 0.000 & 0.092 & Man United & -0.004 & 0.000 & 0.089\\
XG  & Man City & 0.000 & 0.000 & 0.029 & Man United & 0.003 & 0.003 & 0.021\\

\hline
\end{tabular}
\label{tab:descriptive2021}
\end{center}
\end{table}

\subsection{Robustness Checks}

\begin{figure}[h]
    \begin{center}
    \centerline{\includegraphics[width=0.9\textwidth]{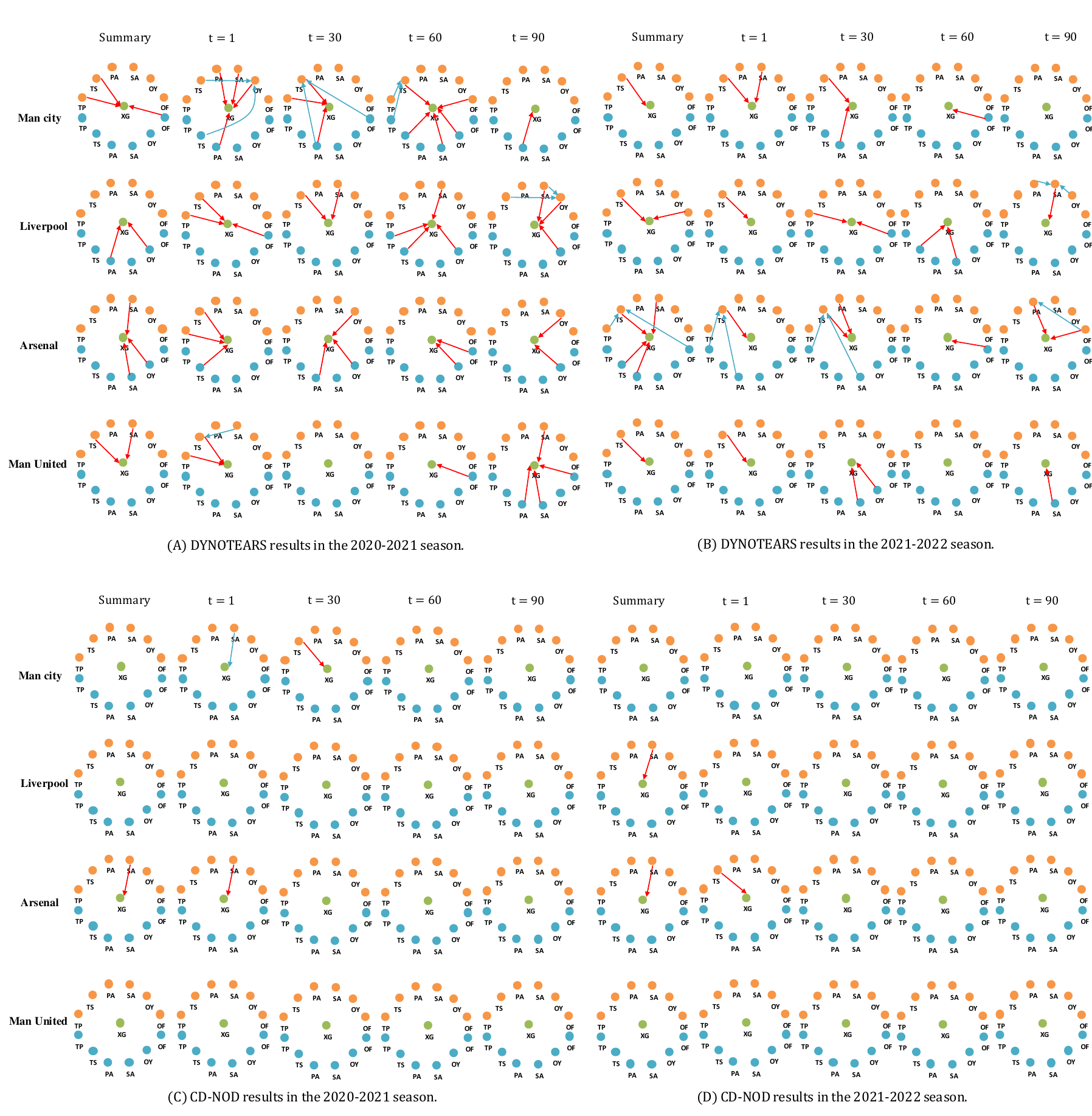}}
    \caption{The model result for 4 representative teams by DYNOTEARS and CD-NOD. Variables: TP - Total Passes, TS - Total Shots, PA - Pass Accuracy, SA - Shot Accuracy, OY - Opponent's yellow cards, OF - Opponent's Fouls. Orange notes represent contemporary variables at time $t$, while blue notes denote the lagged variable at time $t-1$. Orange arrows denote the direct effects on the XG, and blue arrows denote the indirect effects on XG. Panel A and Panel B present the results of DYNOTEARS for the 2020-2021 season and the 2021-2022 season, respectively. Panel C and Panel D present the model results of  CD-NOD during the 2020-2021 season and 2021-2022 season, respectively.
    }
    \label{fig:main_comparison}
    \end{center}
\end{figure}

\textbf{Comparison studies with DYNOTEARS and CD-NOD.} DYNOTEAERS and CD-NOD are evaluated using a sliding-window approach to estimate the time-varying home field advantage, incorporating nearby time points in the estimation process to ensure comparability with our method. Figure \ref{fig:main_comparison} presents the model results by DYNOTEARS and CD-NOD. Panel A and Panel B present the results of DYNOTEARS for the 2020-2021 season and the 2021-2022 season, respectively. Panel C and Panel D present the model results of  CD-NOD during the 2020-2021 season and 2021-2022 season, respectively.

\begin{figure}[!htbp]
    \begin{center}
    \centerline{\includegraphics[width=0.9\textwidth]{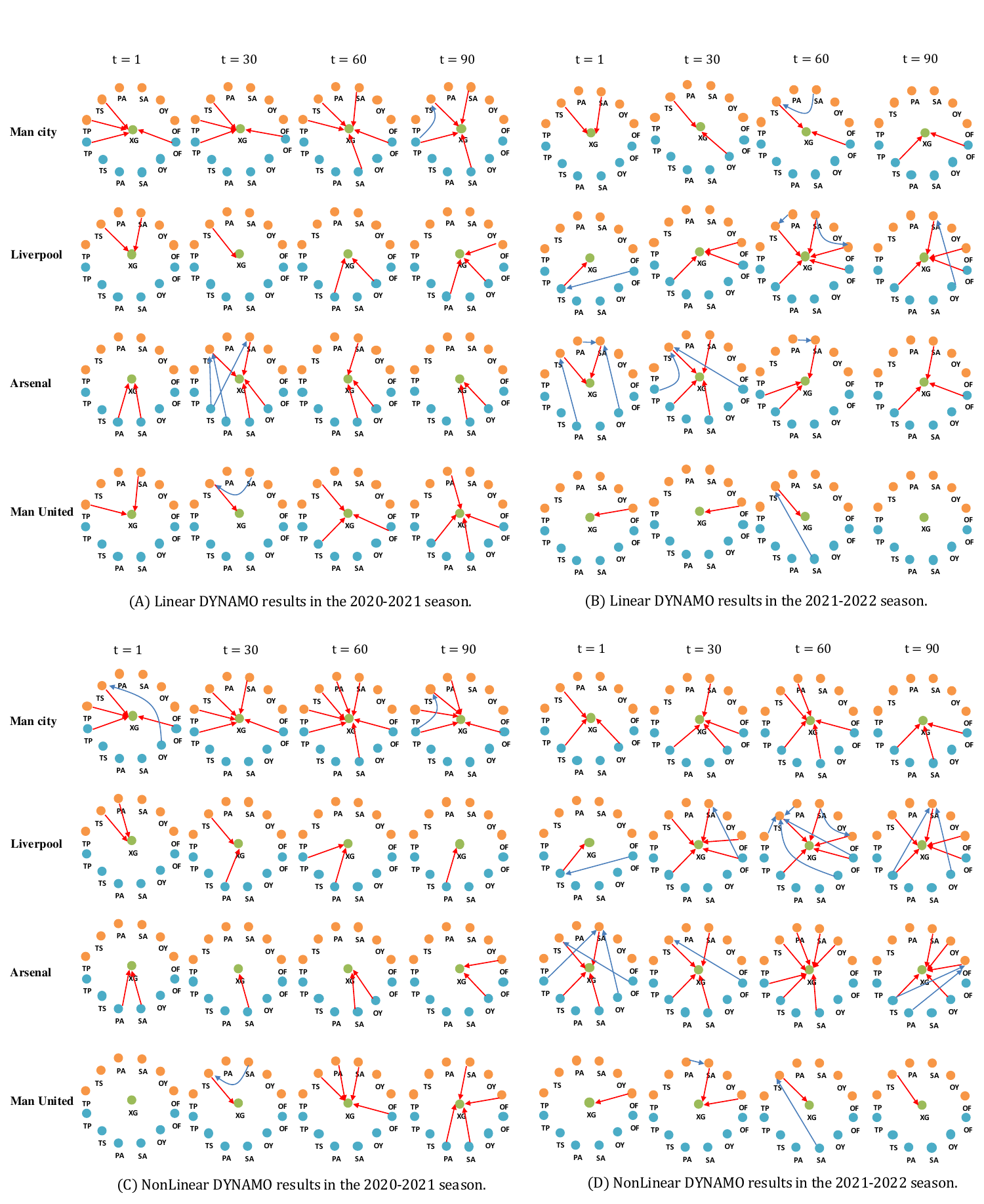}}
    \caption{The robustness checks for 4 representative teams using the second best kernel bandwidth.  Variables: TP - Total Passes, TS - Total Shots, PA - Pass Accuracy, SA - Shot Accuracy, OY - Opponent's yellow cards, OF - Opponent's Fouls. Orange notes represent contemporary variables at time $t$, while blue notes denote the lagged variable at time $t-1$. Orange arrows denote the direct effects on the XG, and blue arrows denote the indirect effects on XG. Panel A and Panel B present the results of linear DYNAMO for the 2020-2021 season and the 2021-2022 season, respectively. Panel C and Panel D present the model results of nonlinear DYNAMO during the 2020-2021 season and 2021-2022 season, respectively.}
    \label{fig:robustness_2ndh}
    \end{center}
\end{figure}

\begin{figure}[!htbp]
    \begin{center}
    \centerline{\includegraphics[width=0.9\textwidth]{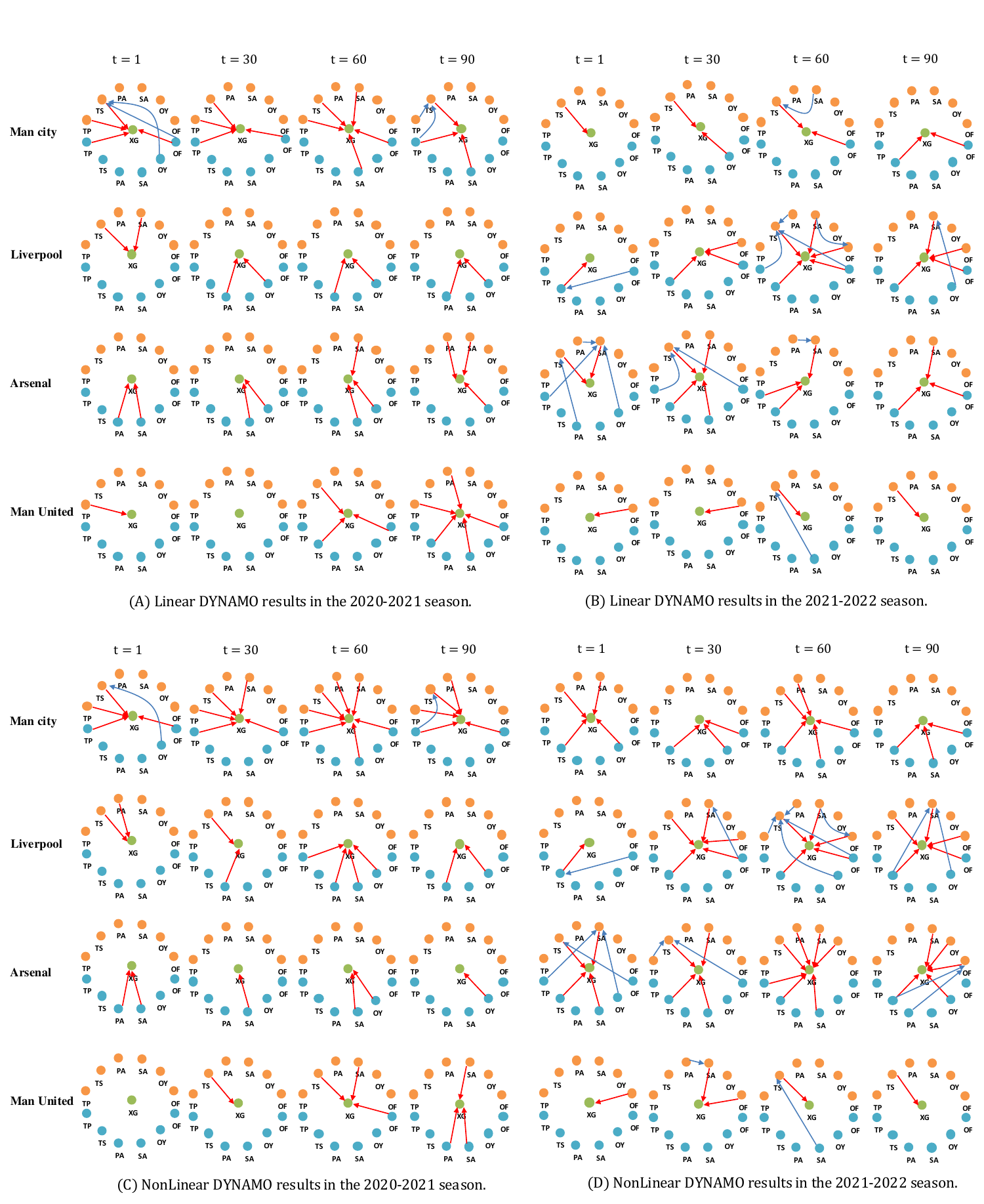}}
    \caption{The robustness checks for 4 representative teams using the third best kernel bandwidth.  Variables: TP - Total Passes, TS - Total Shots, PA - Pass Accuracy, SA - Shot Accuracy, OY - Opponent's yellow cards, OF - Opponent's Fouls. Orange notes represent contemporary variables at time $t$, while blue notes denote the lagged variable at time $t-1$. Orange arrows denote the direct effects on the XG, and blue arrows denote the indirect effects on XG. Panel A and Panel B present the results of linear DYNAMO for the 2020-2021 season and the 2021-2022 season, respectively. Panel C and Panel D present the model results of nonlinear DYNAMO during the 2020-2021 season and 2021-2022 season, respectively.}
    \label{fig:robustness_3rdh}
    \end{center}
\end{figure}

\textbf{Robustness checks with other kernel bandwidths.} Fig.~\ref{fig:robustness_2ndh} and Fig.~\ref{fig:robustness_3rdh} present the results of linear DYNAMO and nonlinear DYNAMO under the second-best bandwidth and the third-best bandwidth selected by Algorithm~\ref{alg:bandwidth}. The best bandwidth, along with the second best and the third best bandwidths, are presented in Table~\ref{tab:bandwidth20} and Table~\ref{tab:bandwidth}. Each element in the table represents the list of bandwidth in model $t=1$, model $t = 30$, model $t = 60$, and model $ t= 90$ respectively. These findings align with our main findings depicted in Fig.~\ref{fig:main_results}.

\begin{table}[!htbp]
    \centering
    \caption{Bandwidth selection for each model during the 2020-2021 season.}
    \begin{tabular}{lllll}
        \bottomrule
          \textbf{Bandwidth} & \textbf{Man. City} & \textbf{Liverpool} &\textbf{ Arsenal} & \textbf{Man. United} \\
        \hline
        Best   & [0.8,0.7,0.6,0.9] & [0.9,0.7,0.9,0.9]  &  [0.8,0.6,0.9,0.9] & [0.9,0.5,0.9,0.8]  \\
        Second best & [0.9,0.8,0.5,0.8] & [0.8,0.5,0.7,0.8] & [0.7,0.7,0.7,0.8] & [0.8,0.6,0.8,0.9]\\
        Third best & [0.7,0.9,0.7,0.7] & [0.7,0.6,0.8,0.7] & [0.9,0.9,0.8,0.6] & [0.7,0.7,0.7,0.7] \\
        \hline
    \end{tabular}
    \label{tab:bandwidth20}
\end{table}

\begin{table}[!htbp]
    \centering
    \caption{Bandwidth selection for each model during the 2021-2022 season.}
    \begin{tabular}{lllll}
        \bottomrule
          \textbf{Bandwidth} & \textbf{Man. City} & \textbf{Liverpool} &\textbf{ Arsenal} & \textbf{Man. United} \\
        \hline
        Best   & [0.7,0.6,0.8,0.7] & [0.9,0.9,0.9,0.9]  &  [0.5,0.9,0.7,0.5] & [0.6,0.5,0.9,0.9]  \\
        Second best & [0.9,0.7,0.7,0.9] & [0.8,0.8,0.8,0.8] & [0.4,0.7,0.8,0.4] & [0.7,0.6,0.8,0.7]\\
        Third best & [0.8,0.8,0.9,0.8] & [0.7,0.7,0.7,0.7] & [0.6,0.8,0.9,0.6] & [0.5,0.7,0.7,0.8] \\
        \hline
    \end{tabular}
    \label{tab:bandwidth}
\end{table}

\begin{figure}[!htbp]
    \begin{center}
    \centerline{\includegraphics[width=0.9\textwidth]{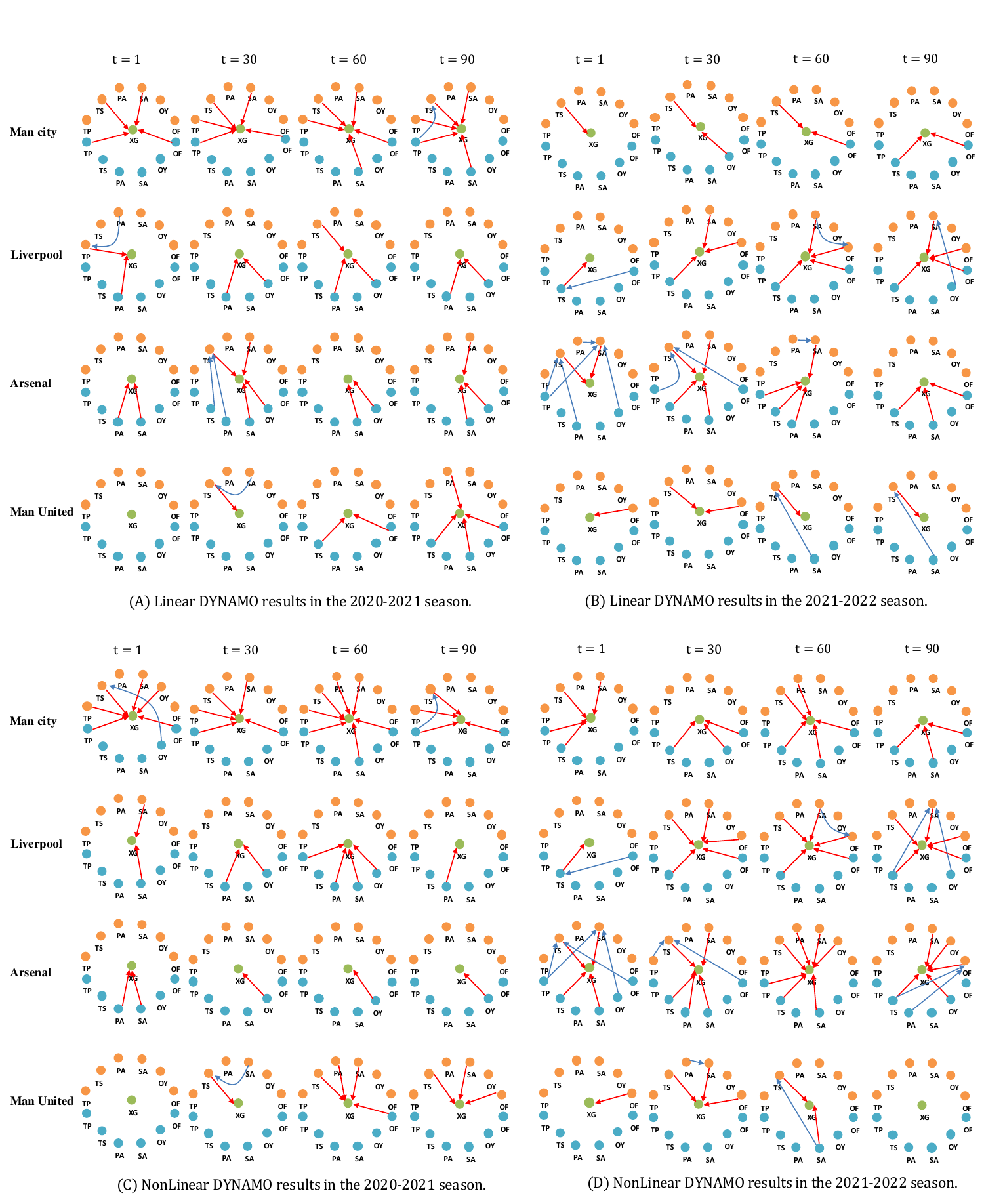}}
    \caption{The robustness checks for 4 representative teams choosing lagged period $L = 2$.  Variables: TP - Total Passes, TS - Total Shots, PA - Pass Accuracy, SA - Shot Accuracy, OY - Opponent's yellow cards, OF - Opponent's Fouls. Orange notes represent contemporary variables at time $t$, while blue notes denote the lagged variable at time $t-1$ and $t-2$. Orange arrows denote the direct effects on the XG, and blue arrows denote the indirect effects on XG. Panel A and Panel B present the results of linear DYNAMO for the 2020-2021 season and the 2021-2022 season, respectively. Panel C and Panel D present the model results of nonlinear DYNAMO during the 2020-2021 season and 2021-2022 season, respectively.}
    \label{fig:robustness_L2}
    \end{center}
\end{figure}

\textbf{Robustness checks with more lagged periods $L = 2$.} Fig.~\ref{fig:robustness_L2} presents the outcomes of linear DYNAMO and nonlinear DYNAMO with lagged periods $L = 2$.  These findings align with our main findings depicted in Section 5.

\subsection{Hyperparameters}
In this section, we present the hyperparameters used in the real data analysis.  We utilize Epanechnikov kernel $K(u) = \frac{3}{4}(1-u^2)I(|u| \leq 1)$ in our linear and nonlinear DYNAMO. Default hyperparameters are employed for all models in this section, including those presented in Fig.~\ref{fig:main_results} in Main Paper, Fig.~\ref{fig:robustness_2ndh}, and Fig.~\ref{fig:robustness_3rdh}. To prevent overfitting of the causal structures, we set a threshold of 0.05, as suggested by previous literature \citep{zheng2018dags, pamfil2020dynotears}. This means that we remove edges where the standardized influence is below 0.05. The specific hyperparameters are as follows:
\begin{itemize}
    \item DYNAMO-linear
        \begin{itemize}
            \item $\lambda_1$ = 0.05
            \item $\lambda_2$ = 0.05
            \item $\eta_{tol}$ = 1e-5
            \item $\text{threshold}$ = 0.05
        \end{itemize}
    \item DYNAMO-nonlinear
        \begin{itemize}
            \item $\lambda_1$ = 0.005
            \item $\lambda_2$ = 0.01
            \item $\eta_{tol}$ = 1e-10
        \end{itemize}
    \item DYNOTEARS
        \begin{itemize}
            \item $\lambda_1$ = 0.05
            \item $\lambda_2$ = 0.05
            \item $\eta_{tol}$ = 1e-5
        \end{itemize}
    \item CD-NOD
        \begin{itemize}
            \item $\alpha$ = 0.05
            \item indep test = fisherz
        \end{itemize}
\end{itemize}

\subsection{Results on Other teams}
In this section, we present model results for the remaining teams. Specifically, we classify teams into four groups based on their average rankings over the two seasons: higher-ranked, middle-ranked, lower-ranked, and relegated teams. The higher-ranked group includes Chelsea, Leicester City, Tottenham Hotspur, and West Ham United. The middle-ranked group comprises Newcastle United, Aston Villa, Wolverhampton Wanderers, and Brighton. The lower-ranked group consists of Southampton, Crystal Palace, Everton, and Leeds United. The relegated group includes Burnley, Fulham, West Bromwich, Sheffield United, Watford, Norwich City, and Brentford. We apply both the linear and nonlinear versions of DYNAMO, using the same hyperparameter selection procedure as in the previous analysis. The results are presented in Figures 5--8.  

\begin{figure}[ht]
    \begin{center}
\vspace{-0.2cm}\centerline{\includegraphics[width=0.9\textwidth]{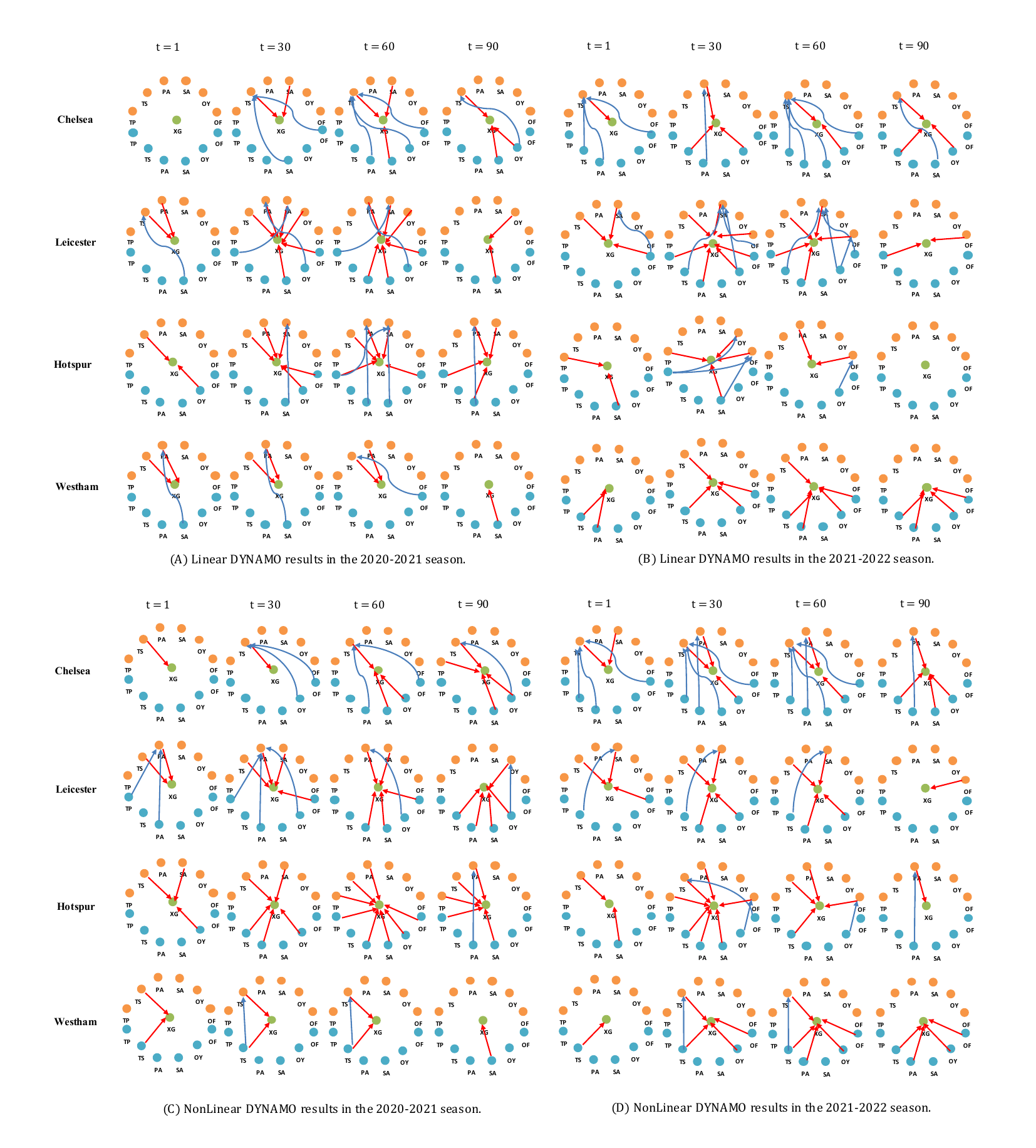}}
    \caption{\footnotesize Model results for 4 higher-ranked teams, Chelsea, Leicester City, Tottenham Hotspur, and West Ham United.  Variables: TP - Total Passes, TS - Total Shots, PA - Pass Accuracy, SA - Shot Accuracy, OY - Opponent's yellow cards, OF - Opponent's Fouls. Orange notes represent contemporary variables at time $t$, while blue notes denote the lagged variable at time $t-1$. Orange arrows denote the direct effects on the XG, and blue arrows denote the indirect effects on the XG. Panel A and Panel B present the results of linear DYNAMO for the 2020-2021 season and the 2021-2022 season, respectively. Panel C and Panel D present the model results of nonlinear DYNAMO during the 2020-2021 season and 2021-2022 season, respectively. }
    \vspace{-0.8cm}
    \end{center}
\end{figure}

\begin{figure}[ht]
    \begin{center}
\vspace{-0.2cm}\centerline{\includegraphics[width=0.9\textwidth]{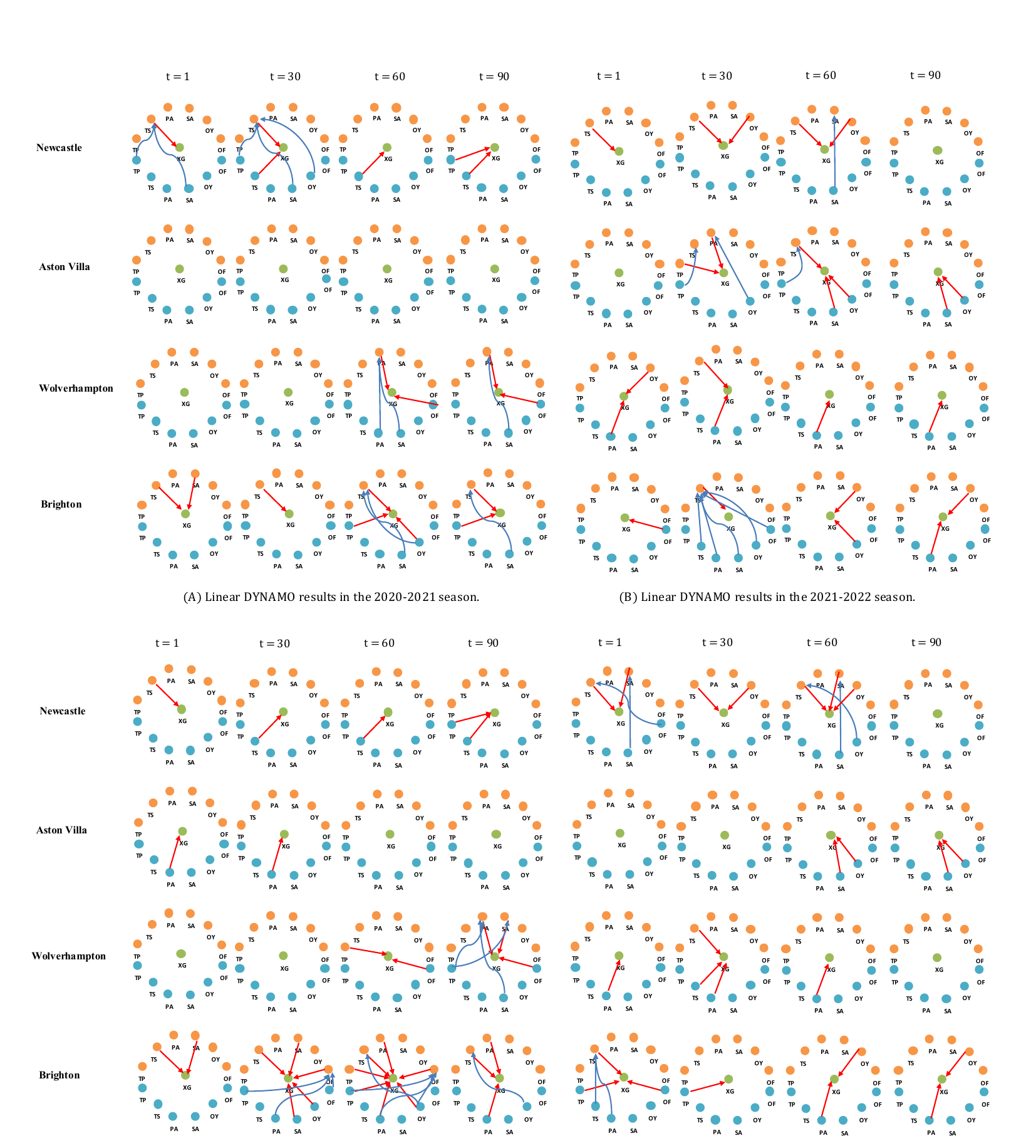}}
    \caption{\footnotesize Model results for 4 middle-ranked teams, Newcastle United, Aston Villa, Wolverhampton Wanderers, and Brighton.  Variables: TP - Total Passes, TS - Total Shots, PA - Pass Accuracy, SA - Shot Accuracy, OY - Opponent's yellow cards, OF - Opponent's Fouls. Orange notes represent contemporary variables at time $t$, while blue notes denote the lagged variable at time $t-1$. Orange arrows denote the direct effects on the XG, and blue arrows denote the indirect effects on the XG. Panel A and Panel B present the results of linear DYNAMO for the 2020-2021 season and the 2021-2022 season, respectively. Panel C and Panel D present the model results of nonlinear DYNAMO during the 2020-2021 season and 2021-2022 season, respectively. }
    \vspace{-0.8cm}
    \end{center}
\end{figure}

\begin{figure}[ht]
    \begin{center}
\vspace{-0.2cm}\centerline{\includegraphics[width=0.9\textwidth]{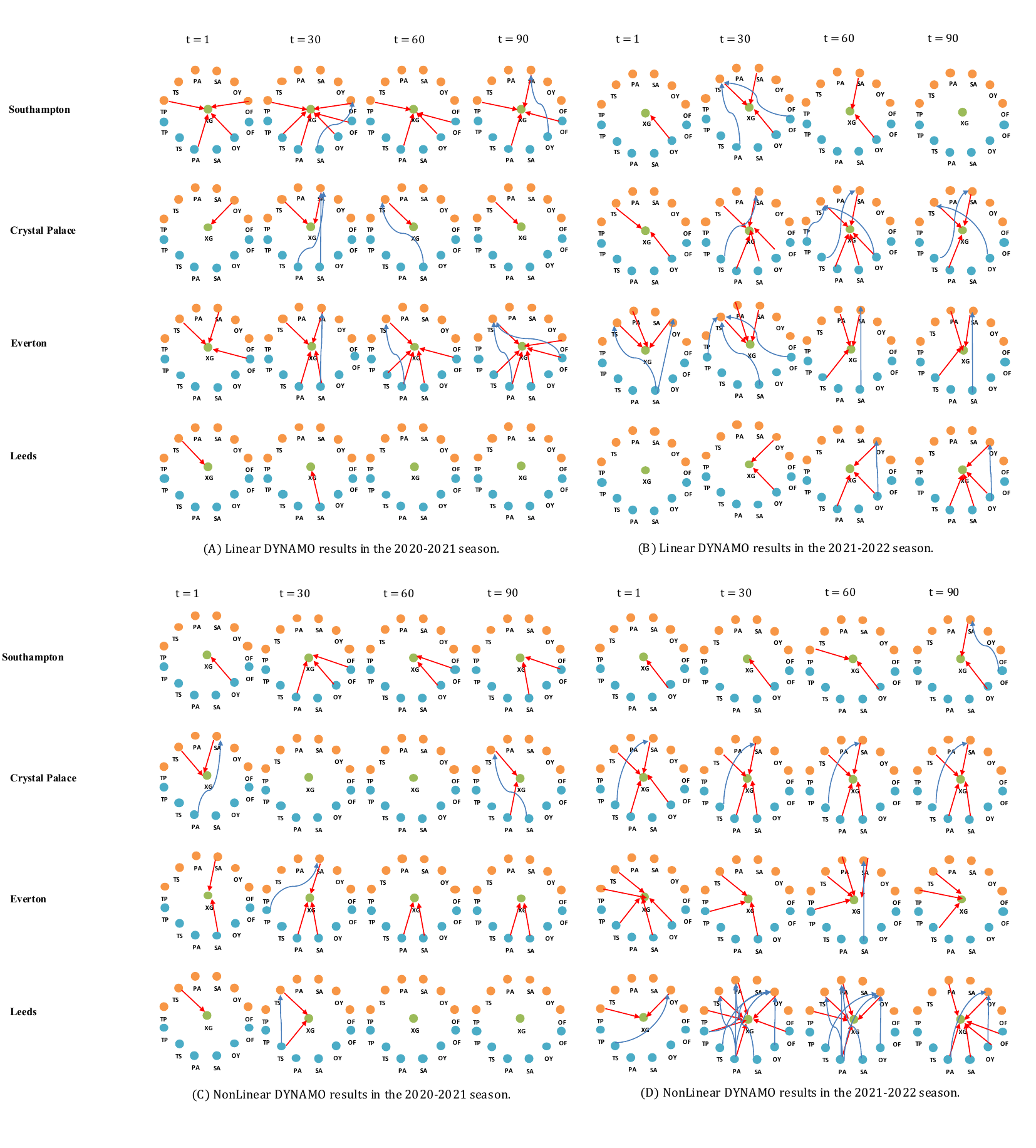}}
    \caption{\footnotesize Model results for 4 lower-ranked teams, Southampton, Crystal Palace, Everton, and Leeds United.  Variables: TP - Total Passes, TS - Total Shots, PA - Pass Accuracy, SA - Shot Accuracy, OY - Opponent's yellow cards, OF - Opponent's Fouls. Orange notes represent contemporary variables at time $t$, while blue notes denote the lagged variable at time $t-1$. Orange arrows denote the direct effects on the XG, and blue arrows denote the indirect effects on the XG. Panel A and Panel B present the results of linear DYNAMO for the 2020-2021 season and the 2021-2022 season, respectively. Panel C and Panel D present the model results of nonlinear DYNAMO during the 2020-2021 season and 2021-2022 season, respectively. }
    \vspace{-0.8cm}
    \end{center}
\end{figure}

\begin{figure}[ht]
    \begin{center}
\vspace{-0.2cm}\centerline{\includegraphics[width=0.9\textwidth]{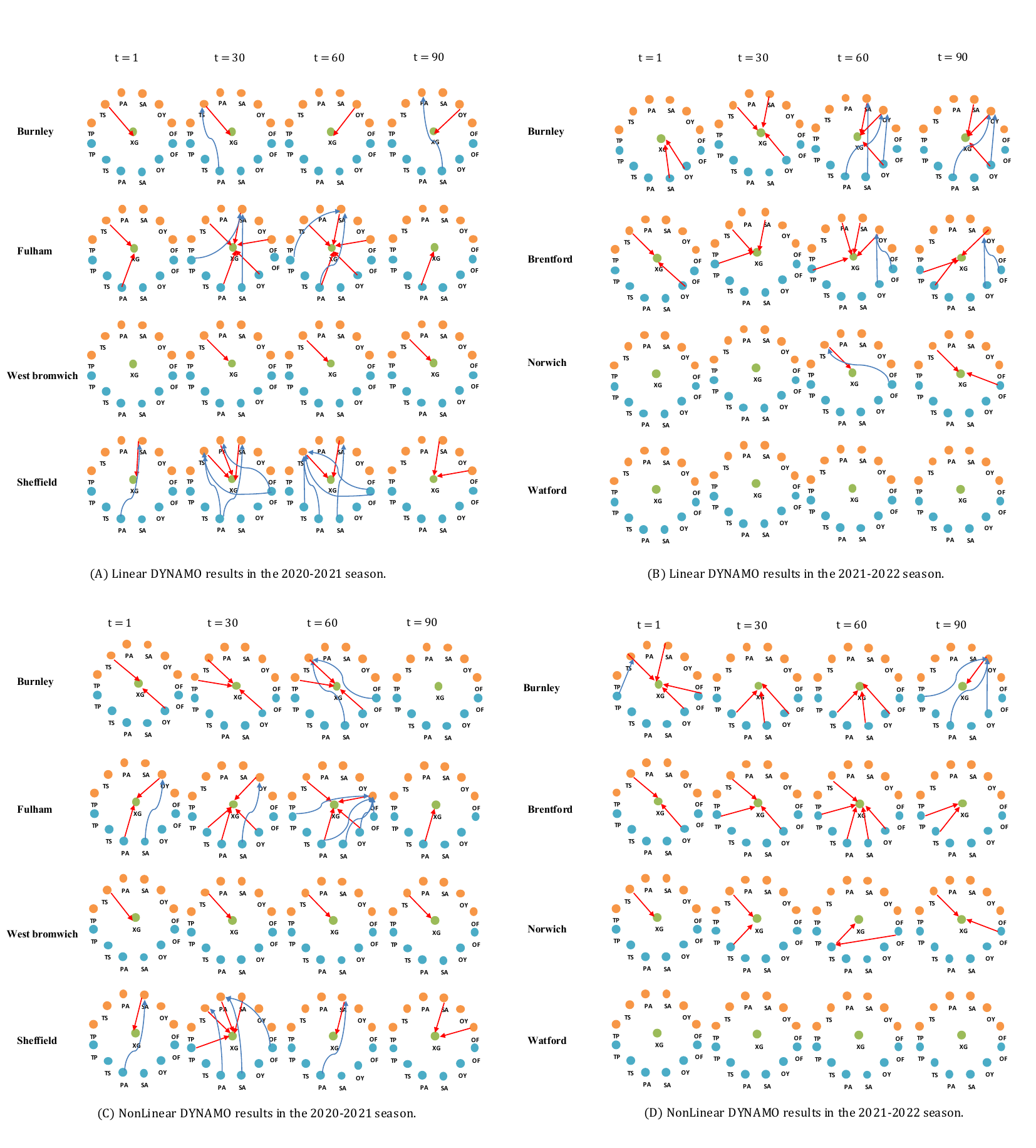}}
    \caption{\footnotesize Model results for 7 relegated teams, Burnley, Fulham, West Bromwich, Sheffield United, Watford, Norwich City, and Brentford.  Variables: TP - Total Passes, TS - Total Shots, PA - Pass Accuracy, SA - Shot Accuracy, OY - Opponent's yellow cards, OF - Opponent's Fouls. Orange notes represent contemporary variables at time $t$, while blue notes denote the lagged variable at time $t-1$. Orange arrows denote the direct effects on the XG, and blue arrows denote the indirect effects on the XG. Panel A and Panel B present the results of linear DYNAMO for the 2020-2021 season and the 2021-2022 season, respectively. Panel C and Panel D present the model results of nonlinear DYNAMO during the 2020-2021 season and 2021-2022 season, respectively. }
    \vspace{-0.8cm}
    \end{center}
\end{figure}

\section{Simulation Extensions}
\label{appx:simulation}

In this section, we detail the experimental setup for data generation and introduce additional experiments covering running times, varied time lengths, and different time points. All experiments were conducted on the same hardware setup: an Intel(R) Xeon(R) Gold 6240 CPU @ 2.60GHz with 125 GB memory.

We leverage Structural Hamming distance (SHD) and F1 score as the evaluation metrics. SHD quantifies the number of changes (i.e., additional edges, missing edges, and reverse edges) between two graphs. F1 score is computed as:
\begin{align}
    F1 = 2 \times \frac{Precision \times Recall}{Precision + Recall},\notag
\end{align}
where $Precision$ represents correctly-identified edges divided by the number of edges in the estimated graph, and $Recall$ denotes the number of correctly-identified edges divided by the number of edges in the ground-truth graph.

\subsection{Data Generation Process}
\textbf{Basic graph generation.} We use the Erdos–Renyi (ER) model with mean degrees of 4 for basic instantaneous graphs and mean degrees of 2 to generate basic lagged graphs. For simplicity, we set the lagged periods $L$ as 1. The process begins by creating a basic causal structure following the aforementioned rules.

\textbf{Time-varying graphs generation.} To introduce progressively evolving causal structures, we initialize a fundamental graph structure using the Erdos–Renyi (ER) model. Subsequently, we adjust its \textbf{non-zeros} weights $w_{ij}(\tau_t)$ (representing instantaneous edges) and its \textbf{non-zeros} weights $a_{ij}(\tau_t)$ (representing lagged edges) as functions of $t$ and a threshold $\gamma$. Specifically, the weight function is defined as
\begin{align}
    w_{ij}(\tau_t) = \cos [ (\delta_{ij}^w+t/\Phi)\pi ] \cdot I(\cos [ (\delta_{ij}^w+t/\Phi)\pi ] > \gamma), \notag \\
    a_{ij}(\tau_t) = \cos [ (\delta_{ij}^a+t/\Phi)\pi ] \cdot I(\cos [ (\delta_{ij}^a+t/\Phi)\pi ] > \gamma),
    \label{eq:nonstationary_ weights}
\end{align}

where $\delta_{ij}^w$ and $\delta_{ij}^a$ are random values drawn from the uniform distribution $U(0,1)$. The parameter $\gamma$ denotes the sparsity threshold, and $\Phi$ determines the changing speed. A smaller $\Phi$ results in a faster alteration in the causal structure. The threshold $\gamma$ ensures that our causal structure $\{\boldsymbol{\theta}(\tau_t)\}$ not only evolves gradually in terms of causal strength but also transforms the entire causal structure as  $t$ progresses from 0 to $T$. For the simulations, we set the $\gamma$ as 0.05, $\Phi$ as $0.5*T$ in the linear model, and $\Phi$ as $0.9*T$ in the nonlinear model.

\textbf{Linear data generation.} To simulate data under linear causal relationships, we utilize the following equation:
\begin{align*}
    \boldsymbol{x}_t = \boldsymbol{x}_t W(\tau_t)  +  Y_t A(\tau_t) + \boldsymbol{\epsilon}_t, 
\end{align*}
where $W(\tau_t)$ represents the instantaneous adjacent matrix with each elements $w_{ij}(\tau_t)$ generated using Eq.~\eqref{eq:nonstationary_ weights}, and $A(\tau_t)$ is the lagged adjacent matrix with each elements $a_{ij}(\tau_t)$ generated similarly. The error term $\boldsymbol{\epsilon}_t $ denotes the vector of standard Gaussian noise variables. 

\textbf{Nonlinear causal structure.} To simulate data under nonlinear relationships, we employ a model incorporating tanh and sigmoid functions.
\begin{align*}
    \boldsymbol{x}_t = tanh(\boldsymbol{x}_t W(\tau_t)  +  Y_t A(\tau_t) ) + sigmoid(\boldsymbol{x}_t W(\tau_t)  +  Y_t A(\tau_t)) + \boldsymbol{\epsilon}_t,
\end{align*}
where
\begin{align*}
    tanh(x) = \frac{e^x - e^{-x}}{e^x + e^{-x}}, \quad sigmoid(x) = \frac{1}{1+e^{-x}}.
\end{align*}
$W(\tau_t)$ and $A(\tau_t)$ represent the time-varying instantaneous and lagged adjacent matrices with each element generated by Eq.~\eqref{eq:nonstationary_ weights}. The noise term $\boldsymbol{\epsilon}_t $ is sampled from a Gaussian distribution. We explore four different numbers of nodes $d \in [5,10,15,20]$. 

\subsection{Additional Experiments}
\label{Appendix_additional_experiments}

\begin{figure}[!htbp]
    \begin{center}
    \centerline{\includegraphics[width=0.8\textwidth]{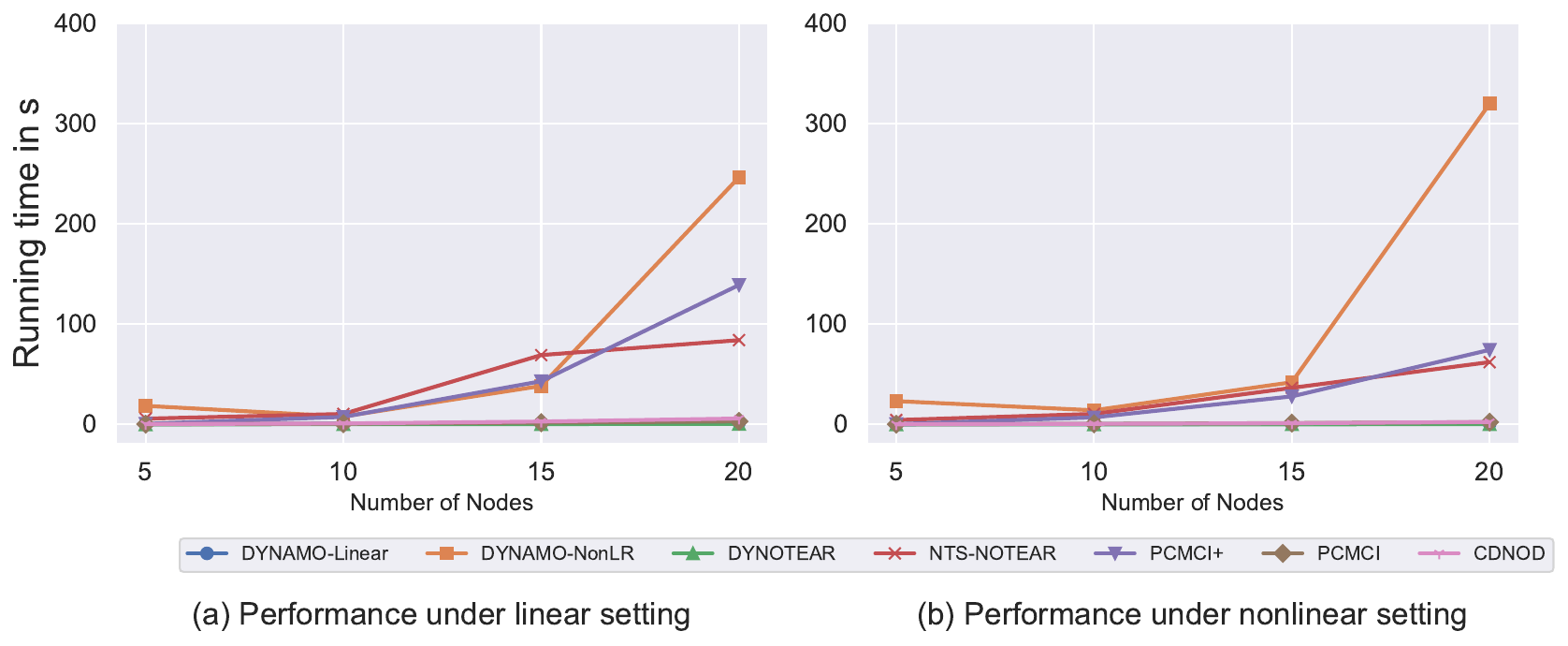}}
    \caption{Comparison studies for running time of different methods over 20 replications. The first column represents the simulation results within a linear time series, and the last column represents the simulation results within a nonlinear time series.}
    \label{fig:Running_time}
    \end{center}
\end{figure}

\textbf{Running time.} We provide running times for different values $d \in [5,10,15,20] $ with $T = 500$. All experiments were conducted on the same Intel(R) Xeon(R) Gold 6240 CPU @ 2.60GHz with 125 GB memory. Fig.~\ref{fig:Running_time} presents the running times of DYNAMO with varying $d$.  The results demonstrate that linear DYNAMO is as fast as the quickest benchmarks,i.e. DYNOTEAS, while nonlinear DYNAMO exhibits the slowest performance. However, in theory, our method offer flexibility to integrate with various loss functions (base learner), which could enhance the computational efficiency of nonlinear DYNAMO.

\begin{figure}[!htbp]
    \begin{center}
    \centerline{\includegraphics[width=\textwidth]{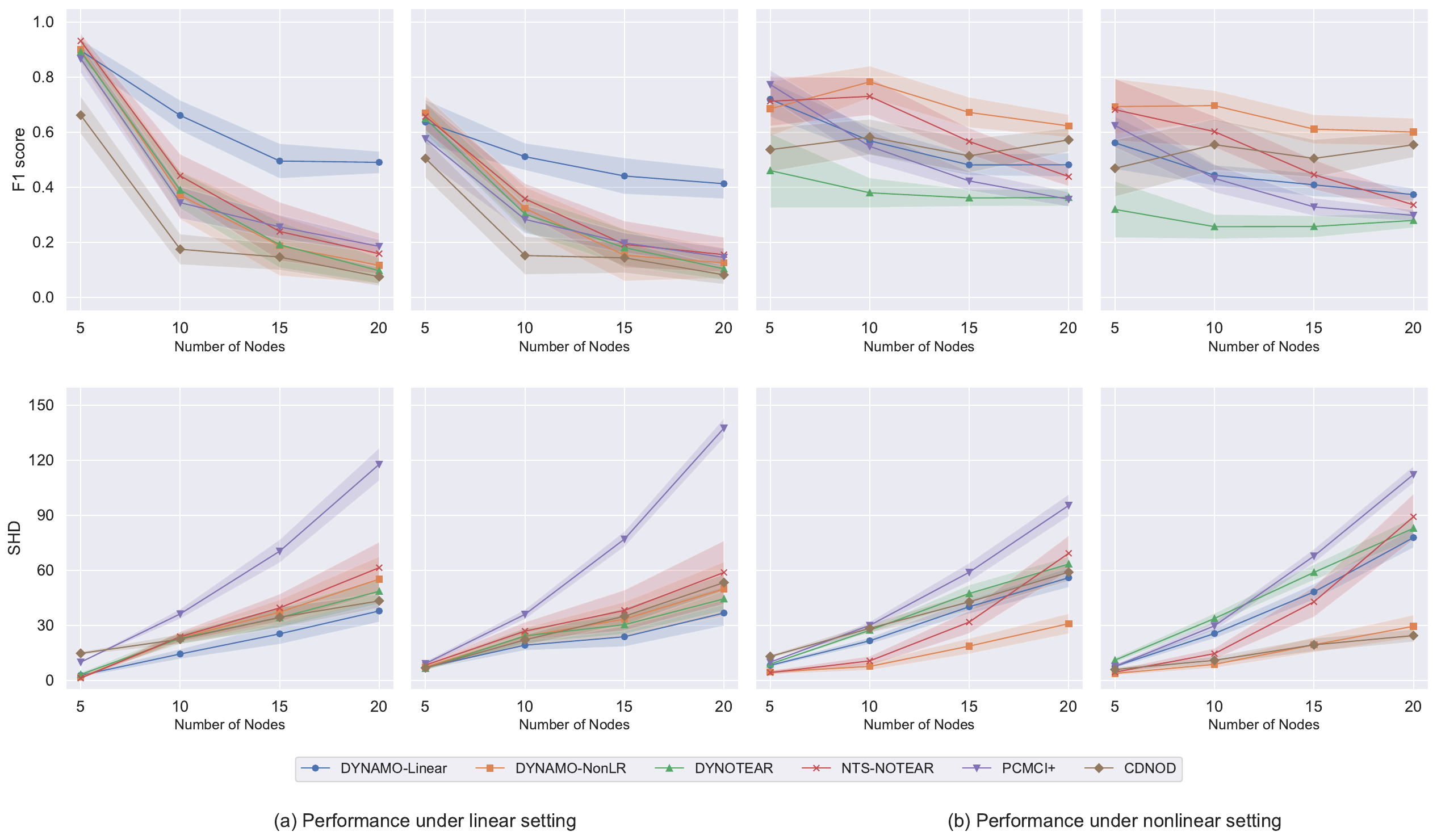}}
    \caption{Comparison studies for linear and nonlinear nonstationary data with $T$ = 500 at time 200 over 20 replications. The first two columns represent the simulation results within a linear time series, and the last two columns represent simulation results within a nonlinear time series. For each setting, the left panel represents the results of instantaneous graphs $W_t$, and the right panel presents the results of lagged graphs $A_t$.}
    \label{fig:T500_t200_n20}
    \end{center}
\end{figure}

\textbf{Experiments at different time stamps $t$}. We present the model results at different time points $t = 200$ with $ T = 500$. While it's impractical to provide results for all time points, we offer insights into our model's performance in the middle of the time length.  Fig.~\ref{fig:T500_t200_n20} presents the model results at time point 200 for $T = 500$ in both linear and nonlinear time series data. These results illustrate our model's superior performance compared to other methods at time point $200$ with a time length $ T = 500$.

\begin{figure}[!htbp]
    \begin{center}
    \centerline{\includegraphics[width=\textwidth]{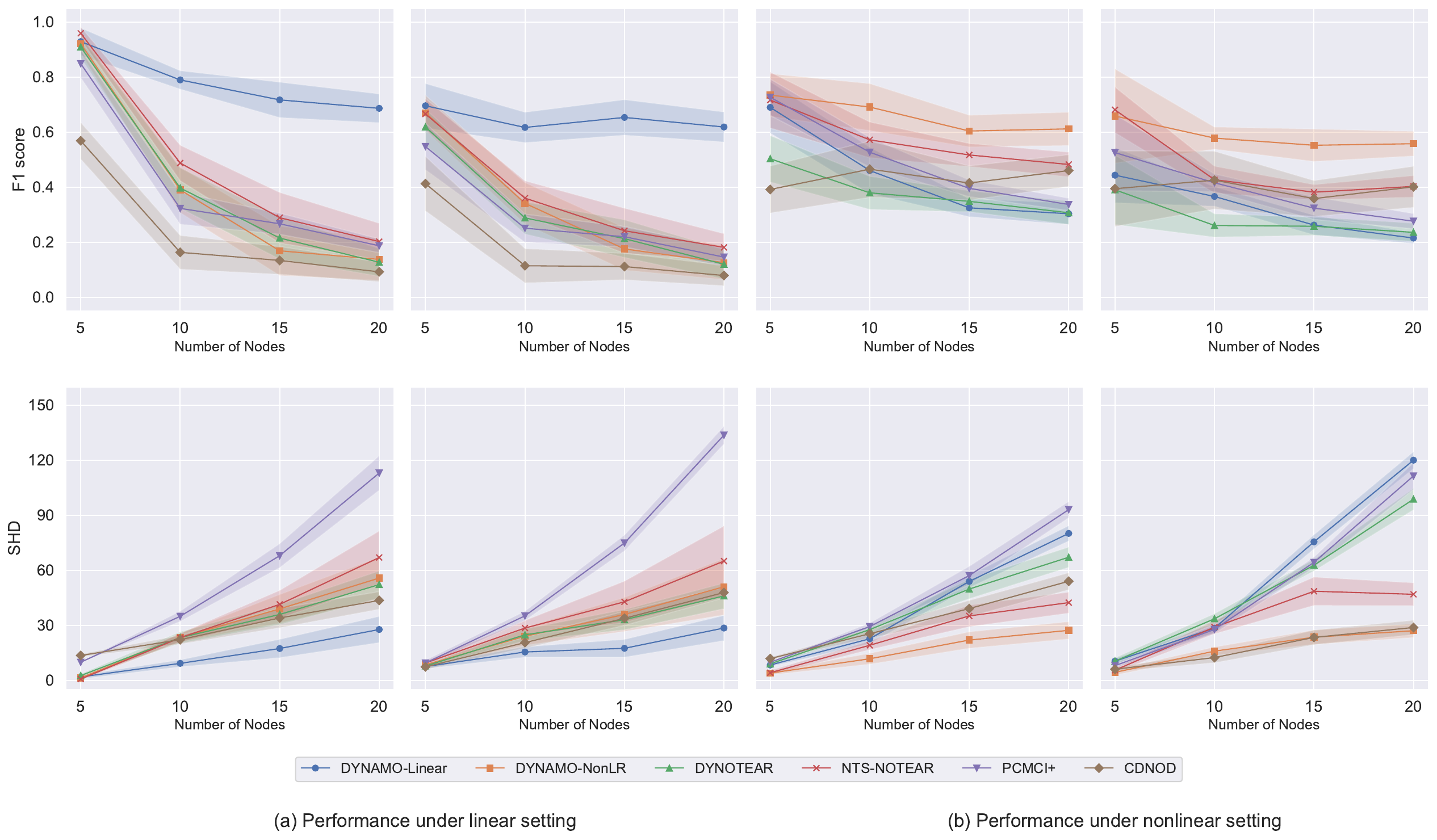}}
    \caption{Comparison studies for linear and nonlinear nonstationary data with $T$ = 300 at time 60 over 20 replications. The first two columns represent the simulation results within a linear time series, and the last two columns represent simulation results within a nonlinear time series. For each setting, the left panel represents the results of instantaneous graphs $W_t$, and the right panel presents the results of lagged graphs $A_t$.}
    \label{fig:T300_t60_n20}
    \end{center}
\end{figure}

\begin{figure}[!htbp]
    \begin{center}
    \centerline{\includegraphics[width=\textwidth]{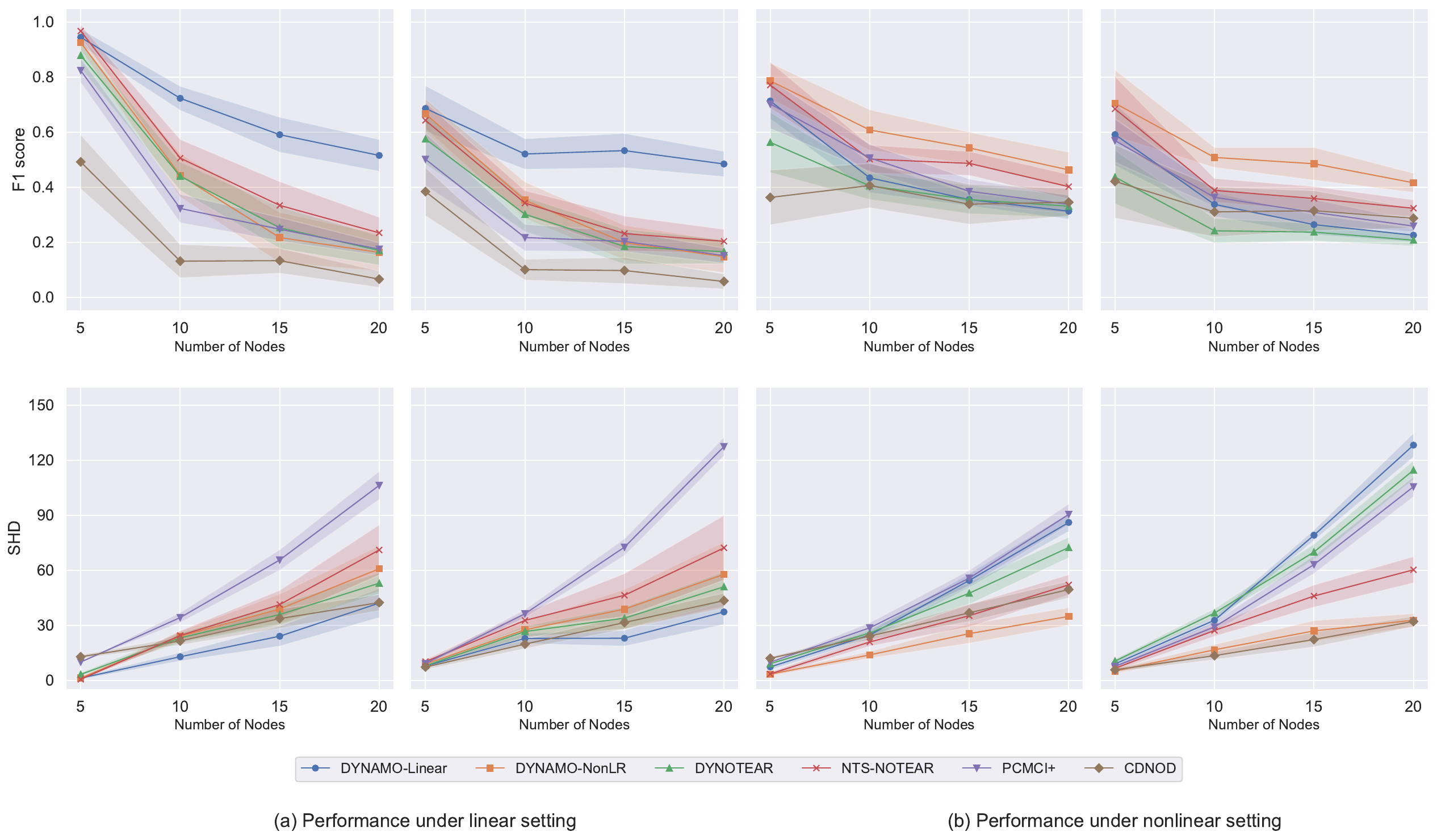}}
    \caption{Comparison studies for linear and nonlinear nonstationary data with $T$ = 200 at time 60 over 20 replications. The first two columns represent the simulation results within a linear time series, and the last two columns represent simulation results within a nonlinear time series. For each setting, the left panel represents the results of instantaneous graphs $W_t$, and the right panel presents the results of lagged graphs $A_t$.}
    \label{fig:T200_t60_n20}
    \end{center}
\end{figure}

\textbf{Experiments with different time lengths $T$}. We further conduct the experiments with different time lengths $T \in \{200,300 \}$. Fig.~\ref{fig:T300_t60_n20}  and Fig.~\ref{fig:T200_t60_n20}  present the model results of time series data with different time length $T = 300$ and $T = 200$ at the same time points $60$, respectively. The results illustrate that the performance of DYNAMO methods continuously improves as the time lengths $T$ increase, while other methods do not gain efficiency with more extended time lengths due to the non-stationary environments.

\begin{figure}[!htbp]
    \begin{center}
    \centerline{\includegraphics[width=\textwidth]{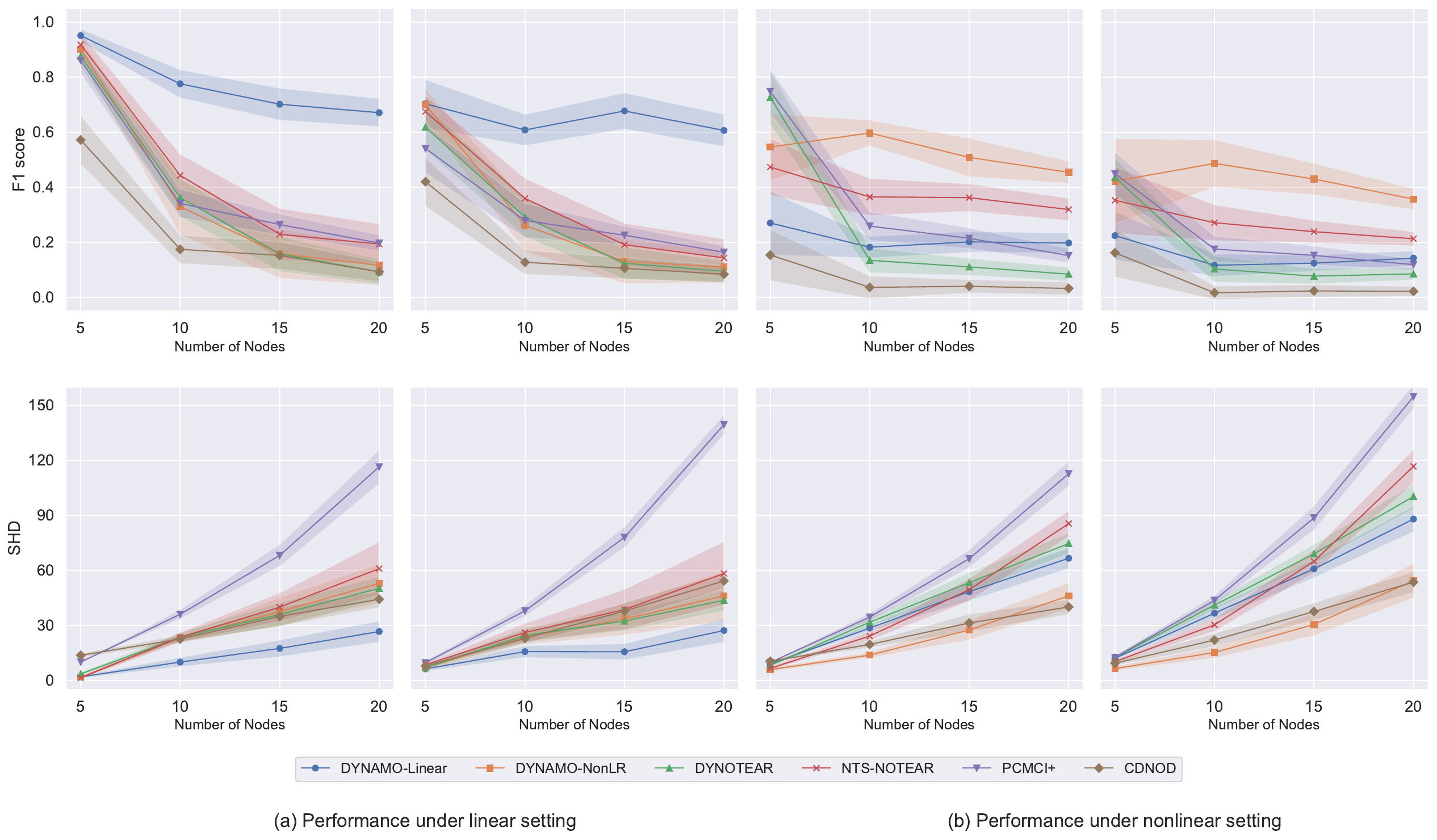}}
    \caption{Comparison studies for linear and nonlinear nonstationary data with faster causal structures on $T$ = 500 at time 60 over 20 replications. The first two columns represent the simulation results within a linear time series, and the last two columns represent simulation results within a nonlinear time series. For each setting, the left panel represents the results of instantaneous graphs $W_t$, and the right panel presents the results of lagged graphs $A_t$.}
    \label{fig:T500_t60_n20_faster}
    \end{center}
\end{figure}

\textbf{Experiments with faster causal structures changes}. We also test the performance of DYNAMO on faster-changing causal structures with $T$ = 500 at time 60 over 20 replications. We set $\Phi$ as $0.3*T$ in the linear model, and $\Phi$ as $0.5*T$ in the nonlinear model. Fig. \ref{fig:T500_t60_n20_faster} presents the performance of DYNAMO methods on faster-changing causal structures. The results illustrate the superior performance of DYNAMO methods as the causal structures change faster.

\begin{figure}[!htbp]
    \begin{center}
    \centerline{\includegraphics[width=\textwidth]{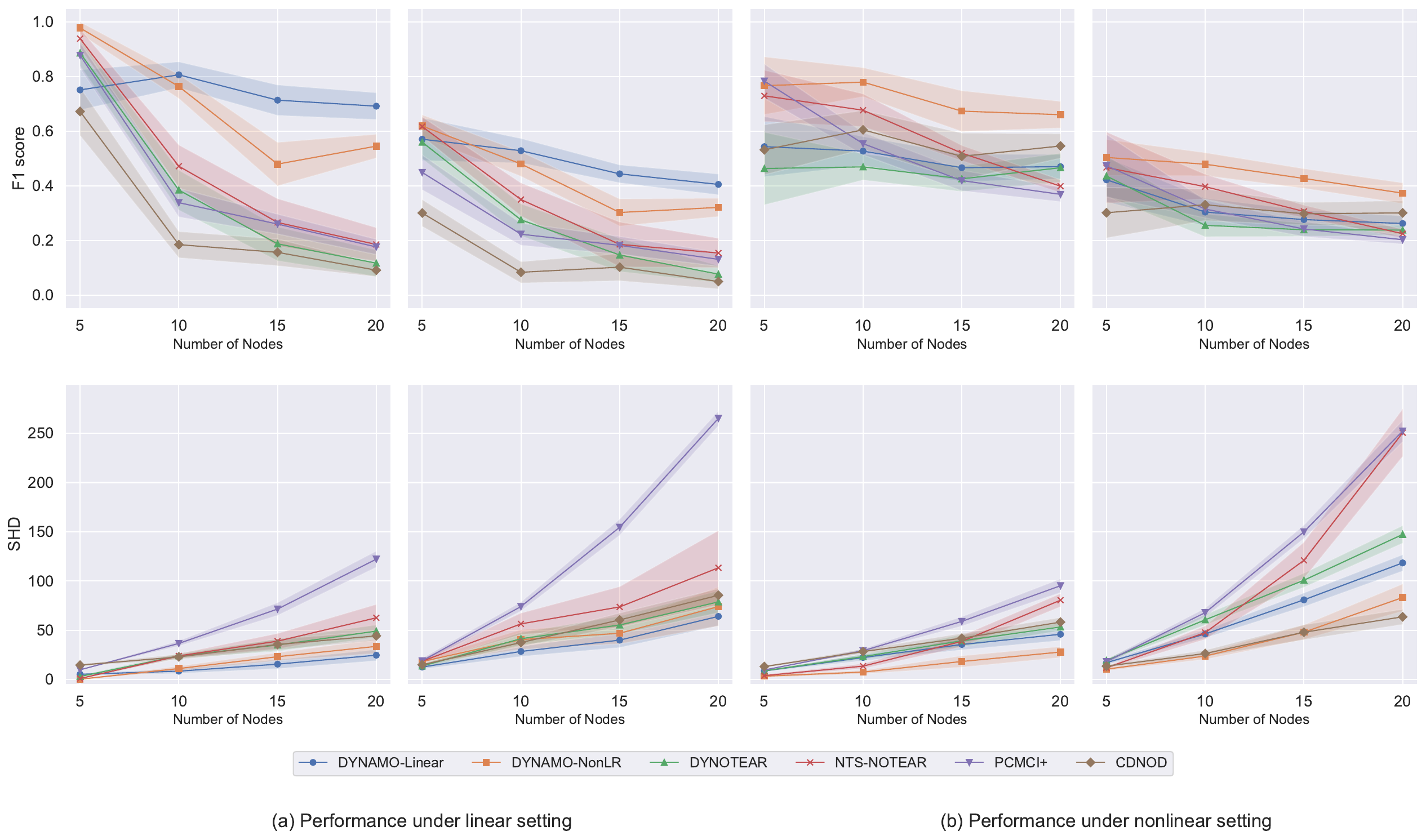}}
    \caption{Comparison studies for linear and nonlinear nonstationary data with more lagged periods $L = 2$ on $T = 500$ at time 60 over 20 replications. The first two columns represent the simulation results within a linear time series, and the last two columns represent simulation results within a nonlinear time series. For each setting, the left panel represents the results of instantaneous graphs $W_t$, and the right panel presents the results of lagged graphs $A_t$.}
    \label{fig:T500_t60_n20_L2}
    \end{center}
\end{figure}

\textbf{Experiments with more lagged periods $L$}. We also test the performance of DYNAMO with more lagged periods. We set $L = 2$ on both linear and nonlinear data simulation strategies. Fig. \ref{fig:T500_t60_n20_L2} presents the performance of DYNAMO methods for linear and nonlinear nonstationary data with more lagged periods $L = 2$ on $T = 500$ at time 60 over 20 replications. The results illustrate the superior performance of DYNAMO methods with more lagged periods $L = 2$.

\begin{figure}[!htbp]
    \begin{center}
    \centerline{\includegraphics[width=\textwidth]{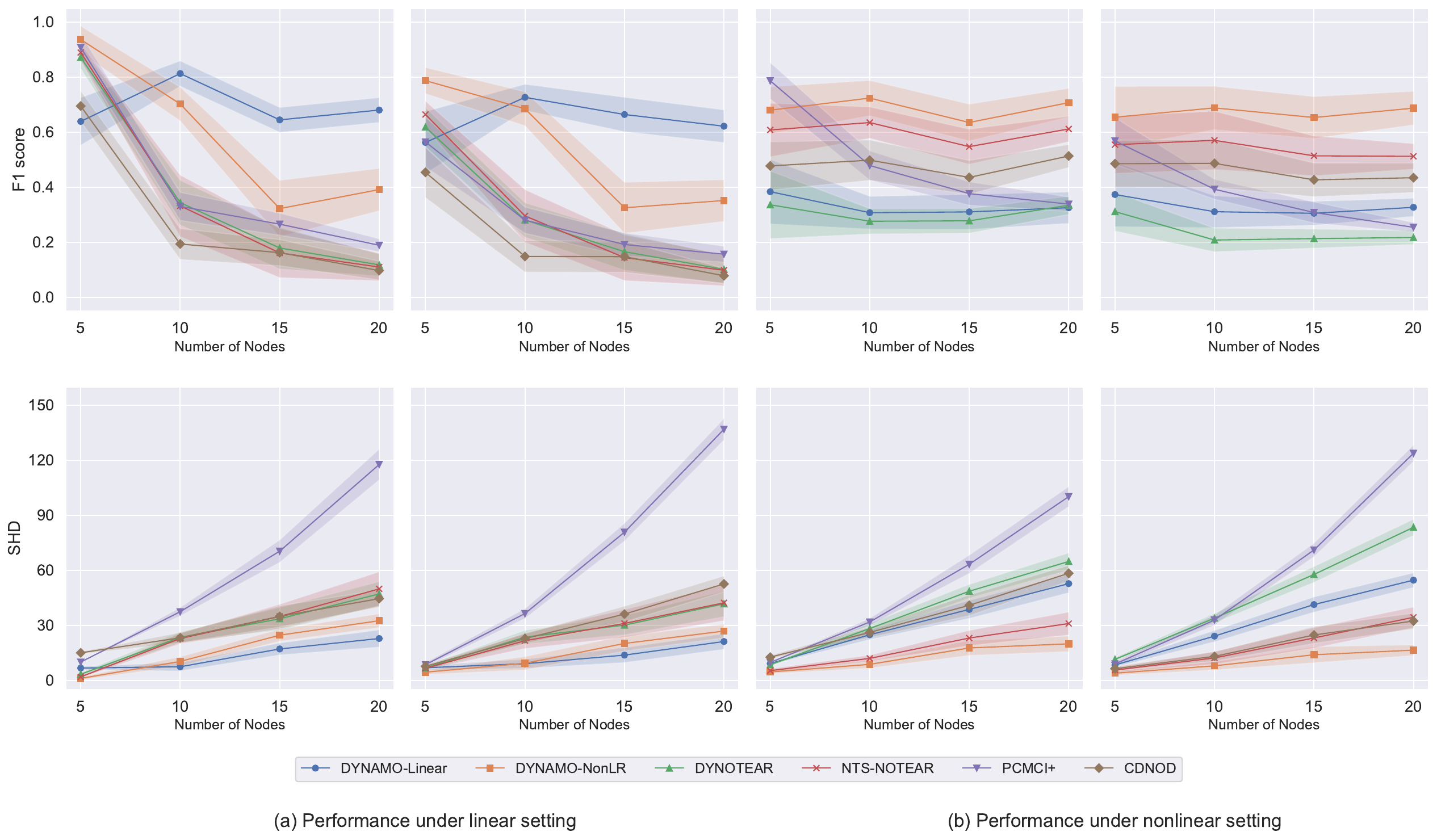}}
    \caption{Comparison studies for linear and nonlinear nonstationary data with non-Gaussian noises on $T = 500$ at time 60 over 20 replications. The first two columns represent the simulation results within a linear time series, and the last two columns represent simulation results within a nonlinear time series. For each setting, the left panel represents the results of instantaneous graphs $W_t$, and the right panel presents the results of lagged graphs $A_t$.}
    \label{fig:T500_t60_n20_uniform}
    \end{center}
\end{figure}

\textbf{Experiments with non-Gaussian noises}. We additionally evaluate the performance of DYNAMO with non-Gaussian noise conditions. We randomly sample noise from either a uniform distribution $U[-0.5,0.5]$ or the standard Gaussian distribution for both linear and nonlinear data simulation strategies. Fig. \ref{fig:T500_t60_n20_uniform} presents the performance of DYNAMO methods for linear and nonlinear nonstationary data with non-Gaussian noises on $T = 500$ at time 60 over 20 replications. The results illustrate the superior performance of DYNAMO methods with different forms of noise.

\subsection{Hyperparameters for simulated data}
\label{subsec:sim_hyperparameters}

All benchmark methods used in the simulation experiments utilize publicly available libraries:  PCMCI+ is implemented in Tigramite, CD-NOD is implemented in causallearn, and NTS-NOTEAR is implemented using code provided by the authors \cite{sun2023nts}. To ensure comparability, we execute the benchmark methods with hyperparameters selected via time series cross-validation using TimeSeriesSplit in the sklearn package of Python. Optimization in DYNAMO employs L-BFGS-B, while other state-of-the-art models use implementations provided by their respective authors. We utilize Epanechnikov kernel $K(u) = \frac{3}{4}(1-u^2)I(|u| \leq 1)$ in our linear and nonlinear DYNAMO. Each experiment is repeated 20 times with 20 random seeds generated using the same seed 24.

\begin{itemize}     
    \item DYNAMO-linear
        \begin{itemize}
            \item $\lambda_1$ = 0.05
            \item $\lambda_2$ = 0.05
            \item $\eta_{tol}$ = 0.001
        \end{itemize}
    \item DYNAMO-nonlinear
        \begin{itemize}
            \item $\lambda_1 \in \{0.005, 0.05,0.05,0.05 \}$, for $T = 500$ and $d \in \{5,10,15,20\}$, respectively.
            \item $\lambda_2 \in \{0.01, 0.01,0.01,0.01 \}$, for $T = 500$ and $d \in \{5,10,15,20\}$, respectively.
            \item $\lambda_1 \in \{0.005, 0.01,0.05,0.05 \}$, for $T = 300$ or $200$, $d \in \{5,10,15,20\}$, respectively.
            \item $\lambda_2 \in \{0.01, 0.05,0.1,0.1 \}$, for $T = 300$ or $200$, $d \in \{5,10,15,20\}$, respectively.
            \item $\eta_{tol}$ = 1e-10
        \end{itemize}
    \item DYNOTEARS
        \begin{itemize}
            \item $\lambda_1$ = 0.05
            \item $\lambda_2$ = 0.05
            \item $\eta_{tol}$ = 0.001
        \end{itemize}
    \item NTS-Notear
        \begin{itemize}
            \item $\lambda_1 \in \{0.005, 0.05,0.05,0.05 \}$, for $T = 500$ and $d \in \{5,10,15,20\}$, respectively.
            \item $\lambda_2 \in \{0.01, 0.01,0.01,0.01 \}$, for $T = 500$ and $d \in \{5,10,15,20\}$, respectively.
            \item $\lambda_1 \in \{0.005, 0.01,0.05,0.05 \}$, for $T = 300$ or $200$, $d \in \{5,10,15,20\}$, respectively.
            \item $\lambda_2 \in \{0.01, 0.05,0.1,0.1 \}$, for $T = 300$ or $200$, $d \in \{5,10,15,20\}$, respectively.
            \item $\eta_{tol}$ = 1e-10
        \end{itemize}
    \item PCMCI+
        \begin{itemize}
            \item $pc-\alpha$ = 0.2
            \item indep test = ParCorr
        \end{itemize}
    \item CD-NOD
        \begin{itemize}
            \item $\alpha$ = 0.05
            \item indep test = fisherz
        \end{itemize}
\end{itemize}

\section{Limitation}
\label{appx:limitation}
The extensive referee bias effects can be estimated using more granular data. While opponents' fouls were sufficient for examining referee bias effects on home field advantage in this study, more detailed information on referee bias (e.g., free kicks, penalty kicks) may offer additional insights. 

The computational efficiency and effectiveness of DYNAMO are significantly influenced by the selection of a base learner. In this paper, the non-linear DYNAMO under NTS-Notear loss exhibits relatively slow running speeds compared to other state-of-the-art models. Exploring strategies to develop computationally efficient base learners for non-stationary causal processes could be beneficial.

Another limitation of our approach is the reliance on choosing between linear and nonlinear DYNAMO, which significantly impacts model results. In real-world datasets, the linearity or nonlinearity of the underlying structure is often unknown. Simulation results have shown that linear DYNAMO performs well in linear datasets but poorly in nonlinear ones, and vice versa for nonlinear DYNAMO. Therefore, future efforts should focus on developing a base learner capable of identifying both linear and nonlinear causal relationships effectively.







\bibliographystyle{chicago}

\bibliography{Bibliography-MM-MC.bib}